\documentclass[twocolumn]{aastex62}

\usepackage{graphicx}
\usepackage{ulem}
\usepackage{url}
\usepackage{threeparttable}
\usepackage{xcolor}
\usepackage{soul}
\usepackage{booktabs}
\usepackage{natbib}
\usepackage{amsmath}
\usepackage{mathrsfs}
\hypersetup{colorlinks, linkcolor=black, anchorcolor=green, citecolor=blue}
\usepackage{txfonts}

\usepackage{rotating}
\usepackage{longtable, needspace}

\usepackage[]{subfigure}

\usepackage{multirow}

\received{}
\revised{}
\accepted{2019 Aug. 10}
\submitjournal{ApJ}

\shorttitle{The chemical structure of young high-mass star-forming clumps: (I) Deuteration}
\shortauthors{Feng et al.}
\begin{document}

\title{The chemical structure of young high-mass star-forming clumps: (I) Deuteration \footnote{}}

\correspondingauthor{Siyi Feng}
\email{siyi.s.feng@gmail.com}

\author[0000-0002-4707-8409]{S. Feng (Chinese Name)}
\affil{National Astronomical Observatory of China, Datun Road 20, Chaoyang, Beijing, 100012, P. R. China}
\affil{CAS Key Laboratory of FAST, NAOC, Chinese Academy of Sciences, P. R. China}
\affil{National Astronomical Observatory of Japan, 2 Chome-21-1 Osawa, Mitaka-shi, Tokyo-to 181-8588, Japan}

\author{P. Caselli}
\affil{Max-Planck-Institut f\"ur Extraterrestrische Physik, Gie{\ss}enbachstra{\ss}e 1,  D-85748,  Garching bei M\"unchen, Germany}

\author{K. Wang  (Chinese Name)}
\affil{Kavli Institute for Astronomy and Astrophysics, Peking University, 5 Yiheyuan Road, Haidian District, Beijing 100871, China}

\author{Y. Lin  (Chinese Name)}
\affil{Max Planck Institute for Radio Astronomy, Auf dem H\"ugel 69, D-53121 Bonn, Germany}

\author{H. Beuther }
\affil{Max-Planck-Institut f\"ur Astronomie, K\"onigstuhl 17,  D-69117,  Heidelberg, Germany}

\author{O.  Sipil\"a}
\affil{Max-Planck-Institut f\"ur Extraterrestrische Physik, Gie{\ss}enbachstra{\ss}e 1,  D-85748,  Garching bei M\"unchen, Germany}
%% Note that the \and command from previous versions of AASTeX is now
%% depreciated in this version as it is no longer necessary. AASTeX 
%% automatically takes care of all commas and "and"s between authors names.

%% AASTeX 6.2 has the new \collaboration and \nocollaboration commands to
%% provide the collaboration status of a group of authors. These commands 
%% can be used either before or after the list of corresponding authors. The
%% argument for \collaboration is the collaboration identifier. Authors are
%% encouraged to surround collaboration identifiers with ()s. The 
%% \nocollaboration command takes no argument and exists to indicate that
%% the nearby authors are not part of surrounding collaborations.

%% Mark off the abstract in the ``abstract'' environment. 
\begin{abstract}

The chemical structure of high-mass star nurseries is important for a general understanding of star formation. Deuteration is a key chemical process in the earliest stages of star formation {because its efficiency is sensitive to the environment.}
Using the IRAM-30\,m telescope at 1.3--4.3\,mm wavelengths, we have imaged two parsec-scale high-mass protostellar clumps (P1 and S) that  show different evolutionary stages but are located in the same giant filamentary {infrared dark cloud} G28.34+0.06.
Deep spectral images at subparsec resolution reveal the dust and gas physical structures of both clumps. 
We find that (1) the low-$J$ lines of $\rm N_2H^+$, HCN, HNC, and $\rm HCO^+$ isotopologues are subthermally excited;
and (2) the deuteration of $\rm N_2H^+$ is more efficient than that of $\rm HCO^+$, HCN, and HNC by an order of magnitude. The deuterations of these species are enriched toward the chemically younger clump S compared with P1, indicating that this process favors the colder and denser environment ($\rm T_{kin} \sim14 K$, $\rm N(NH_3) \sim 9\times 10^{15}\,cm^{-2}$).
In contrast, single deuteration of $\rm NH_3$ is insensitive to the  environmental difference between P1 and S;
and (3) single deuteration of $\rm CH_3OH$ ($\rm > 10\%$) is detected toward the location where CO shows a depletion of $\sim10$.
This comparative chemical study between P1 and S links the chemical variations to the environmental differences and shows chemical similarities between the early phases of high- and low-mass star-forming regions.
\end{abstract}


%% From the front matter, we move on to the body of the paper.
%% Sections are demarcated by \section and \subsection, respectively.
%% Observe the use of the LaTeX \label
%% command after the \subsection to give a symbolic KEY to the
%% subsection for cross-referencing in a \ref command.
%% You can use LaTeX's \ref and \label commands to keep track of
%% cross-references to sections, equations, tables, and figures.
%% That way, if you change the order of any elements, LaTeX will
%% automatically renumber them.
%%
%% We recommend that authors also use the natbib \citep
%% and \citet commands to identify citations.  The citations are
%% tied to the reference list via symbolic KEYs. The KEY corresponds
%% to the KEY in the \bibitem in the reference list below. 

%______________________________________________________________
\section{Introduction}

{ 
Unlike high-mass protostellar sources, which are characterized by a high temperature and rich spectra, 
dense regions ($\rm n\ge10^3-10^5~cm^{-3}$) where protostellar objects have not yet formed
 are less well characterized  \citep[e.g., ][]{teyssier02,rathborne06,butler09,ragan09}.  The main reason for this is the low temperature  \citep[$\rm T<20\,K$, e.g., ][]{sridharan05,pillai06,wangy08,wienen12,chira13}, which translates into a lower degree of excitation in gas-phase molecular lines.
In low-mass prestellar cores and high-mass starless cores \citep[cores have the mass reservoir to form a high-mass star in the prestellar phase,][]{mckee03,tan14}, heating or feedback activities are not present, so the signatures of the early kinematic and chemical properties are not yet destroyed. Therefore, the natal molecular clouds, which harbor low-mass prestellar cores, high-mass starless cores, as well as high-mass young stellar objects, usually have low luminosity, showing high dust extinction even at 70\,$\mu$m wavelength \citep[e.g., ][]{ragan12,ragan13}. The so-called {infrared dark clouds}  \citep[IRDCs, e.g., ][]{cyganowski08,robitaille08,carey09} have provided excellent road maps for studying these dense and cold clumps, especially the initial conditions for high-mass star formation (HMSF, \citealt{jimenez10, butler12,peretto13,henshaw16}).

The column density ratio (relative abundance) between the deuterated %($\rm N_{deu}$) 
and hydrogenated isotopologues %($\rm N_{hyd}$) 
of the same species, denoted $D\rm(hyd)=\chi(deu/hyd)$,   is one of the most important chemical tools for diagnosing whether a particular core is in the prestellar or protostellar phase. 

Formed at the birth of the Universe, deuterium (D) is  slowly destroyed in the interiors of the stars, and mainly exists in molecular clouds in the form of  HD \citep[e.g., ][]{millar89,ceccarelli14}.
Atomic deuterium can be unlocked from the HD reservoir by the cosmic ray-driven ion-molecule chemistry at low temperature and high densities, starting from the isotopic exchange reactions with $\rm H_3^+$ and leading to $\rm H_2D^+$, $\rm D_2H^+$, and $\rm D_3^+$ \citep[e.g., ][]{caselli02b,crapsi05,vastel06,chenhr10}. The deuterated ions can then donate deuterons to other species; this process is called deuterium fractionation.

{%In the initial stages of star formation,  gravitational collapse increases gas density. 
The low-mass prestellar cores and high-mass cores in the earliest star-forming stage provide low-temperature and high column density environments, where gaseous heavy-element-bearing molecules such as CO start to freeze out onto the cold dust grains. As a result,  the deuterium fractionations of several molecules are enhanced \citep[][]{bacmann03,crapsi05,ceccarelli07} toward these environments, showing higher D-fraction $D\rm(hyd)$ than the isotopic ratio  $\mathcal{R}_{\rm D/H}$ in the environment with little star formation \citep[e.g., $\rm \sim10^{-5}$ in the diffuse interstellar medium (ISM), ][]{oliveira03,linsky06,prodanovic10} by several orders of magnitude.
After the young stellar object(s) are formed and warm the natal clouds up, molecules in the grain mantles especially CO, evaporate into the gas phase, destroying the deuterated isotopologues, so D-fraction of several species should decrease. 
}

%{\color{red}Paola: [please note that ÒfractionationÓ refers to the process, while ÒfractionÓ should be used instead when referring to abundance ratios]}
{ Comparing the D-fraction of various species between low-mass prestellar and protostellar objects \citep[e.g., ][]{crapsi05,emprechtinger09}, as well as between the high-mass clumps at the cold, young phase and more evolved phases \citep[e.g., ][]{pillai06,sakai08,chenhr10,fontani11,miettinen11,vasyunin11,fontani14,gerner15,kong15}, previous observations found enhanced $D \rm (N_2H^+)$ at low temperature, which is consistent with theoretical predictions \citep[e.g.,][]{caselli12,Albertsson_ea13}. }
In contrast to the low-temperature enhancement of $D \rm (N_2H^+)$, recent  single-dish pointing observations \citep[e.g., ][]{fontani15} show that, statistically, $D(\rm CH_3OH)$ is a good probe of the earliest protostellar phases, while $D(\rm NH_3)$ does not show  significant changes when the high-mass protostellar objects evolve. The deuterium fractionation efficiency seems species dependent, which may be due to their different gas-grain forming pathways.

The above conclusion is, however, not solid  because of the limited number of species observed by previous works. {  We also note that  high-mass starless objects, which are used to compare with high-mass protostellar objects, may not be well classified. In fact, the evolutionary status of these sources classified by  the traditional %IR classification method, i.e., based on their 
spectral energy distribution (SED) method is affected by large uncertainty due to the lack of continuum emission at IR wavelengths. 
Recent line observations at high angular resolution and high sensitivity have detected young stellar objects in candidates of ``starless clumps", invalidating the prestellar nature of these regions. For example, outflows have been detected toward young high-mass cores with bolometric luminosity-to-mass-ratios less than $\rm 1L_{\odot}/M_{\odot}$  \citep[][]{feng16b,tan16}.}
 More high-mass protostellar objects are detected toward  70\,$\mu$m dark clouds  \citep[e.g., ][]{csengeri17,sanhueza17}, which have been called as ``high-mass starless clump" candidates in previous works. These examples suggest that truly high-mass starless objects have a short lifetime or do not exist.
%{\color{red}Paola, can we use this term everywhere, instead of ÒprestellarÓ?
%The definition of ÒprestellarÓ cores for the moment only applies to low-mass objects.
%Ke, change to ``young stellar objects in "prestellar" clumps, invalidating the prestellar nature of these regions. This suggests that truly high-mass prestellar objects have a short lifetime or do not exist.(ref)"}

 Therefore,  conclusions from the previous observations of deuterated molecules may be affected by the wrong assumptions about % may be biased by the selection rule of the sample, such as imprecision in 
 the evolutionary phase of individual sources and  unresolved star formation activities in  single-dish pointing observations. A more accurate deuterium fractionation study requires  the exploration of more species toward a sample of sources, for which we have well characterized  their physical structures (i.e., the internal sources, the temperature profile, and the density profile) as well as knowledge of the dynamics (i.e., the star-forming activities and the evolutionary phases).}\\

\noindent {\bf \large Targets: }
G28.34 P1 and S is a pair of  high-mass clumps ($\rm H_2$ column density $\rm 10^{24}\,cm^{-2}$) 
  located in the filamentary IRDC G28.34+0.06 (distance to the Galactic center ${\it D}_{\rm GC}=$4.7\,kpc,\citealp{wangk18}).  Using  the IRAM-30\,m, NOEMA, SMA, and ALMA, we have %characterized their physical structures and 
  found that this pair of sources comprises different evolutionary stages {\citep[e.g.,][]{wangk12,feng16b}.}

Clump {S} is located at  the 70\,$\mu$m extinction peak of the filament.  On a scale of 0.8\,pc, it shows low dust temperature  (14--16 K, \citealp{wangy08}), low luminosity ($\rm 10\,L_{\odot}$, \citealt{ragan13}), and high CO depletion \citep{feng16a},  indicating that it is extremely young. However, on a scale of 0.08\,pc, we found extended bipolar red-shifted and blue-shifted lobes from SiO (2--1) and (5--4) emissions, indicating a young outflow \citep[i.e., less energetic than the outflows from more evolved high-mass protostellar objects by a factor of at least 10, ][]{feng16b,kong18}.

Clump {P1} is an IR-bright source \citep[$\rm \sim10^3\,L_{\odot}$, 18--28\,K, ][]{wangk11} located $\rm 1.5^{'}$ northeast of S.  Several intense CO outflows ($\rm 10^{-5}\,M_{\odot}\,yr^{-1}$,\citealt{wangk11,zhang15}),  water maser, and $\rm CH_3OH$ maser detections indicate that the young stellar objects embedded in P1 are more evolved than those embedded in S \citep{wangk12,wangk15b}. 

{ The P1 and S clump pair, showing evolutionary difference in the same molecular cloud, provides a good laboratory for studying the initial conditions of HMSF. In particular, by characterizing their dust and gas properties, we will be able to investigate the effects of  environmental variations on the physio-chemical processes at different evolutionary phases. }

In this paper, we present a detailed D-fraction study toward the G28.34 P1--S region incorporating data from IRAM-30\,m observations. In Section~\ref{obs} we summarize the observations and the data quality. 
We present the beam-averaged spectra toward P1 and S in Section~\ref{id} and show the maps of line spatial distributions in Section~\ref{distribution}. 
We characterize the physical structures of P1 and S in Section ~\ref{physic}, and present the D-fraction maps of six species in Section~\ref{column}. 
%, and discuss the uncertainties in Section~\ref{uncertainty}. 
%We compare the  D-fractions in different species, and compare our results with the previous studies quantitatively in Section~\ref{variation}. 
In Section~\ref{codepletion} we discuss  the spatial correlation between the CO depletion and deuterium fractionation  for different species. 
Finally, a summary of our main results can be found Section~\ref{conclusion}.

%  {\color{red}Referee: The authors mention a set of observations used to characterized G28.34, i.e. IRAM-30m, NOEMA, SMA and ALMA data. Nevertheless, only IRAM-30m and NOEMA data are discussed in the paper.\\
%  \bf Siyi, the observations other than IRAM-30m are discussed in Wang et al. 2011, Zhang et al.2015, Feng et al. 2016b, and references mentioned in section 1.
%}

 %%%%%%%%%%%%%%%
\section{Observations}\label{obs}
\subsection{IRAM-30m}
From 2014 May  to 2017 September, we conducted an line imaging survey of our targets with the IRAM-30\,m telescope at 1.3--4.3\,mm. Observations were performed in the on-the-fly (OTF) mode, mapping two $1.5'\times1.5'$ regions, centered at $\rm 18^h42^m50^s.740$, $\rm -04^{\circ}03^{'}15^{''}.300$ (J2000) and $\rm 18^h42^m46^s.597$, $\rm -04^{\circ}04^{'}11^{''}.930$ (J2000) for P1 and S, respectively. 
The broad bandpass of EMIR covers the 16\,GHz bandwidth simultaneously for each spectral tuning. { By superpositioning different spectral tunings, the frequency range covers 70.718--78.622, 82.249--101.981,
130.698--138.481, 151.818--175.481, and 217.019--224.800\,
GHz.} Using the FTS200 backend, we achieve a frequency resolution of 0.195\,MHz (corresponding to $\rm 0.637\,km\,s^{-1}$ at 93.173\,GHz).
The angular resolution of the IRAM-30\,m telescope is, for example, $\sim11.9\arcsec$ at 218.760\,GHz (Table~\ref{tab:dehyd}).
We had good  weather conditions during the observations ({the radiometer opacity $\rm \tau$ at 255GHz is 0.16--0.38}), and we used Saturn, 1749+096, or 1741-038 for pointing and focus.
Using the corresponding forward efficiency ($\rm F_{eff}$) and a main-beam efficiency ($\rm B_{eff}$) at individual frequencies, \footnote{http://www.iram.es/IRAMES/mainWiki/Iram30\,mEfficiencies}  we converted the data from
antenna temperature ($\rm T_{A}^*$) to main-beam temperature ($\rm T_{mb}=F_{eff}/B_{eff}\times T_{A}^*$).   
We used the GILDAS\footnote{http://www.iram.fr/IRAMFR/GILDAS} software package for data reduction and  line identification. \footnote{The ``Weeds'' is an extension of GILDAS for line identification  \citep{maret11}.} The $\rm1\sigma$ rms $\rm T_{mb}$ in the line free channels are 10.4--12.9\,mK at 1.3--1.4\,mm, 2.9--7.6\,mK at 1.7--2.0\,mm, 11.4--16.4\,mK at 2.1--2.3\,mm, 2.8--6.6\,mK at 2.9--3.6\,mm, and 3.1--3.6\,mK at 3.8--4.3\,mm.

{ 
\subsection{Archival data}\label{arch}
Furthermore,  with the aim of characterizing the physical structure of the P1--S region, we use the following archival data:

The continuum data are obtained from {\it Herschel} Infrared
Galactic Plane (Hi-GAL) survey at 70/160/250/500\,$\mu$m (\citealp{molinari10}, obsID: 1342218694),  from {\it CSO/SHARC-II} at 350\,$\mu$m (PI: H. B. Liu), as well as from the combination of {\it Planck} and James Clerk Maxwell telescope {\it (JCMT) -SCUBA2} at 850\,$\mu$m (Program ID: M11BEC30). The data quality is described in \citet{lin17}.

The $\rm NH_3\,(\it J, \it K \rm)$=(1,1) and (2,2) lines are observed with the Very Large Array (VLA; \citealp{wangy08}) and Effelsberg-100\,m single-dish telescope \citep{pillai06}. {Continuum emission at 1.3\,cm toward our source is not detected.} After a combination in the UV-domain  \citep[][]{wangy08,wangk18}, the data achieve an angular resolution of $\rm \sim5\arcsec$ and a velocity resolution of $\rm 0.6\,km\,s^{-1}$. 
%The combination process and the data quality is described in. 
}
%{\color{red} Beuther: I would add here a paragraph outlining which archival data are taken, from NH3 to ATLASGAL, Herschel and SHARC data ....\\
%Paola, cite and describe Table 1}

%%%%%%%%%%%%%%%%%%%%%%%%%%%%%
\section{Results}\label{result}
In this paper, we focus on the study of D-fraction by analyzing several pairs of  hydrogenated and deuterated molecules that are detected in our IRAM-30\,m line imaging survey. 

A line is considered detected if its main-beam temperature is $\rm >4\sigma$.  Table~\ref{tab:dehyd} lists all the detected lines  in this paper. Their spectroscopic parameters are taken from the Cologne Database for Molecular Spectroscopy \citep[CDMS, \footnote{https://cdms.astro.uni-koeln.de}][]{mueller02,muller05} or from the Jet Propulsion Laboratory  \citep[JPL, \footnote{http://spec.jpl.nasa.gov}][]{pickett98}.

\begin{table*}[]
%\scalebox{0.9}{
\caption{Transitions Detected %from pairs of hydrogenated and deuterated molecules 
Toward G28.34\,P1--S
}\label{tab:dehyd}
\centering
\begin{tabular}{llllllll}
%\begin{longtable*}{lllllll}

\hline\hline
Mol. &Freq.  &Transition    &$\rm S\mu^2$$^a$         &$\rm E_{\it u}/k_B$$^a$   &$n_{crit}^e$  &$n_{eff}^f$ &Beam   \\
         &(GHz) &                  &($\rm D^2$)             &  (K)             &$\rm (\times10^5\,cm^{-3})$     &$\rm (\times10^5\,cm^{-3})$   &($\rm \arcsec$)      \\    
\hline
%HCN   &88.632      &J=1--0$^b$   & 26.8    &4.2  &10.1 &29.3\\
$\rm H^{13}CN$   &86.340      &J=1--0$^c$   & 26.7    &4.1  &9.3 &2.2--3.5  &30.0\\
$\rm H^{13}CN$   &172.678     &J=2--1$^c$   & 53.4    &12.4  &89.0 &8.6--19.0  &15.0\\
$\rm HC^{15}N$   &86.055     &1--0   & 8.9    &4.1 &9.8  &$--$ &30.1\\
DCN  &72.415    &J=1--0$^c$    & 26.8    &3.5  &5.5  &$--$ &35.8\\
\hline
%HNC &90.664     &1--0   & 9.3    &4.3 &2.8     &28.6\\
$\rm HN^{13}C$   &87.091    &1--0      &7.3    &4.2  &1.9  &$--$  &29.8\\
$\rm H^{15}NC$  &88.866     &1--0     &7.3    &4.3  &2.0 &$--$   &29.2\\
DNC   &76.306     &1--0    & 9.3    &3.7  &1.7  &$--$ &34.0\\
\hline
%$\rm HCO^+$  &89.189    &1--0   &15.2   &4.2  &1.5    &29.1\\
$\rm H^{13}CO^+$  &86.754    &J=1--0$^b$     &15.2   &4.2   &1.4 &0.3--0.4  &29.9\\
$\rm H^{13}CO^+$  &173.507  &J=2--1$^b$     &30.4   &12.5 &28.4  &1.0--1.7  &15.0\\
$\rm HC^{18}O^+$ &	85.162   &1--0       &15.2   &4.1  & 1.0  &$--$   &30.5\\
$\rm DCO^+$  &72.039    &J=1--0$^b$  & 14.5   &3.5 &2.6   &$--$   &36.0\\
\hline
$\rm N_2H^+$ &93.173     &J=1--0$^c$   &104.0     &4.5 &1.5 &0.1  &27.8\\
$\rm N_2D^+$ &77.109    &J=1--0$^c$  &104.0   &3.7 &0.5  &$--$  &33.6\\
$\rm N_2D^+$ &154.217   &J=2--1$^c$  &208.1     &11.1&9.2  &$--$   &16.8\\
\hline
$\rm NH_3$$^d$  &23.694  	&$\rm 1_{1}\,0a\text{--}1_{1}\,0s$   &6.6  &23.8 &0.1 &0.01 &\\
$\rm NH_2D$  &85.926   	&$\rm 1_{1,1}\,0s\text{--}1_{0,1}\,0a$   &28.6  &20.7  &39.0 &$--$  &30.2\\
\hline
$\rm CH_3OH$  &76.510  &$\rm 5_{0 , 5}\text{--}4_{1 , 3}\,E$   &1.9 &47.9 &0.4 &$--$  &33.9\\
$\rm CH_3OH$ &84.521   &$\rm 5_{-1, 5}\text{--}4_{0 , 4}\,E$$^g$  &3.1    &40.4 &3.0 &$--$  &30.7\\
$\rm CH_3OH$ &95.169   &$\rm 8_{0, 8}\text{--}7_{1, 7}\,A$$^g$  &7.2 &83.5 &1.6 &$--$  &27.3\\
$\rm CH_3OH$ &95.914   &$\rm 2_{1, 2}\text{--}1_{1, 1}\,A$  &1.2  &21.4 &0.5 &$--$  &27.0\\
$\rm CH_3OH$ &96.739   &$\rm 2_{-1, 2}\text{--}1_{-1, 1}\,E$   &1.2    &12.5 &82.5 &$--$  &26.8 \\
$\rm CH_3OH$ &96.741   &$\rm 2_{0,2}\text{--}1_{0,1}\,A$ &1.6  &7.0   &0.5 &$--$    &26.8\\
$\rm CH_3OH$ &96.745   &$\rm 2_{0 , 2}\text{--}1_{0 , 1}\,E$  &1.6 &20.1&3.4  &$--$  &26.8\\
$\rm CH_3OH$ &157.246  &$\rm 4_{0 , 4}\text{--}4_{-1, 4}\,E$   &4.2 &36.3 &60.0 &$--$  &16.5\\
$\rm CH_3OH$ &157.271  &$\rm 1_{0 , 1}\text{--}1_{-1, 1}\,E$  &1.5 &15.4 &1.8 &$--$  &16.5\\
$\rm CH_3OH$ &157.272  &$\rm 3_{0 , 3}\text{--}3_{-1, 3}\,E$  &3.3 &27.1 &32.5 &$--$   &16.5\\
$\rm CH_3OH$ &157.276  &$\rm 2_{0 , 2}\text{--}2_{-1, 2}\,E$  &2.4 &20.1&1678.4 &$--$   &16.5\\
$\rm CH_3OH$ &170.061  &$\rm 3_{2 , 1}\text{--}2_{1 , 1}\,E$  &3.1 &36.2 &1963.1 &$--$  &15.3\\
$\rm CH_3OH$ &218.440  &$\rm 4_{2 , 2}\text{--}3_{1 , 2}\,E$$^g$  &3.5 &45.5 &1.6 &$--$  &11.9\\

$\rm CH_2DOH$   &89.408    &$\rm 2_{0,2}\text{--}1_{0,1}\,e0$ &1.2  &6.4 &0.2  &$--$   &29.0\\

$\rm ^{13}CH_3OH$  &94.405  &$\rm 2_{-1 , 2}\text{--}1_{-1 , 1}\,E$   &1.2  &12.4 &82.5 &$--$  &27.5\\
$\rm ^{13}CH_3OH$  &94.407  &$\rm 2_{0 , 2}\text{--}1_{0 , 1}\,A$   &1.6  &6.9 &0.4 &$--$  &27.5\\
\hline

%%%%%%%%%%%%%%%%%%%%%%%%%%%%%
$\rm  H_2CO$  &  72.838 &$\rm 1_{0,1}\text{--}0_{0,0} $  &5.4  &3.5 &1.6 &0.3--0.5 &35.6\\
$\rm  H_2CO$  &  218.222 &$\rm 3_{0,3}\text{--}2_{0,2} $  &16.3  &21.0 &28.8 &2.9--7.7 &11.9\\
$\rm  H_2CO$  &  218.476 &$\rm 3_{2,2}\text{--}2_{2,1}  $  &9.1  &68.1 &12.1 &$--$ &11.9\\
$\rm  H_2CO$  &  218.760 &$\rm 3_{2,1}\text{--}2_{2,0} $   &9.1  &68.1 &25.9 &$--$ &11.9\\
\hline \hline
\multicolumn{8}{l}{{\bf Note.} $^a$ Line spectroscopic parameters are given according to catalogs including the JPL and CDMS;}\\
\multicolumn{8}{l}{~~~ ~~~ ~~$^b$ Hyperfine splittings are recorded in JPL and CDMS but not resolved in our observations,}\\
\multicolumn{8}{l}{ ~~~ ~~~  ~~~ so the sum of $\rm S\mu^2$ is used for the rotational transitions to calculate the total column density;}\\
\multicolumn{8}{l}{~~~ ~~~ ~~$^c$ Hyperfine splittings are resolved in our observations, and only the sum of $\rm S\mu^2$ is  needed for the }\\
\multicolumn{8}{l}{ ~~~ ~~~  ~~~ rotational transitions to calculate the total column density;}\\
\multicolumn{8}{l}{~~~ ~~~  ~~$^d$ Data are from the combination of VLA and Effelsberg, with a combined synthesized beam }\\
\multicolumn{8}{l}{~~~ ~~~  ~~~  as $\rm \small 6.04\arcsec\times3.57\arcsec$ (position angle as $\rm -10.20^\circ$);}\\
\multicolumn{8}{l}{~~~ ~~~  ~~$^e$ The critical density of each transition $n_{crit}$ is derived from the Einstein coefficient $A_{ij}$ }\\
\multicolumn{8}{l}{ ~~~ ~~~  ~~~  and the collision rate $C_{ij}$ at 5--20\,K given by LAMDA \citep{schoier05}. We assume that }\\
\multicolumn{8}{l}{ ~~~ ~~~  ~~~  the deuterated lines have the same $C_{ij}$ as their hydrogenated counterparts;}\\
\multicolumn{8}{l}{~~~ ~~~  ~~$^f$ The effective excitation density at kinetic temperature of 10--15\,K from \citet{shirley15},}\\
\multicolumn{8}{l}{ ~~~ ~~~  ~~~   ``$--$" labels the non-recorded value;}\\
\multicolumn{8}{l}{~~~ ~~~ ~~$^g$ Known interstellar Class I methanol maser transitions \citep[see][ and the references therein]{leurini16}.}\\

\end{tabular}
%}
%\end{longtable*}
\end{table*}

\subsection{Line profiles}\label{id}
Lines list in Table~\ref{tab:dehyd} in general have high critical densities ($\rm >10^4\,cm^{-3}$), high effective excitation density \citep{shirley15}, and low $\rm E_{\it u}/k_B$ (except for some $\rm CH_3OH$ lines), 
and thus are dense and cold gas tracers. 
These lines are observed at different angular resolutions, allowing us to trace the gas properties of each clump on scales from 0.8\,pc down to 0.2\,pc.

To improve the signal-to-noise ratio (S/N) as well as to gauge the gas kinematics directly  from the observations, 
we extract the beam-averaged spectra of each line from the 870\,$\mu$m dust continuum peak P1 and S, without smoothing their native velocity or angular resolutions. %(Figure~\ref{tab:dehyd}).

For $\rm HCO^+$, we note that the   $\rm ^{12}C$ line is affected by self-absorption, as also found by \citet{feng16a}. For HCN and HNC, the main isotopologue $\rm ^{12}C$ line profiles show that multiple velocity components are present, as emission from various Galactic spiral arms is detected \citep{beuther07c}.  Because we detect the $\rm ^{13}C$, $ \rm ^{18}O$, or/and   $ \rm ^{15}N$ isotopologue lines of these species, we exclude the $\rm ^{12}C$ lines from discussion in this paper.

%{\color{red}Referee: Did you extract the spectra from a region with the same beam dimensions?\\
%\bf Siyi: To improve the signal-to-noise ratio (S/N) as well as to gauge the gas kinematics directly from the observations, the spectra shown in this Section and Figure 1are extracted  from the observations, without smoothing their native velocity or angular resolutions. However, to estimate the molecular column densities, we smooth the lines for each species to the same angular resolution as described in Section 3.3.  }

\begin{figure*}
%\begin{sidewaysfigure*}

%\rule{5cm}{10cm}
\centering
\includegraphics[angle=0,scale=0.65] {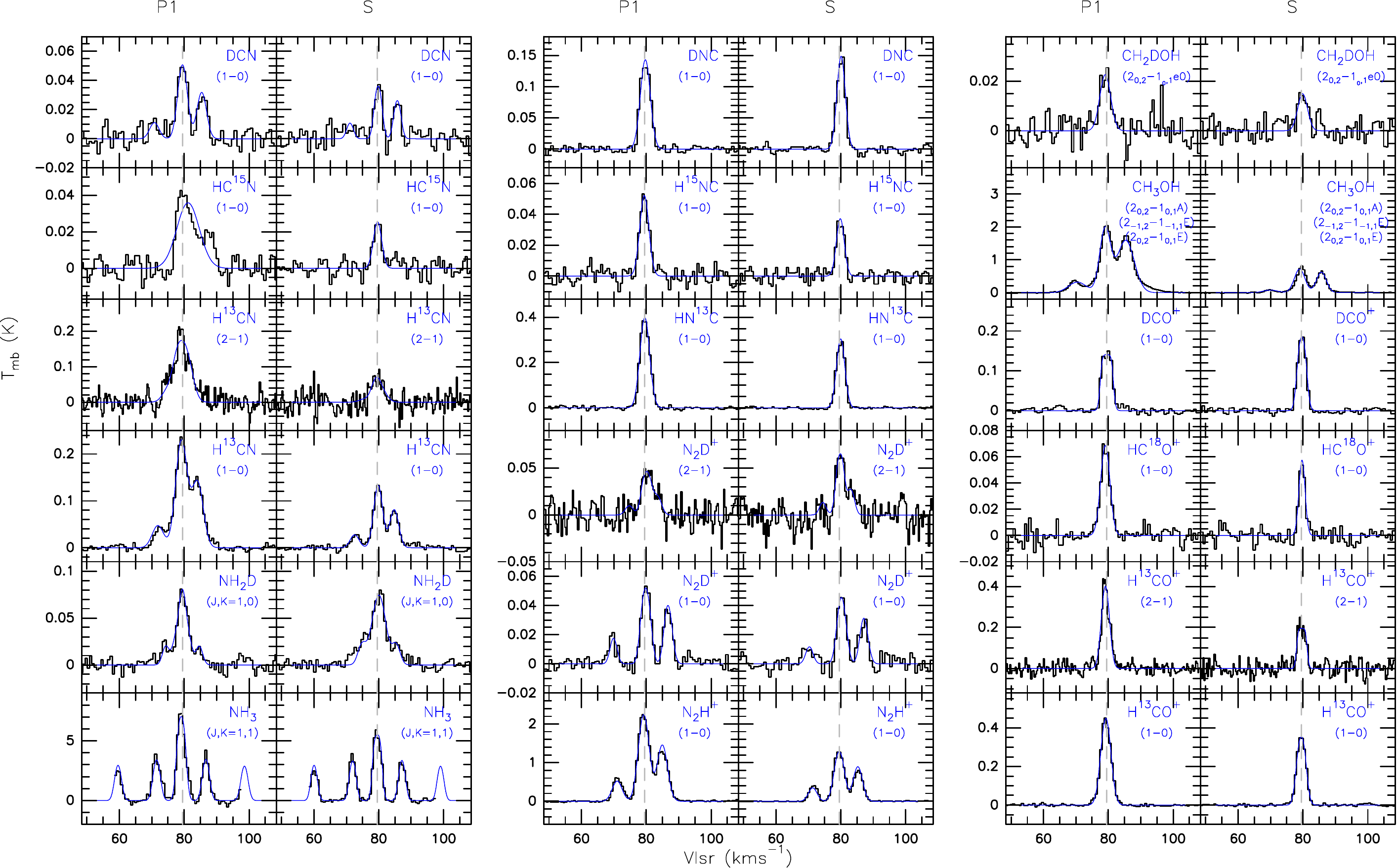}

\caption{Beam-averaged spectra of identified lines from deuterated-hydrogenated isotopologue pairs extracted from P1 and S. The positions are labeled at the top of each column. All lines are extracted from images that we regridded to the same pixel size,  but whose native angular and velocity resolution we kept as in the observations (see beam information in Table \ref{tab:dehyd}). The best-fit parameters by using the GAUSS or HFS methods are listed in Table~\ref{tab:linedeu} and Table~\ref{tab:lineh2coch3oh}, and the synthetic spectrum for each line is plotted in blue. The system velocity ($\rm 79.5\,km\,s^{-1}$) is shown as a gray dashed line in each panel.
 %from GAUSS fits includes the velocity at line intensity peak $\rm V_{peak}\,(km\,s^{-1})$, the FWHM line width $\rm \Delta V\,(km\,s^{-1})$, and the integrated intensity for a single component $\rm Area\,(K\,km\,s^{-1})$. The best-fit parameters from HFS fits include the $\rm V_{lsr}\,(km\,s^{-1})$, the optical depth $\rm \tau_0$, the FWHM line width $\rm \Delta V\,(km\,s^{-1})$, and the measured intensity $\rm A\tau_0\,(K\,km\,s^{-1}$) of the main line in resolved hyperfine splittings.\\
{\color{red}
%Ke, show vsys as vertical line to guide eyes
%referee: remove the text from the figures, these are already too small and difficult to read as it is a 42 panels figure. In addition, instead of plotting everything, the authors should select the most relevant tracers and remove (or put in an appendix) the others. In this way the figures could be placed larger. For instance, HNC, HCN, and HCO+ present self-absorption features, i.e. they are not useful for the analysis the authors performed. Of course the authors can assume that all the lines are optically thin (i.e., second bullet point in Section 3.5), but considering that in some cases the self-absorption is well visible, I suggest - at least - to discuss this point in the paper, using lower-limits in the column densities, or an upper-limit in the deuteration.
}
}\label{velpro}
%\end{sidewaysfigure*}
\end{figure*}

The spectra of key transitions detected from the hydrogenated-deuterated isotopologue pairs toward P1 and S are shown in Figure~\ref{velpro}; the fits have been done by using   the GILDAS package. 
All these transitions show a single velocity component at the given velocity resolution. If hyperfine splitting is resolved, we fit it by using the HyperFine-Structure (HFS)  method, %\footnote{Fitting for the optical depth of the main line is only valid when $\rm \tau>0.1$.}, 
and derive the excitation temperature\footnote{We use a subscript to specify the excitation temperature derived from different $J$ levels ($\rm T_{\Delta J}$), ($J$, $K$) levels ($\rm T_{\Delta J, \Delta K}$), and $F$ levels ($\rm T_{\Delta F}$, $\rm T_{\Delta F1, \Delta F}$). } $\rm T_{\Delta F}$ of the transition, if optically thick ({Table~\ref{tab:tex},} see \citealp{feng16a} for details). 
%{\color{red}why the authors define the excitation temperature as $T_{\Delta F}$, and not as usual as $T_{ex}$?}
For the hyperfine splittings that are neither recorded in JPL/CDMS  nor resolved in our observations, the line is fit by using the Gaussian (GAUSS) method. 
In the case that hyperfine splitting is recorded but is unresolved from our observations, we compare the fitting results by using the HFS and 
GAUSS methods, and find that broadening of the line width due to the hyperfine splitting is negligible.
%{\color{red}(Paola, did you take into account the broadening due to the hyperfine splitting? No)}
The best-fit parameters are listed in Table~\ref{tab:linedeu}--\ref{tab:lineh2coch3oh}.

In general, { the molecular lines we study in this work show the same centroid velocities ($\rm \sim79.5\,km\,s^{-1}$) toward both P1 and S.
Their line widths, in terms of the full width at half maximum (FWHM), }%line widths of the lines %we study in this work 
are in the  3--5\,$\rm km\,s^{-1}$ range toward P1, and slightly ($\rm <20\%$) narrower toward S. 
{In each clump, lines from different isotopologues of the same species have a similar FWHM, indicating that they trace the same bulk of gas.
Specifically,  when we compare the beam-averaged line profile of these molecules between S and P1, the hydrogenated molecular lines show stronger emission at their centroid velocities  toward P1 than toward S by a factor of 1.5--2, which is consistent with the nature that P1 is chemically and physically more evolved. 
In contrast,  the emission intensities  at the centroid velocities of the deuterated molecular lines do not follow a unique enhancement trend from S to P1. }
The diversity of the measured D-fraction enrichment is discussed later in Section \ref{distribution} and Section~\ref{variation}.

\subsection{Line spatial distribution maps}\label{distribution}
\begin{figure*}
 \centering
\includegraphics[angle=0,scale=0.9]{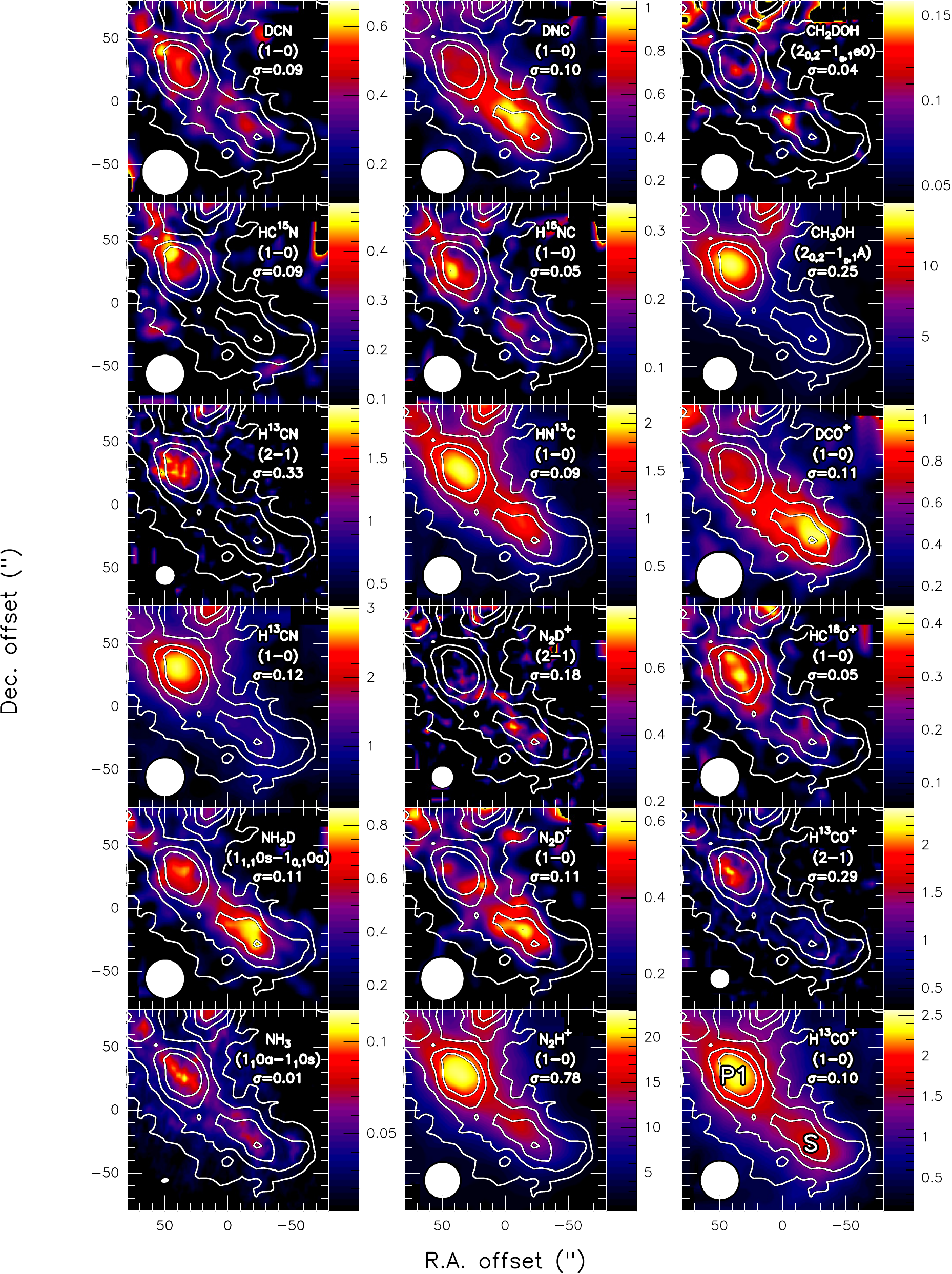}
\caption{Molecular line distributions shown as color maps. The line intensities are integrated over their entire velocity dispersion (listed in  Table~\ref{tab:linedeu}) and shown in units of main-beam temperature as $\rm K\,km\,s^{-1}$. The transition of each line and   $\rm 1\sigma$ rms ($\rm K\,km\,s^{-1}$) of each image   are given, with the angular resolution shown as a beam in white in the bottom left corner of each panel. % The pixels where the line shows $\rm<3\sigma$ integrated intensity are blanked in each panel.  
The white contours show the continuum emission from $870\,\mu$m APEX data  at an angular resolution of 18.2\arcsec \citep{schuller09}, starting from $5\sigma$ and increasing by $5\sigma$ ($\rm \sigma=0.042\,Jy\,beam^{-1}$).  
The P1 and S denominations are shown in the bottom right panel.
{\color{red}
%referee: also, in this case, the not-relevant tracers should be removed, in particular, the species which are below the 3-sigma detection. A couple of comments: the NH3 panel misses the beam size, and ii) the authors should consider to mask everything below 3-sigma for better readability of the plots. Once removed some of the tracers the authors should make these figures larger and place them horizontally.
}
}\label{molint}
\end{figure*}

To compare the spatial distribution of  deuterated and hydrogenated molecules, the line  intensities are  integrated over the same velocity range  (listed in  Table~\ref{tab:linedeu}), \footnote{In the case that a particular line shows resolved hyperfine splittings, we integrate its intensity over the entire velocity range, which covers all the hyperfine components.} and they are shown  as color maps in Figure \ref{molint}.
To compare the spatial correlation between the dust and gas distribution,  the continuum emission is overlaid as white contours on each of the line maps. Because P1 and S are dark at wavelengths shorter than 70\,$\mu$m \citep{feng16c}, we chose the 870\,$\mu$m APEX continuum emission \citep{schuller09} as representative of the dust distribution.

%{\color{red}Referee: I suppose that the superposition of the 870 micron APEX contours is plotted to show the dust distribution, but this is not written in the text. I suggest starting this sentence by explaining why the knowledge of the dust distribution is important and then explain why the 870 micron APEX observations were chosen.\\
%Siyi, I am not sure if the sentence in bold is sufficient to explain or not.}

In clump P1, 
all the lines of hydrogenated molecules show their
emissions at a {\color{black}local maximum} position  coincident with the 870\,$\mu$m continuum peak, while in clump S,  these molecules show emission at a {\color{black}local maximum}  position $\rm \sim20\arcsec$ northeast (NE) to the 870\,$\mu$m continuum peak. 

The deuterated molecules present their strongest emissions at different locations within the P1--S region: 
(1) Most lines, for example $\rm N_2D^+$\,(1--0), DNC\,(1--0), and  $\rm NH_2D\,(1_{1,1}0s\text{--}1_{0,1}0a)$, show the strongest emission at $\sim\rm 20\arcsec$ offset NE of clump S, where $\rm ^{13}CO$, $\rm C^{18}O$, and $\rm C^{17}O$ show high depletion \citep[][also see Section~\ref{codepletion}]{feng16a}; 
(2) DCN (1--0) shows the strongest emission toward P1; 
(3) Although we are not able to confirm the location where $\rm CH_2DOH\, (2_{0,2}\text{--}1_{0,1})$ shows the strongest emission due to $\rm S/N<4$ in most pixels, the emission maximum appears at  $\rm \sim25\arcsec$ offset NE of clump S.

We note that an offset of 20\arcsec~ corresponds to  2/3  of the beam at 3\,mm.
Because these lines are observed simultaneously, the different locations of their individual emission maxima is likely the result of their different distributions, rather than any uncertainty caused by the pointing error between the IRAM-30\,m (for molecular lines) and APEX (for the dust continuum) telescopes.
Hereafter, we denote this location (20\arcsec~ NE of S) as ``Soff".

%{\color{red}Referee; - general comment for the Figures: in most of the figures, it is very difficult to read both the captions inside the plots, both axes and the numbers related to the colour bar. In general, the size of the text of the figures must be at least the same as the text of the paper. In addition, it is suggested to change the colour scale, using one that favours the readability of images (e.g. viridis, inferno, ...).

%Overall, I think that if the authors will focus on a logical re-shape of the draft, following the above suggestions, but also organizing better the information they want to give to the community, the work will improve a lot. The number of figures must be drastically reduced, the quality improved, and the text shortened and carefully revised.

%\bf Siyi: We remove 60\% of the subfigures, and enlarge the text, axes, and the color scale (especially in Figure 2). We reorganize Section 3.3 and 3.4 in a more logical way.
%}

%\section{Analysis}\label{analysis}
\section{Physical structure}\label{physic}
We have detected deuterated isotopologues of six species: HCN, HNC, $\rm HCO^+$, $\rm N_2H^+$, $\rm NH_3$, and $\rm CH_3OH$. They are believed to have different gas-grain formation routes \citep[e.g.,][]{fontani15}.
To understand how  different environments affect the deuterium fractionation process, it is essential to characterize the  physical structure of the source, i.e., via reconstructing the gas and dust temperature as well as density profiles.

\subsection{Dust temperature and density maps}
Following the iterative SED fitting procedure described in \citet{lin17}, {we establish a reliable blackbody model and obtain the dust opacity index $\beta$ map at an angular resolution of 22\arcsec~ using the archival {\it Herschel} PACS 70, 160, {\it Herschel} SPIRE 250 maps, the $\rm 350$\,$\mu$m map by combining {\it Herschel} SPIRE and CSO/SHARC-II, as well as the $\rm 850$\,$\mu$m map by combining the deconvolved Planck 353\,GHz and JCMT-SCUBA2 data (Section~\ref{arch}). 
Then, assuming the $\beta$ map has no local variation from 22\arcsec~ to 10\arcsec~ resolution, we apply the Monte Carlo method to fit the continuum data from {\it Herschel} PACS at 70 $\mu$m and from {\it CSO/SHARC-II} at 350\,$\mu$m.  At an angular resolution of 10\arcsec, we obtain the maps of $\rm H_2$ column density  $\rm N(H_{2})$ and dust temperature T(dust) simultaneously toward P1--S.}
%To establish a reliable black-body model for the flux at different wavelengths in the P1--S region, and to obtain the dust opacity index $\beta$ pixel by pixel, we also use the continuum images at 70, 160, 250, 350, 500, and 850\,$\mu$m {\it Herschel}, {\it CSO/SHARC-II}, {\it Planck}, and  {\it JCMT -SCUBA2} (Section~\ref{arch}).  
The detailed fitting procedure is also given in Appendix \ref{blackbody}.
%%%%%%%%
%{\color{red}Sec. 4.1 is still confusing. It is not clear to me if the authors performed a two frequencies fit, i.e. they are fitting a black body with two points, or if they are doing a fit with more points as explained some sentences later. This part should be rewritten in a clearer way.}
%%%%%%%%%

\begin{figure*}[b]
\begin{center}
\begin{minipage}[c]{.4\textwidth}
\subfigure[]{\includegraphics[height=5cm] {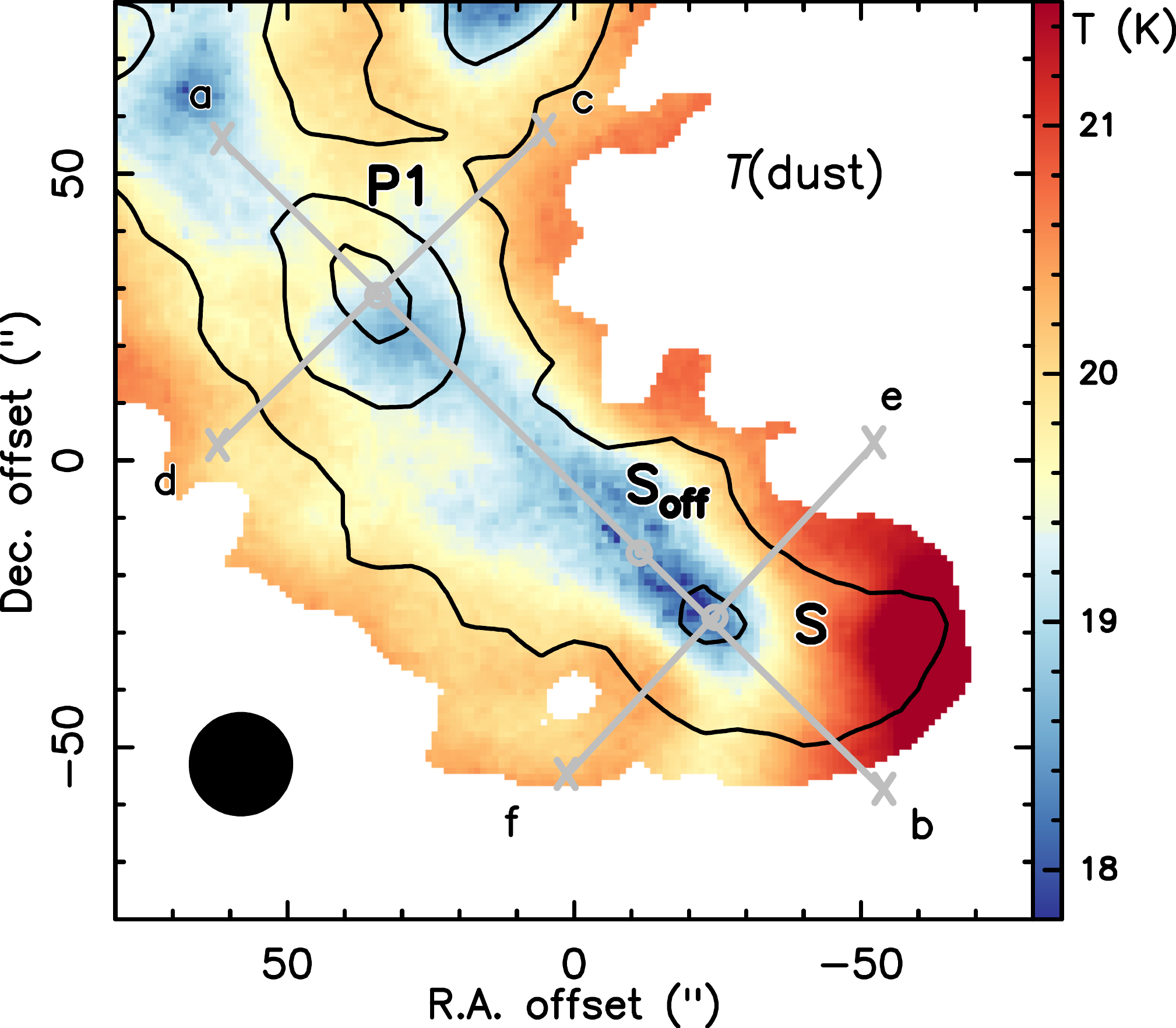}}
\end{minipage}
%\begin{minipage}[c]{.28\textwidth}
%\subfigure[]{\includegraphics[height=4cm] {errmap-TDust-SED-eps-converted-to.pdf}}
%\end{minipage}
\begin{minipage}[c]{.5\textwidth}
\subfigure[]{\includegraphics[height=5cm] {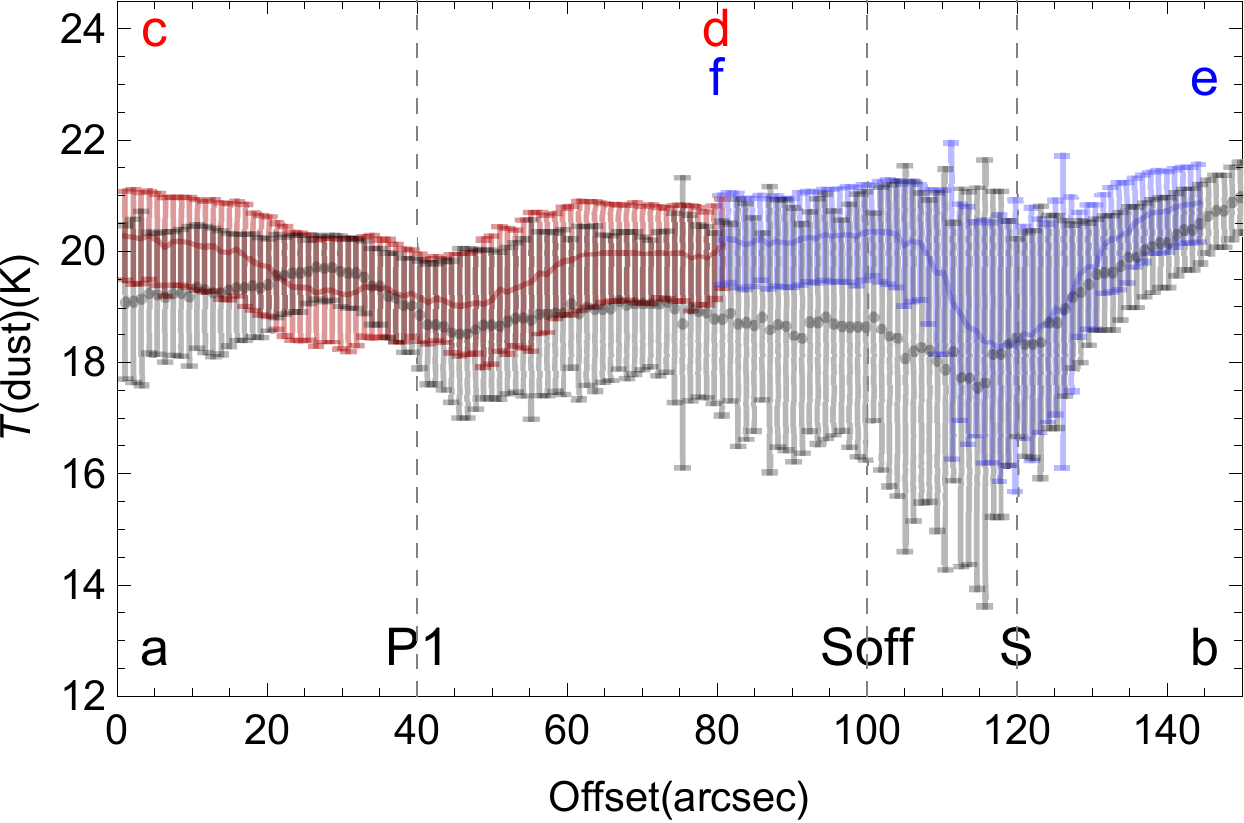}}
\end{minipage}
\\
\begin{minipage}[c]{.4\textwidth}
\subfigure[]{\includegraphics[height=5cm] {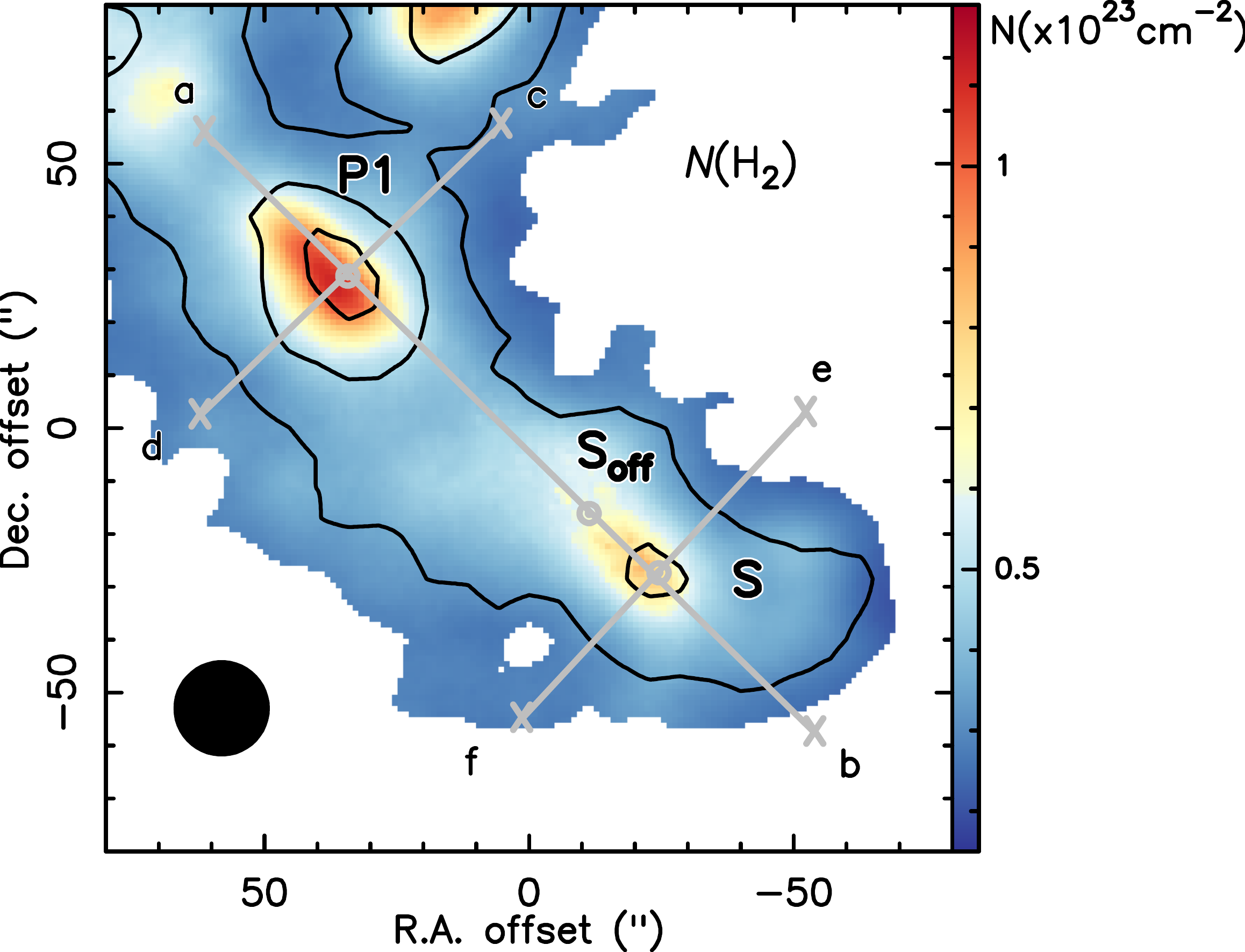}}
\end{minipage}
%\begin{minipage}[c]{.28\textwidth}
%\subfigure[]{\includegraphics[height=4cm] {errmap-NDust-SED-eps-converted-to.pdf}}
%\end{minipage}
\begin{minipage}[c]{.5\textwidth}
\subfigure[]{\includegraphics[height=5cm] {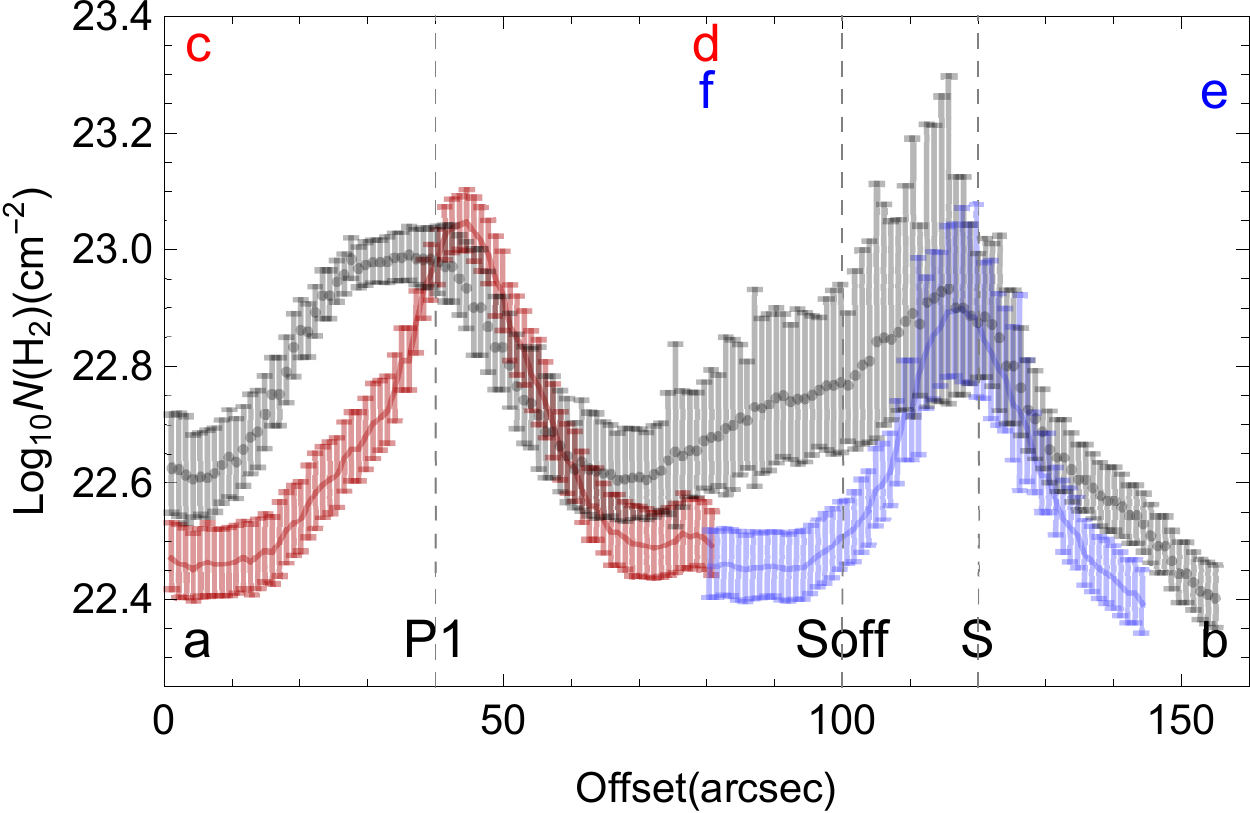}}
\end{minipage}

\end{center}

\caption{Dust temperature and $\rm H_2$ column density maps for G28.34\,P1--S. Maps are derived by iterative SED fits to the data from {\it Herschel} PACS at 70 $\mu$m and from {\it CSO/SHARC-II} at 350\,$\mu$m, achieving an angular resolution of 10\arcsec \citep{lin17}. The continuum images at 70, 160, 250, 350, 500, and 850\,$\mu$m from  {\it Herschel}, {\it CSO/SHARC-II}, {\it Planck}, and  {\it JCMT-SCUBA2} are used to establish a reliable blackbody model pixel by pixel. 
%To establish a reliable black-body model, and to obtain the dust opacity index pixel by pixel, we also use the continuum images from {\it Herschel} at 70/160/250/500\,$\mu$m, from the combination of  {\it Herschel} and {\it CSO/SHARC-II} at 350\,$\mu$m, as well as from the combination of {\it Planck} and  {\it JCMT -SCUBA2} at 850\,$\mu$m.
{\it Panel I}:  dust temperature map.
{\it Panel II}: dust temperature profile extracted along the the filamentary elongation, plotted in black, with the ends labeled as ``a" and ``b"; dust temperature profiles extracted along the  perpendicular direction to the filamentary elongation is plotted in red (from point ``c" over the 870\,$\mu$m continuum peak P1 to point ``d") or blue (from point ``e" over the 870\,$\mu$m continuum peak  S to point ``f"). 
{\it Panel III}: $\rm H_2$ column density map.
{\it Panel IV}: $\rm H_2$ column density profile in the directions of a--b (in black), c--d (in red), and e--f (in blue). 
The pixels with $\rm<5\sigma$ continuum emission at 870\,$\mu$m are blanked. The black contour of the 870\,$\mu$m continuum emission starts from $\rm 10\sigma$ and increases in steps of  $\rm 10\sigma$ ($\rm \sigma=0.042\,Jy\,beam^{-1}$).
The angular resolution achieved in SED fits is shown in black in the bottom left corner of  {\it Panels I, III}.
}\label{dust}
\end{figure*}

As shown in Figure~\ref{dust}I and \ref{dust}III, P1 and S cannot be modeled as spherical sources at a linear resolution of 0.2\,pc; instead,  they  show 2D Gaussian structures in the plane of the sky. Therefore, instead of extracting the temperature and column density profiles averaged along the radii of each clump, hereafter we extract the physical and chemical parameters along three directions  in the plane of the sky: (1) the filamentary elongation along P1 and S, (2) perpendicular to the filament elongation with the center on 870\,$\mu$m continuum peak P1, and (3) perpendicular to the filament elongation with the center on 870\,$\mu$m continuum peak S. Given that the separation between P1 and S is only  $\rm \sim80\arcsec$ (1.8\,pc) in the plane of the sky, the gas and dust properties along the P1--S ridge of the filament may be influenced by both clumps, while those in the perpendicular direction and centered on a particular clump are likely less impacted by the distant clump. To visualize the mutual effects of clumps P1 and S on each other, we set the starting and end point of each direction (points labeled as  ``a", ``b", ``c", ``d", ``e", ``f")  as an offset of $\rm \pm40\arcsec$ to the center P1 or S.

%%%%%
Inspecting Figure~\ref{dust}, we note a larger uncertainty of the $\rm H_2$ column density and  dust temperature toward S than toward P1. This is a compromise made to obtain the maps at a higher angular resolution by including the {\it Herschel} 70\,$\mu$m data in the SED fits. Nevertheless, our fits indicate that the P1--S region is dense\footnote{The average volume number density $n$ of the gas can be estimated from the source kinematic distance $D_{GC}$, the angular resolution $\theta$, and the $\rm H_2$ column density $\rm N(H_2)$ as $n=\frac{N\rm (H_2)}{\theta {\it D}_{GC}}$.} ($\rm \sim10^5\,cm^{-3}$) and cold ($\rm \sim19\,K$), with the $\rm H_2$ column density ($\rm \sim1.2\times10^{23}\,cm^{-2}$) toward P1  higher than that toward S by a factor of 2.

 We also note a decrease ($\rm \sim$4\,K) of the dust temperature from the outskirts of the filament to the center. Because the dust temperature derived from SED fitting traces the weighted average temperature along the line of sight \citep[e.g., ][]{sokolov18}, such a decrease can be explained considering that the heating from the protostellar objects P1 and S to their gas envelopes is comparably less efficient than the interstellar radiation impinging on the outskirts of this filament.

\subsection{Gas temperature and density maps derived from $p$-$\rm NH_3$}\label{gasmap}
The dust continuum map suffers from the contamination of the foreground and background dust.
In contrast to this, we can  integrate the line intensity of a gaseous species over its velocity dispersion around the system velocity of the target source, so that the gas emission map  toward the source can be well characterized by excluding the foreground/background contamination (Section~\ref{id}). 
%{\color{red}Paola, how do you know that the observed molecular line is not contaminated by foreground and background emission? You need to specify that the line selected is specifically tracing high density gas, otherwise it is not clear (e.g. CO low J lines are definitely contaminated by foreground and background emission!)}

{ 
CO isotopologues are relatively more abundant than the other species. Our observations imaged $\rm C^{18}O$\,(2--1) and $\rm C^{17}O$\,(2--1) lines with high S/N. They show similar line profiles to the lines we study in this work \citep[see ][]{feng16a}. Although 2--1 transitions of CO isotopologues are easily thermalized near their critical densities ($\rm 2\times10^4\,cm^{-3}$ at 10--20\,K, derived using data from the Leiden Atomic and Molecular Database, 
LAMDA\footnote{http://home.strw.leidenuniv.nl/$\sim$moldata}), CO is known to freeze out at an environment of $\rm T<20\,K$ and density slightly higher than a few $\rm 10^4\,cm^{-3}$ \citep[e.g., ][]{caselli99,fontani06, aikawa08}. Therefore, showing high depletion \citep[][see also Section~\ref{codepletion}]{feng16a}, it cannot be used as a high-density tracer.
}

%{\color{red} Sec. 4.2 Did the authors apply a subtraction of the background here? If not, can the authors explain why?\\
%Re: we are not sure if the referee refer to (1) a background subtraction to dust emission or (2) a continuum subtraction to the NH3 lines. For (2), there is no detectable centimeter continuum at the NH3 lines, because free-free emission in IRDCs, especially in the two clumps we observed, is negligible (Rosero et al. 2014ApJ...796..130R). NH3 line emission around the cloud systematic velocity is assured to originate from the clumps. For (1), all line-of-sight dust emission are projected, and there is no perfect way to determine background/foreground emission (Wang et al. 2015MNRAS.450.4043W). Therefore, we choose to use NH3 emission around Vlsr.\\
%Sec. 4.2. It would be very useful to the readers if the authors could report on the SNR. Same in Sec. 6.}
%\item {p-NH3:}
Compared with CO, N-bearing species are more resilient to depletion \citep[e.g.,][]{caselli99, bergin02,caselli02c,jorgensen04}. The (1,1) and (2,2) inversion transition lines of $\rm NH_3$, with their critical density of (1--2)$\rm\times10^3\,cm^{-3}$ \citep{shirley15} in the temperature range of 20--100\,K, provide a sensitive gas thermometer in the cold environment \citep[e.g., ][]{ho83,walmsley83,crapsi07,rosolowsky08,juvela11}. Given that the dipole transitions between its different K ladders are forbidden, the ortho ($K=3n$) and para ($K\rm \neq3\it n$)  spin states of $\rm NH_3$ behave as distinct species. An accurate gas kinetic temperature should be derived from either para ($p$) or ortho ($o$) lines \citep[e.g.,][]{ragan11,battersby14,wangk14,bihr15,svoboda16}.

\begin{figure*}
\begin{center}
%\begin{minipage}[r]{.25\textwidth}
%\subfigure[]{\includegraphics[height=4cm] {NH3-11-Tex-eps-converted-to.pdf}}
%\subfigure[]{\includegraphics[height=4cm] {err-NH3-11-Tex-eps-converted-to.pdf}}
%\end{minipage}
%\begin{minipage}[c]{.25\textwidth}
%\subfigure[]{\includegraphics[height=4cm] {NH3-11-Trot-eps-converted-to.pdf}}
%\subfigure[]{\includegraphics[height=4cm] {err-NH3-11-Trot-eps-converted-to.pdf}}
%\end{minipage}
\begin{minipage}[l]{.4\textwidth}
\subfigure[]{\includegraphics[height=5cm] {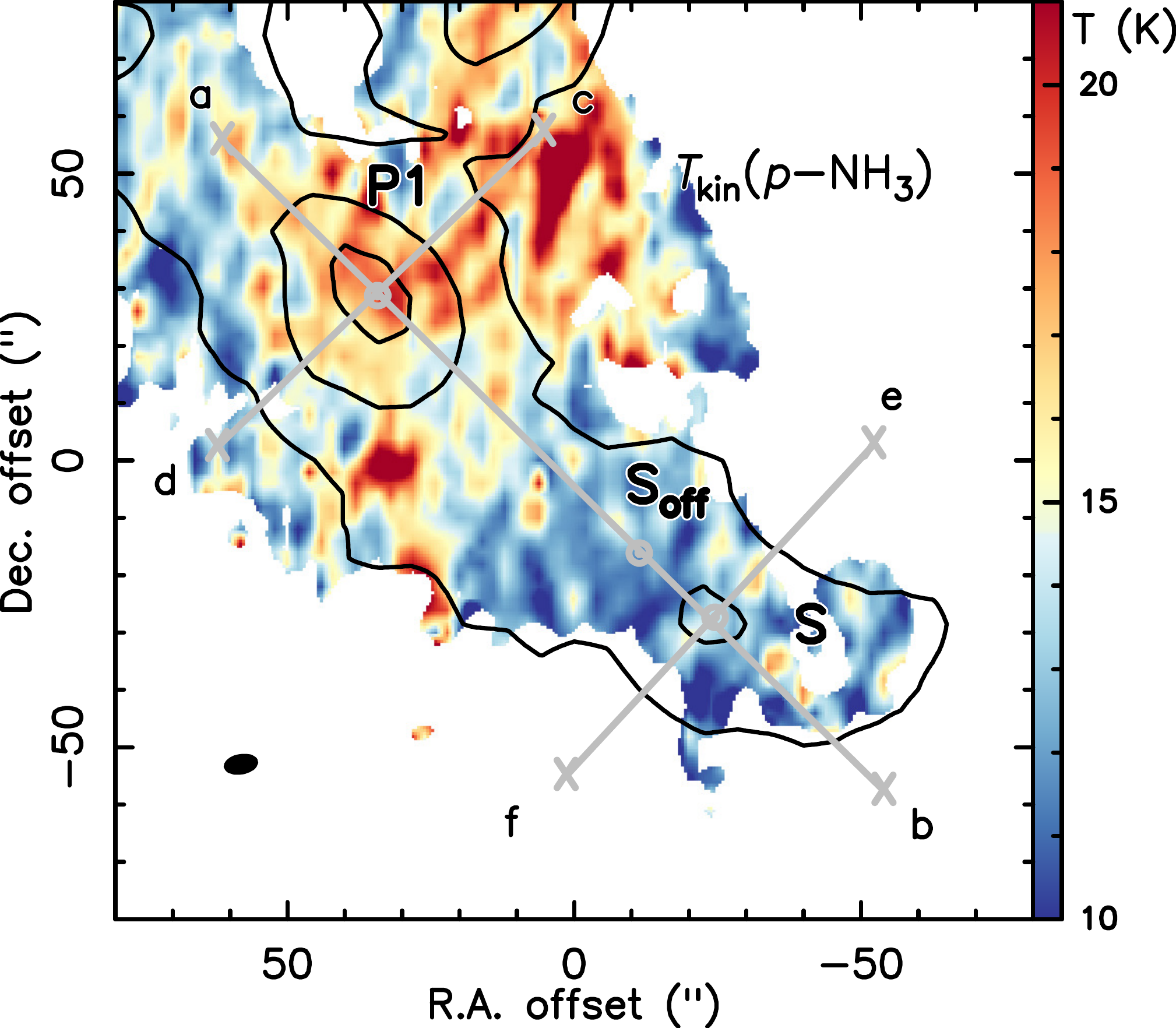}}
%\subfigure[]{\includegraphics[height=4cm] {err-NH3-11-Tk-eps-converted-to.pdf}}
\end{minipage}
%\\
%-----------------------------------------------------------------------------------------------------------------------------------------\\
%\begin{minipage}[r]{.25\textwidth}
%\subfigure[]{\includegÄraphics[height=4cm] {NH3-col-lower-eps-converted-to.pdf}}
%\subfigure[]{\includegraphics[height=4cm] {err-NH3-col-lower-eps-converted-to.pdf}}
%\end{minipage}
\begin{minipage}[c]{.5\textwidth}
\subfigure[]{\includegraphics[height=5cm] {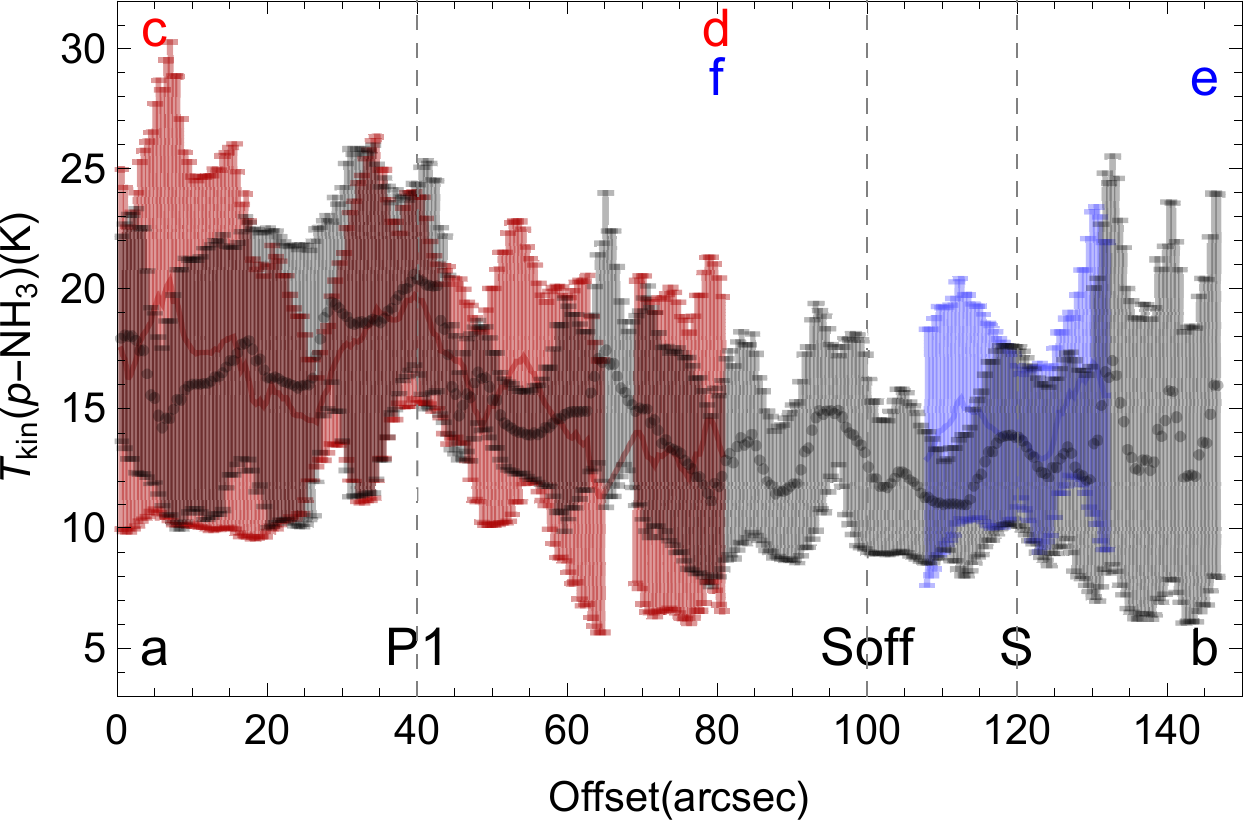}}
%\subfigure[]{\includegraphics[height=4cm] {err-NH3-col-upper-eps-converted-to.pdf}}
\end{minipage}
\begin{minipage}[l]{0.4\textwidth}
\subfigure[]{\includegraphics[height=5cm] {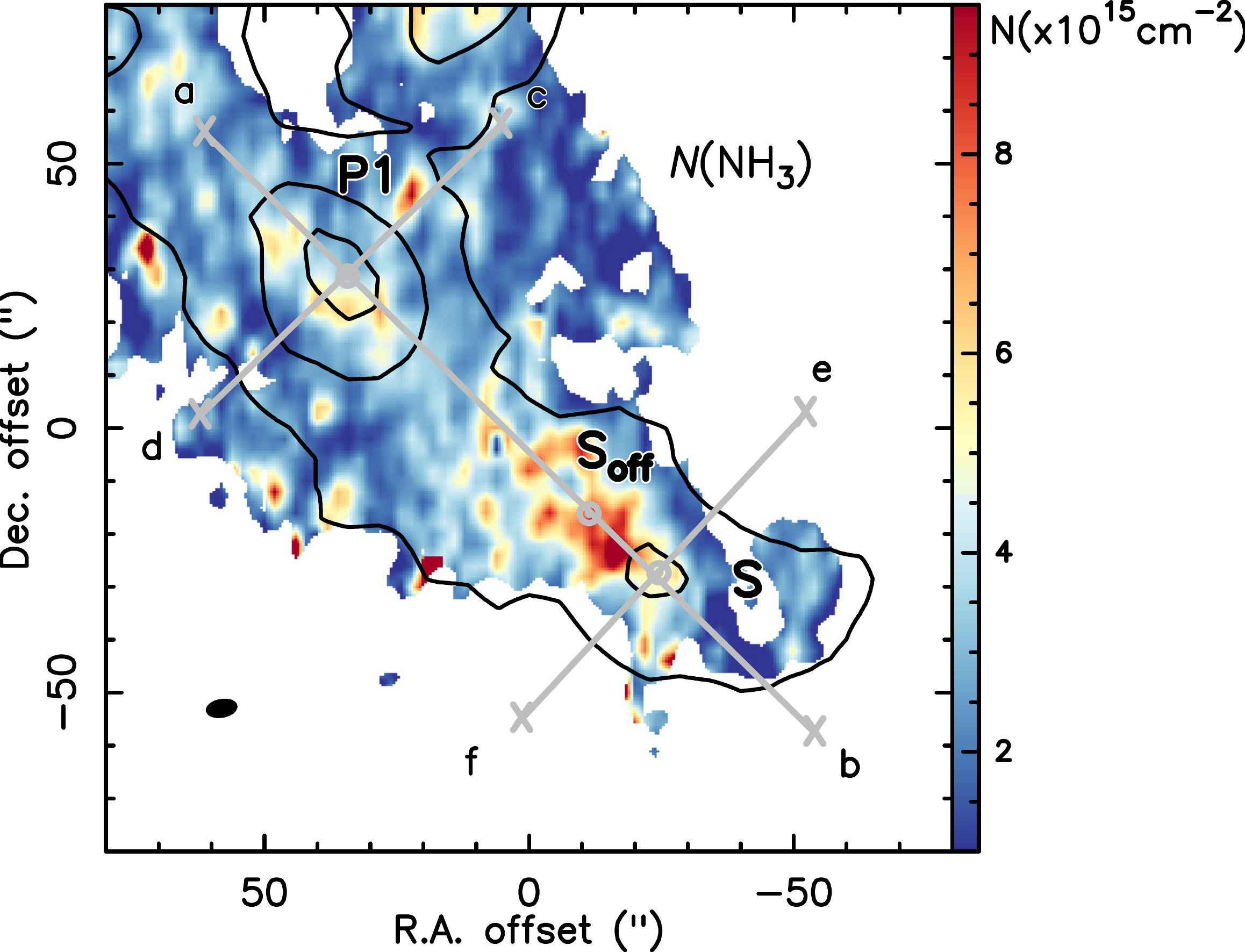}}
\end{minipage}
\begin{minipage}[l]{0.5\textwidth}
\subfigure[]{\includegraphics[height=5cm] {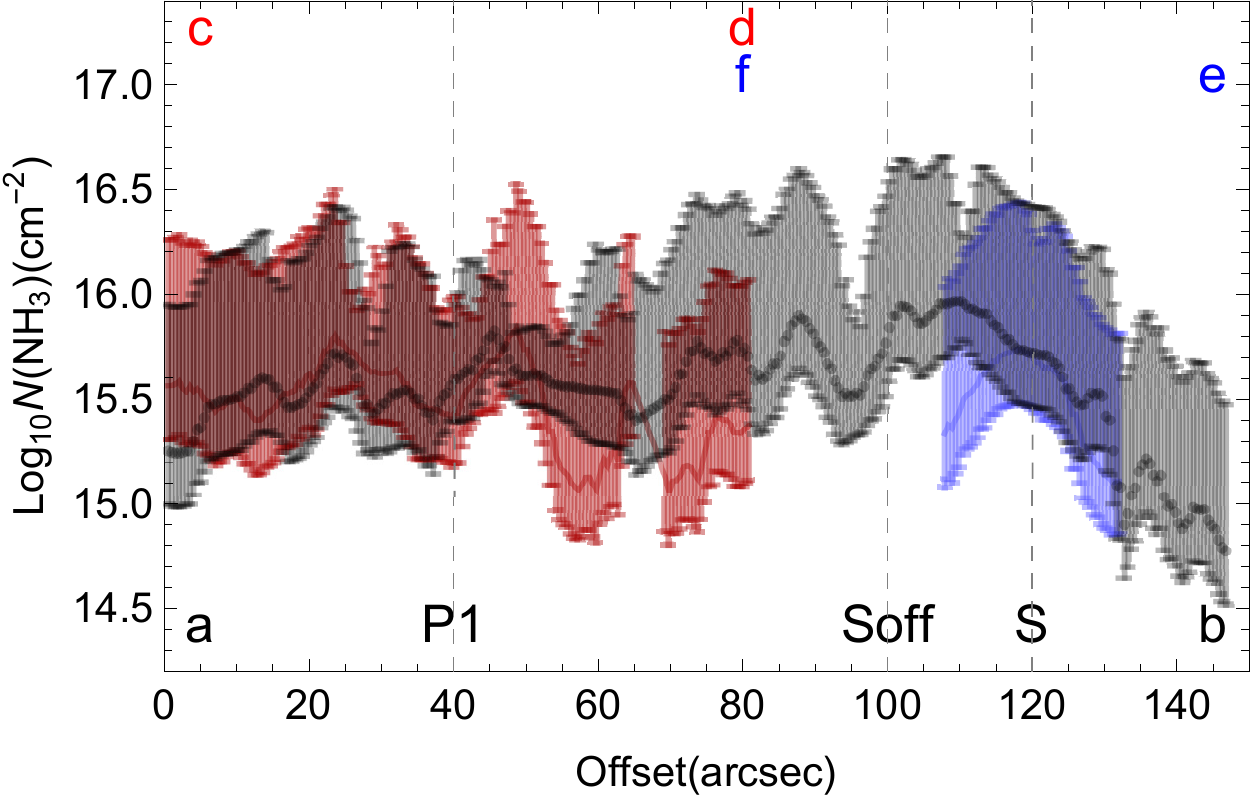}}
\end{minipage}
\end{center}
\caption{Gas kinetic temperatures and $\rm NH_3$ column densities derived from the multilevel system of $p$-$\rm NH_3$.
{\it Panel I}:  the gas kinetic temperature $\rm T_{kin}$ map.  
{\it Panel II}:  the gas kinetic temperature profiles in the directions of a--b (in black), c--d (in red), and e--f (in blue);
{\it Panel III}: the $\rm NH_3$ column density map with assumptions of no beam dilution  and an OPR of 1;  
{\it Panel IV}: the $\rm NH_3$ column density  profiles in the directions of a--b (in black), c--d (in red), and e--f (in blue). 
The black contour of 870\,$\mu$m continuum emission starts from $\rm 10\sigma$ and increases in steps of  $\rm 10\sigma$ ($\rm \sigma=0.042\,Jy\,beam^{-1}$).
The labeled positions are the same as those in Figure~\ref{dust}. 
{The pixels where the $\rm NH_3\,(\it J, \it K \rm)=(1,1)$ line shows $\rm<4\sigma$ integrated intensity are blanked in each panel.  }
The synthesized beam of the VLA-Effelsberg combined data is shown in the lower left corner of  {\it Panel I, III}.   
%{\color{red}Fig.3, I don't understand why the author place the text in the panel (right panels) instead of having in the axis label?! For instance the log10 (that should be anyway written as log10) is log10 without the quantity, better to move N(H2) in the axis label. Same for Tdust and for the other figures where the authors did the same.}
}\label{texnh3}
\end{figure*}

At an angular resolution of $\rm \sim5\arcsec$  and a velocity resolution of $\rm 0.6\,km\,s^{-1}$, the $\rm NH_3\,(\it J, \it K \rm)$=(1,1) and (2,2) lines obtained from the combination of  the VLA and Effelsberg-100\,m show one velocity component in the P1--S region. 
Applying the Monte Carlo fitting tool HfS developed by \citet{estalella17} to our VLA-Effelsberg combined data, \footnote{We compare the fitting results to our $\rm NH_3$\,(2,2)/(1,1) data by using HfS, CLASS/GILDAS, and PySpecKit \citep{ginsburg11}. Although the values of the fitting parameters from these softwares agree with each other, CLASS and PySpecKit seem to underestimate the uncertainty of the parameters  \citep{estalella17}.} we derive the value and uncertainty maps  for the following parameters: 
(1) the excitation temperature $\rm T_{(\it \Delta F_1,\Delta F)}$ for the (1,1) hyperfine splitting, derived from the main and satellite lines with different %angular momentum
quantum numbers ($\rm \Delta \overrightarrow{F_1}$, $\rm \Delta  \overrightarrow{F}$);
(2) the rotation temperature $\rm T_{(\it \Delta J,\Delta K)}$, derived from  the (1,1) and (2,2) inversion states, which are not radiatively coupled;
(3) the kinetic temperature $\rm T_{kin}$ (Figure~\ref{texnh3}I), based on the \citet{maret09} approximation, by taking the  collision transitions into account;
(4) the total column density $\rm N(NH_3)$ (Figure~\ref{texnh3}III), derived from  the (1,1) and (2,2) inversion states,  with an assumption of ortho-to-para ratio (OPR) of 1.

We found that $\rm T_{(\it \Delta F_1,\Delta F)}$ derived from (1,1) hyperfine splitting lines is close to the rotation temperature $\rm T_{(\it \Delta J,\Delta K)}$  (Table~\ref{tab:tex}) and the gas kinetic temperature $\rm T_{kin}$ at each pixel, the 3--5\,K difference between them is within the systematic uncertainty given by the Monte Carlo approach. Therefore, we can reasonably assume that $p$-$\rm NH_3$ lines are thermalized and under local thermodynamic equilibrium (LTE) conditions, and hence the constant excitation temperature approximation \citep[the CTEX method,][]{caselli02b} is valid for the multilevel system of $\rm NH_3\,(\it J, \it K \rm)=(1,1)$  
in our target region.

{On the one hand, we find that the gas kinetic temperature derived from $p$-$\rm NH_3$  (Figure~\ref{texnh3}II) has a larger dynamic range than the dust temperature ($\rm \sim19\,K$, Figure~\ref{dust}II) along the filament elongation \citep[see also in][]{wangk18}, with P1 ($\rm 20\pm3$\,K)  slightly warmer than S  ($\rm 14\pm3$\,K).
On the other hand, $\rm NH_3$ shows the column density maximum toward ``Soff" ($\rm \sim 10^{16}\,cm^{-2}$), higher than that toward S and P1 by a factor of 2 and 3, respectively (Figure~\ref{texnh3}IV), which is opposite to the trend found with the dust column density (Figure~\ref{dust}IV). Therefore, gas temperature and volume density may play an important role for the gas chemistry variations along the filament, from P1 to S.
}
\\

\begin{table}
\caption{Excitation temperature derived from hyperfine transitions or from line ratios
}\label{tab:tex}
\scalebox{0.9}{
\begin{tabular}{ccccc}
\hline\hline
Line   &T    &P1    &Soff   &S\\
&             &(K)    &(K)     &(K)\\
\hline
$p$-$\rm NH_3$(1,1)   &$\rm T_{(\it \Delta F_1,\Delta F)}$   &$\rm 20.0\pm 10.1 $ &$\rm 10.4\pm 13.2$  &$\rm 10.2\pm 1.0 $ \\
$p$-$\rm NH_3$(2,2)/(1,1)      &$\rm T_{(\it \Delta J,\Delta K)}$      &$\rm 18.0\pm 2.6$  &$\rm 13.5\pm 2.9$  &$\rm 14.0\pm 3.9 $\\
\hline
$o$-$\rm NH_2D$(1,0)      &$\rm T_{\Delta F}$    &$\rm 3.0\pm 2.4$  &$\rm 5.7\pm 2.9$  &$\rm 3.0\pm 2.4 $ \\
\hline
$\rm N_2H^+$(1-0)       &$\rm T_{\Delta F}$   &$\rm 8.1\pm 5.1$   &$\rm 6.0\pm 2.8$  &$\rm 5.2\pm 2.4$\\
$\rm N_2D^+$(2-1)/(1-0)       &$\rm T_{\Delta J}$   &$\rm 4.5\pm 1.2$   &$\rm 4.0\pm 0.5$   &$\rm 3.8\pm 0.5$ \\
\hline
$\rm H^{13}CN$(1-0)       &$\rm T_{\Delta F}$   &$\rm 8.7\pm 6.9 $    &$\rm 9.5\pm 7.4$ &$8.6\pm 5.2$  \\
$\rm H^{13}CN$(2-1)      &$\rm T_{\Delta F}$    &$\rm 3.5\pm 2.8 $  &$\rm 8.2\pm 5.6$ &$\rm --^a$ \\
$\rm DCN$(1-0)      &$\rm T_{\Delta F}$    &$\rm 4.2\pm 3.3 $ &$\rm 3.2\pm 1.8$   &$\rm 3.5\pm 1.5$  \\
$\rm H^{13}CN$(2-1)/(1-0)       &$\rm T_{\Delta J}$   &$\rm 3.8\pm 0.5$  &$\rm 3.2\pm 1.7$  &$\rm 3.0\pm 0.3$ \\
\hline
$\rm H^{13}CO^+$(2-1)/(1-0)       &$\rm T_{\Delta J}$    &$\rm 4.1\pm 0.3$ &$\rm 3.6\pm 0.3 $  &$--^a$\\
\hline

\multicolumn{5}{l}{{\bf Note.} $^a$ $\rm S/N<4$}\\
\end{tabular}
}

\end{table}

\subsection{Gas temperature from $p$-$\rm H_2CO$ and $\rm CH_3OH$}\label{h2cotemp}
%{\color{red} Section 3.3.4: The results of this paragraph are not used anywhere in the paper. This should be included in paragraph 3.3.3 (Gas temperature and column density from p-NH3 AND p-H2CO) and drastically shorten.}

Taking advantage of the broad bandwidth of our IRAM-30\,m observations, we  detected four $p$-$\rm H_2CO$ lines, three A-type $\rm CH_3OH$ lines and  eleven E-type $\rm CH_3OH$ lines, with different ($J$, $K$) levels and $\rm E_{\it u}/k_B$ spanning the range of 7--84\,K (Table~\ref{tab:dehyd}).
Both species are likely formed on the surface of dust grains by successive hydrogenation of CO \citep{watanabe02,woon02,hidaka04}, and are believed to be  precursors of large complex organic molecules \citep{barone15}.
In addition to $\rm NH_3$, both species are usually taken as typical thermometers \citep[e.g.,][]{mangum93,caselli93,johnston03,leurini04,leurini07,giannetti17}. Using different ($J$, $K$) level lines with higher critical density and effective excitation density than $\rm NH_3$\,(1,1) and (2,2), we can trace the temperature of denser gas %which is so low that the $\rm NH_3$ lines $\rm (E_{\it u}/k>23\,K)$ are not sensitive to anymore 
\citep[e.g., ][]{ao13,ginsburg16,tang18}.  %{\color{red}Paola, do not agree, because NH3 from crapsi 2007 detect temperature as low as 6K.}

For each species, we assume that all  lines are optically thin and under LTE. After smoothing all lines to the same angular resolution (Figure~\ref{h2cospec} and \ref{ch3ohspec}), {we use  the rotational diagram (RD) method \citep[see, e.g., ][]{feng15} to derive the rotation temperature ($\rm T_{rot}$) map as well as the molecular total column density ($\rm N_{rot}$) map  for a particular species}.
Although A-type and E-type $\rm CH_3OH$ cannot be interconverted in chemical reactions,  the partition functions of both types are the same \citep{rabli10}. Therefore, we do not separate them in the RD calculation.
$\rm T_{rot}$ of $p$-$\rm H_2CO$ and $\rm CH_3OH$ extracted from P1, S, and Soff in Table~\ref{tab:lvgfit} are  consistent with $\rm T_{kin}$ derived from $p$-$\rm NH_3$ (Figure~\ref{texnh3}).

Furthermore, we note that 
lines at 1\,mm, 2\,mm, and 3\,mm with different $\rm E_{\it u}/k_B$  may in fact trace gas from different layers of the source envelope. 
%{\color{red}Paola, can not you smooth before RD?}
Given that some low-$J$ lines might be optically thick, and the lines with high critical densities ($\rm >10^6\,cm^{-3}$) might be in non-LTE, we use the large velocity gradient (LVG) approximation to fit the $p$-$\rm H_2CO$, A-$\rm CH_3OH$, and E-$\rm CH_3OH$ lines individually.  

Using the statistical equilibrium radiative transfer code RADEX \citep{vandertak07} and a related solver (Fujun Du's myRadex\footnote{See https://github.com/fjdu/myRadex.}) as well as the MultiNest algorithm \citep{feroz07,feroz08,feroz13}, we derive the probability density function (PDF) of parameters, including the $\rm H_2$ number density $n$, molecular column density $\rm N_{col}$,  the gas kinetic temperature $\rm T_{kin}$, and the filling factor, toward P1, S, and Soff (see Figure~\ref{a-ch3ohlvg} as an example).
%{\color{red} show an example}

When we assume that all lines from the same species have the same  filling factors, our  results show that the filling factors are close to unity for both species. While A-/E-$\rm CH_3OH$ seem to trace a denser region ($\rm >10^5\,cm^{-3}$) compared to $p$-$\rm H_2CO$ ($\rm \sim 5\times10^4\,cm^{-3}$), $\rm T_{kin}$ from these tracers is consistent, showing a decrease from P1 (28--50\,K) to Soff (16--24\,K) and S (19--26\,K). 
 
We note that $\rm T_{kin}$ measured from  A-/E-$\rm CH_3OH$ and $p$-$\rm H_2CO$ at P1, S, and Soff are in general higher than their $\rm T_{rot}$. One possible reason for this is that these $\rm CH_3OH$ and $\rm H_2CO$ lines are subthermally excited in our sources \citep[e.g., ][]{kalenskii16}. %these molecular lines are sensitive to shocks \citep[e.g.,][]{tafalla10} {\color{red}Paola, explain}. 
Moreover, in P1 and S shock environment, some $\rm CH_3OH$ lines may be masering (e.g., at 84.521, 95.169,  and 218.440\,GHz in Table~\ref{tab:dehyd}), and this phenomenon cannot be solved by RADEX \citep{vandertak07}.
Furthermore,  $n$, $\rm N_{col}$, and $\rm T_{kin}$ are three free parameters in our LVG fit, and  the best-fit result has  an intrinsic degeneracy in the parameter combination. Therefore, we are not able to tell the precise value of individual parameters from the multiparameter fits.
% {\color{red}, rephrase: show strong degeneracy on their PDF toward all positions.}

Nevertheless, the molecular column densities toward individual positions from LVG are  consistent with the estimation under LTE assumption, showing an increase from S and Soff to P1 by a factor of 3--5. Therefore, we believe that the LTE-RD analysis provides a good approximation to LVG, in obtaining the gas temperature and molecular column density maps of $\rm H_2CO$ and $\rm CH_3OH$.

\begin{table*}

\caption{Mean and standard deviation of parameters from LVG  and LTE RD fittings
}\label{tab:lvgfit}
\scalebox{1}{
\begin{tabular}{cc|ccc|cc}
\hline\hline
& &\multicolumn{3}{c|}{LVG}  &\multicolumn{2}{c}{LTE}\\

Mol.   &Position    &$n\,\rm(10^5\,cm^{-3})$    &$\rm T_{kin}\,(K)$    &$\rm N_{col}\,(10^{14}\,cm^{-2})$  &$\rm T_{rot}\,(K)$    &$\rm N_{rot}\,(10^{14}\,cm^{-2})$\\
\hline
$p$-$\rm H_2CO$       &P1   &$\rm 0.5\pm0.1$    &$\rm 51.6\pm6.2$      &$\rm 0.3\pm0.0$  &$\rm 19.3\pm 6.5 $  &$\rm 0.8\pm 0.7$ \\
$p$-$\rm H_2CO$      &S     &$\rm 0.5\pm0.1$    &$\rm 26.4\pm2.0$      &$\rm 0.1\pm0.0$  &$\rm 14.3\pm 3.6 $   &$\rm 0.2\pm 0.1$ \\
$p$-$\rm H_2CO$       &Soff   &$\rm 0.5\pm0.2$    &$\rm 19.5\pm1.5$      &$\rm 0.1\pm0.0$  &$\rm 13.0\pm 2.9$ &$\rm 0.2\pm 0.1$ \\
\hline
A-$\rm CH_3OH$     &P1   &$\rm 4.8\pm0.9$    &$\rm 28.1\pm1.3$      &$\rm 4.9\pm0.4$  &\multirow{ 2}{*}{$\rm 23.4\pm 7.8$}  &\multirow{ 2}{*}{$\rm 2.9\pm 2.1$}\\
E-$\rm CH_3OH$          &P1   &$\rm 2.1\pm0.1$    &$\rm 40.6\pm2.8$      &$\rm 2.8\pm0.1$\\
A-$\rm CH_3OH$     &S    &$\rm 5.1\pm0.8$    &$\rm 22.4\pm0.8$      &$\rm 1.1\pm0.1$ &\multirow{ 2}{*}{$\rm 20.1\pm 5.8 $}  &\multirow{ 2}{*}{$\rm 0.8\pm 0.5$}\\
E-$\rm CH_3OH$           &S   &$\rm 1.3\pm0.5$    &$\rm 19.3\pm1.5$      &$\rm 0.9\pm0.1$\\
A-$\rm CH_3OH$     &Soff  &$\rm 2.9\pm0.4$    &$\rm 24.2\pm1.2$      &$\rm 1.0\pm0.1$  &\multirow{ 2}{*}{$\rm 19.2\pm 0.5$}  &\multirow{ 2}{*}{$\rm 0.7\pm 0.5$}\\
E-$\rm CH_3OH$          &Soff   &$\rm 1.2\pm0.0$    &$\rm 16.0\pm1.1$      &$\rm 0.9\pm0.1$\\
\hline

\end{tabular}
}

\end{table*}

\section{Molecular deuterium fraction maps}\label{column}
{ The environmental differences along the filament from P1 to S are small but not negligible. To quantify the effects of environmental changes on the deuterium fractionation of different species, molecular column densities need to be measured precisely.}

According to Figure~\ref{velpro},  isotopologue lines with the same $J$-level  from the same species have the same $\rm E_u/k_B$, and they show similar line profiles toward the same source.
 Therefore, after smoothing them to the same angular resolution, we assume that they trace the same gas, and that the excitation effect is excluded for any variations at different pixels.  Moreover,  most lines show extended emission in the plane of the sky, so we assume that they do not suffer from beam dilution.

\begin{figure*}
\begin{center}
\includegraphics[width=18cm] {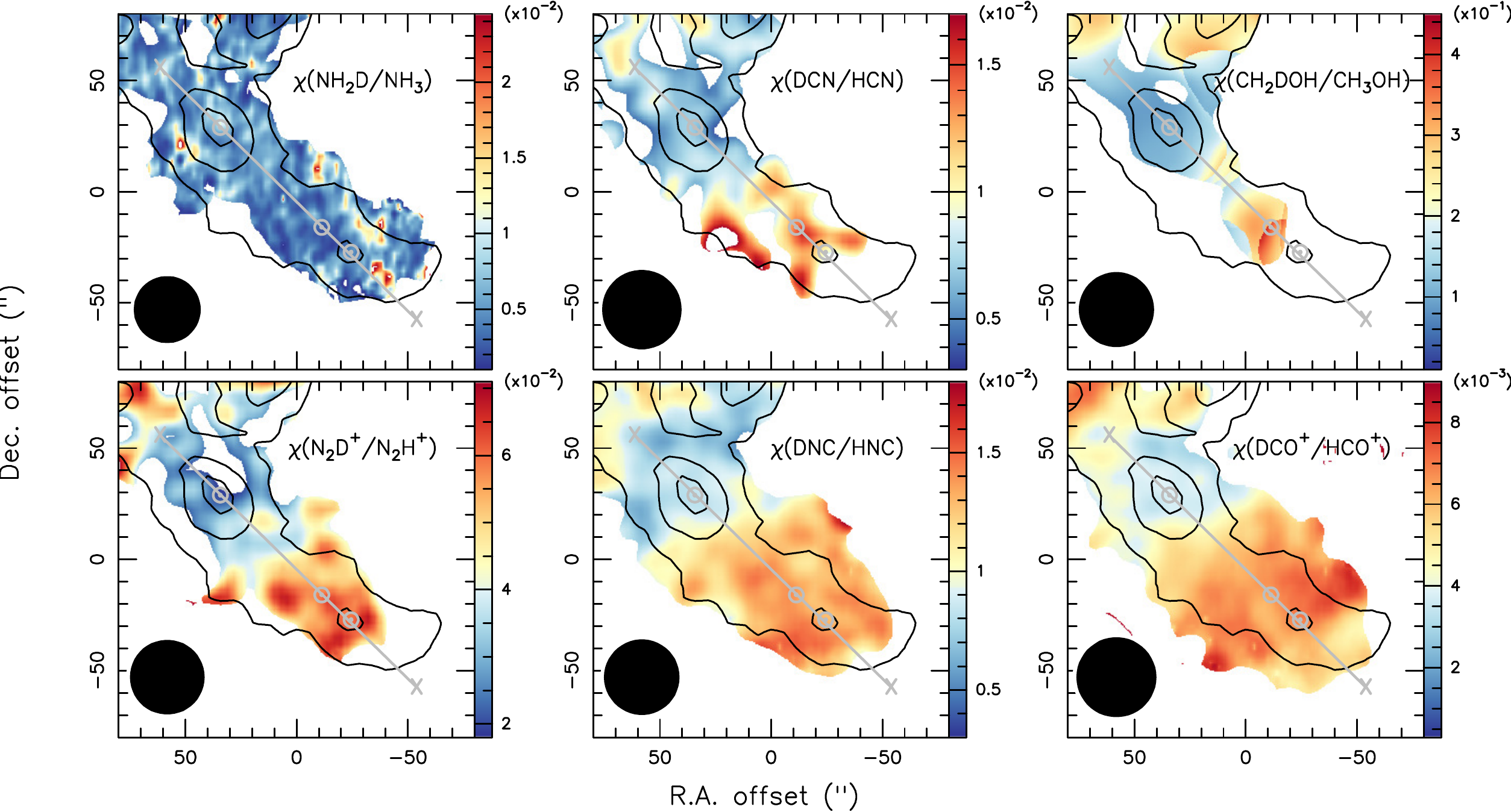}
%\begin{minipage}[c]{.9\textwidth}
%\subfigure[]{\includegraphics[height=5cm] {X-NH2D-fromTk-2-NH3-eps-converted-to.pdf}}
%\subfigure[]{\includegraphics[height=5cm] {X-N2Dp-fromTk-2-N2Hp-eps-converted-to.pdf}}
%\subfigure[]{\includegraphics[height=5cm] {X-DCN-fromTk-2-H13CN-eps-converted-to.pdf}}
%\end{minipage}
%\\
%\begin{minipage}[c]{1\textwidth}
%\subfigure[]{\includegraphics[height=5cm] {X-DCOp-fromTk-2-H13COp-eps-converted-to.pdf}}
%\subfigure[]{\includegraphics[height=5cm] {X-CH2DOH-fromTrot-2-CH3OH-eps-converted-to.pdf}}
%\subfigure[]{\includegraphics[height=5cm] {X-DNC-fromTk-2-HN13C-eps-converted-to.pdf}}
%\end{minipage}

\end{center}
\caption{Column density ratio (relative abundance) maps for the deuterated isotopologues with respect to their hydrogenated isotopologues. 
The column densities of HCN, HNC, and $\rm HCO^+$ are converted from those of $\rm H^{13}CN$, $\rm HN^{13}C$, and $\rm H^{13}CO^+$ by assuming $\rm ^{12}C/{^{13}C}\sim 42.9$.
The black contour of 870\,$\mu$m continuum emission starts from $\rm 10\sigma$ and increases in steps of  $\rm 10\sigma$ ($\rm \sigma=0.042\,Jy\,beam^{-1}$).
The gray line along the filamentary elongation as well as the position marks on it  are the same as  in Figure~\ref{dust}. 
The pixels are blanked where the lines used for calculation have $\rm <3\sigma$ integrated intensity.
The angular resolution is shown as the black circle in the bottom left corner.
%{\color{red} convert into 12C map}
}\label{deutertionall}
\end{figure*}

Note that most rotational transitions in our study show resolvable hyperfine splittings. Previous studies simplified the estimation of the total molecular column density of one species, based on an assumption that all of its transitions have a constant excitation temperature {(i.e., ``CTEX" approximation)}. Unlike the $\rm NH_3$\,(1,1) line which is in LTE toward the P1--S region (Section~\ref{gasmap}), this method may not tell the precise column densities of the dense gas tracers if they are in non-LTE.  For example, low-$J$  lines of $\rm HCO^+$, HCN, and $\rm N_2H^+$ isotopologues are not sensitive to the excitation temperature changes between 5 and 10\,K {(\citealp[see the discussion in][]{shirley13} and \citealt{mangum15})}. {Therefore, when the measured excitation temperature falls into this range, it might be only a lower limit of the real excitation temperature. %{\color{red}}why lower limit}

For the following species,  we measure the  excitation temperatures of the detected transitions, and compare their measured D-fractions by using CTEX method with those measured by using $\rm T_{kin}$.
%(see the equations in Appendix~\ref{molcalculation}):

%\begin{itemize} 
\subsection {$\rm NH_2D$ Column Density and Deuterium Fraction of $\rm NH_3$}
Our observations only cover one  $o$-$\rm NH_2D$ ($J ,K$)=(1,0) line, with hyperfine splitting marginally resolved at a velocity resolution of $\rm 0.7\,km\,s^{-1}$. This line has a high critical density ($\rm 4\times10^6\,cm^{-3}$ at 5--20\,K), so it might not be thermalized in our sources. Moreover, HFS fitting indicates that the main line of this transition is optically thin toward P1 ($\rm \tau\sim0.4$) and S ($\rm \tau\sim0.9$). 
{For a single transition, the observed brightness temperature is a function of $\rm T_{\Delta F}$, the unknown filling factor, and the  line optical depth. Therefore, the $\rm T_{\Delta F}$ derived from fitting the line profile of a single transition ($\rm <6\,K$ with large uncertainty, Table~\ref{tab:tex}) may be underestimated. }

A plausible column density map of $\rm NH_2D$ can be given using $\rm T_{kin}$ from $\rm NH_3$ fitting. Here, we assume  the OPR of $\rm NH_2D$ and $\rm NH_3$ as statistical values of 3 and 1, respectively \citep[e.g.,][]{fontani15, harju17}, and derive the D-fraction from $\rm NH_3$ to $\rm NH_2D$ by converting the column densities of $o$-$\rm NH_2D$ and $p$-$\rm NH_3$ into total ($o$+$p$) column densities.

The single D-fraction of $\rm NH_3$, shown as $\chi \rm (NH_2D/NH_3)$ in Figure~\ref{deutertionall}, seems to be almost uniform, as $\rm (5\pm3)\times10^{-3}$ in the entire P1--S region, indicating small sensitivity to the gas temperature variation when $\rm T_{kin} <22\,K$.

\subsection {Deuterium Fraction of $\rm N_2H^+$, $\rm HCN$, $\rm HCO^+$, and $\rm HNC$}
 
The low-$J$ transitions of $\rm N_2H^+$, $\rm HCN$, $\rm HCO^+$, and $\rm HNC$ isotopologues are typical dense gas tracers (with critical densities of $\rm10^{5}\text{--}10^{6}\,cm^{-3}$).  In this work, we derive the  column densities of the hydrogenated molecules by assuming that 
$\rm N_2H^+$\,(1--0) and the lines from  $\rm ^{13}C$ isotopologues are optically thin at pixels where $\rm S/N>4$  for the following reasons:
(1) The main hyperfine line of $\rm N_2H^+$\,(1--0) has a small optical depth toward P1 and S ($\rm \tau$=0.1--0.3).
(2) The 1--0 lines of  $\rm ^{13}C$ isotopologues show a homogeneous intensity ratio with respect to their less abundant  isotopologue lines $\rm HC^{15}N$\,(1--0), $\rm H^{15}NC$\,(1--0), and $\rm HC^{18}O^+$\,(1--0)  in the entire P1--S region (see the discussion in Section~\ref{uncertainty}). 
 (3) Compared with the $\rm ^{15}N$ or $\rm ^{18}O$ lines, the $\rm ^{13}C$ lines show a higher S/N toward each pixel. 

The hyperfine splittings of $\rm N_2H^+$\,(1--0), $\rm H^{13}CN$\,(1--0),  $\rm H^{13}CN$\,(2--1), and DCN\,(1--0) are marginally resolved at our velocity resolution, and blended into three Gaussian peaks (Fig.~\ref{velpro}). For these, we derive the excitation temperature  ($\rm T_{\Delta F}$) for each single line using the HFS method. 
Moreover, we detected two low-$J$ transitions (2--1 and 1--0) of $\rm N_2D^+$ and $\rm H^{13}CN$, so the excitation temperatures  ($\rm T_{\Delta J}$) of both isotopologues can be derived using lines at different $J$ levels. Listing $\rm T_{\Delta F}$ and $\rm T_{\Delta J}$ in Table~\ref{tab:tex}, we find they are comparable for all species, implying that the CTEX method is a good approximation to estimate the column density of these molecules.

Note that the detected lines are all low-$J$ transitions, with $\rm E_{\it u}/k_B$ spanning the range of 4--12\,K. Therefore, like the case of $\rm NH_2D$\,($J$,$K$)=(1,0), the excitation temperatures that  fall in the range of 3--10\,K may be underestimated. 

We estimate the column densities of individual molecules using the maps of $\rm T_{\Delta J}$ (as a lower limit) and $\rm T_{kin}$ derived from $p$-$\rm NH_3$, separately. For $\rm HN^{13}C$, there is only one line detected in this work, so the $\rm T_{\Delta J}$ we use as approximation is derived from $\rm H^{13}CN$ lines. 
The molecular column densities derived from two sets of temperature maps are similar within the systematic uncertainty. 

Furthermore, using the isotopic ratio $\mathcal{R}_{\rm ^{12}C/^{13}C}=7.5{\it D}_{GC}+7.6=42.9$  \citep{giannetti14}, we convert the column density map of the $\rm ^{13}C$ isotopologues to those of the $\rm ^{12}C$ isotopologues, and show the D-fraction maps in Figure~\ref{deutertionall}.
Although the $\rm T_{kin}$ map  shows a larger dynamic range than the $\rm T_{\Delta J}$ maps, the D-fraction maps derived from both sets of temperatures show the same enhancement from P1 to S by a factor of 2--3.
%{\color{red}(Siyi, is there any ref. to confirm that these species are formed exclusively in the gas phase?)}

\subsection {Deuterium Fraction of $\rm CH_3OH$}
Because the LTE and non-LTE analysis give similar $\rm CH_3OH$ column density values toward P1, S, and Soff, and the $\rm T_{rot}$ map of $\rm CH_3OH$ agrees with the $\rm T_{kin}$ map derived from $p$-$\rm NH_3$ (Section~\ref{h2cotemp}), we estimate the $\rm CH_2DOH$ column density by assuming that its only detected line is optically thin, and that the $\rm T_{rot}$ map derived from the $\rm CH_3OH$ lines can be applied to $\rm CH_2DOH$.
The map of $\rm CH_2DOH$ relative abundance ratio with respect to $\rm CH_3OH$, shown as $\chi \rm (CH_2DOH/CH_3OH)$ in Figure~\ref{deutertionall}, indicates an enhancement peak at Soff.

}

%\end{itemize}

\subsection{Error budget}\label{uncertainty}
During our observations with the EMIR receiver, the uncertainty in the absolute flux calibration\footnote{See http://www.iram.fr/GENERAL/calls/s17/30mCapabilities.pdf}, \footnote{See also {\it Calibration of spectral line data at the IRAM 30m radio telescope}  by C. Kramer (1997).} was $\rm <10\%$ \citep[][]{carter12}. Comparing different line-fitting methods, we found that  the  uncertainty in the integrated intensity measurement is  $\rm <30\%$, as the result of the line blending  at the given velocity resolution, as well as the velocity range we select to integrate the lines (Table~\ref{tab:linedeu}). 
In calculating the molecular column densities and relative abundance ratios, uncertainties are given by following the error propagation formulae. \footnote{See calculations in https://bit.ly/2J92r10}
%{\color{red}Paola,[I do not understand why you list the various things above; can you please just explain how do you estimate this uncertainty?]}
More importantly, the following assumptions we used in this work may lead to larger systematic uncertainties:

%{\color{red} Siyi, I am not sure if the list or the subsection of a better way to write the following part of this section.}
%{\color{red}check if bold face or italic is allowed for emphasis}
\begin{enumerate}  

\item {\it The LTE condition.} The excitation temperatures ($\rm T_{\Delta F}$ and $\rm T_{\Delta J}$) derived for the dense gas tracers, i.e., low-$J$ lines  from $\rm H^{13}CN$ (DCN), $\rm H^{13}CO^{+}$ ($\rm DCO^{+}$), and $\rm N_2H^+$ ($\rm N_2D^+$), are 5--10\,K (Table~\ref{tab:tex}), which is close to their $\rm E_{\it u}/k_B$ but lower than the $\rm T_{kin}$ derived from $\rm NH_3$ (15--20\,K, Figure~\ref{texnh3}), $\rm H_2CO$, and $\rm CH_3OH$ ($\rm >16\,K$, Table~\ref{tab:lvgfit}), indicating that these lines are subthermally excited. %Since high-$J$ lines of these species may not be excited in high-mass or low-mass infrared dark clouds, a 
A feature termed ``low excitation temperature"  is commonly reported \citep[e.g., ][]{caselli02b,crapsi05,miettinen11,fontani12,gerner15}.
A total column density map for each molecule, based on either the $\rm T_{\Delta J}$ or $\rm T_{kin}$ maps, has an uncertainty of a factor of 2 at each pixel. Nevertheless, the uncertainty does not change the molecular column density gradient along the filamentary elongation from P1 to S. Moreover, such uncertainty can be canceled out when we derive the D-fraction map from the  relative abundance ratio of the deuterated molecules with respect to their hydrogenated isotopologues. 

\item {\it Optically thin lines.} We assume that the lines from the $\rm^{15}N$, $\rm^{18}O$, $\rm ^{17}O$, and  $\rm ^{13}C$ hydrogenated molecules as well as the D-lines of species such as HCN, HNC, $\rm HCO^{+}$ are optically thin. The $\rm ^{13}C$ lines, although possibly optically thick toward  prestellar cores \citep[e.g.,][]{padovani11},  are used for deriving the D-fraction maps, which is a compromise  for the high-fidelity  images covering the entire P1--S region.  To test whether the optically thin assumption is valid  for the $\rm ^{13}C$ lines, we measured the relative abundance ratio $\chi \rm (H^{15}NC/HN^{13}C)$,  $\chi\rm (HC^{15}N/H^{13}CN)$, and $\chi\rm (HC^{18}O^+/H^{13}CO^+)$, and found no variations toward the pixels where both $\rm ^{13}C$ line and $\rm^{15}N$/ $\rm^{18}O$ lines show $\rm >3\sigma$ detections   (Figure~\ref{isoabundance}). Therefore, even if the $\rm ^{13}C$- lines toward P1--S are optically thick { and hence the D-fractions in Figure~\ref{deutertionall} are upper limits,} their optical depths are likely similar toward all pixels in the region we are interested in. Therefore, any gradient found in D-fraction along the filament elongation from P1 to S is a chemical effect rather than an optical depth effect.

%For $\rm CH_3OH$ and $p$-$\rm H_2CO$, we are not able to test the optical depth as well as excitation condition of individual lines. Nevertheless, LVG fittings (Figure~\ref{h2colvg}, \ref{a-ch3ohlvg}, and \ref{e-ch3ohlvg}) of both species show consistent column densities with the results from their individual rotation diagrams. 

\item {\it The unity beam-filling factor.} Due to the unknown distribution of the molecular lines along the line-of-sight, we assume that all lines we used to derive the molecular column densities have no beam dilution.
For $\rm NH_3$, we found that  its column density map from ``HfS fitting" of the (1,1) and (2, 2) lines shows negligible differences when assuming large or no beam dilution, indicating that the emission of the (1,1) and (2,2) lines is spatially extended at an angular resolution of $\rm \sim5\arcsec$.
However,  this assumption may not be true for some lines even if we only consider the  spatial distribution in the plane of the sky. Taking $\rm CH_2DOH\,(2_{0,2}-1_{0,1})$ as an example, we find that the region where it is detected  with $>3\sigma$ emission shows a narrow ($\rm \sim10\arcsec$ in width) filamentary morphology. %, {\color{red}and the extension perpendicular to the filament elongation is smaller than the beam of this line. (rephrase)}
Therefore, the column density of this species may be underestimated.

\item {\it The fractionation of $\rm ^{12}C/^{13}C$.} In order to compare the D-fraction of each species derived here with values found in the literatures, we convert the column densities of the $\rm ^{13}C$ isotopologues into those of the  $\rm ^{12}C$ isotopologues using the isotopic ratio $\mathcal{R}_{\rm ^{12}C/^{13}C}=42.9$ as a conversion factor. % between the $\rm ^{12}C$- and $\rm ^{13}C$- isotopologues. %relative abundance ratio between the $\rm D$-isotopologues with respect to their $\rm ^{13}C$-isotopologues into the D-fraction of the $\rm ^{12}C$-isotopologues. 
%The conversion factor between the $\rm ^{12}C$- and $\rm ^{13}C$- isotopologues  we used is the isotopic ratio $\mathcal{R}_{\rm ^{12}C/^{13}C}=42.9$, which 
This is a statistic value according to the source distance to the Galactic center  \citep{giannetti14}. However, in an environment like that of the P1--S region, i.e., with dynamic ages of $\rm 10^5\,yr$ of the source \citep{feng16b}, gas temperatures in the range of of 10--20\,K, and a number density in the range of $\rm 10^4-10^5\,cm^{-3}$, $\rm ^{12}C\Leftrightarrow {^{13}C}$ conversion via isotopic exchange reactions is active. For example, model predictions of the $\mathcal{R}_{\rm ^{12}C/^{13}C}$ in gaseous HCN, HNC, and $\rm H^{13}CO^+$ are $\rm \sim80$, $\rm \sim70$, and $\rm \sim 47$, respectively, in such an environment \citep{furuya11}. Our D-fraction estimates for these species might be overestimated by a factor of 1.5--2.

\item {\it The OPR for $\rm NH_3$ and $\rm NH_2D$.} We assume $\rm OPR\sim1$ for $\rm NH_3$ and $\rm OPR\sim3$ for $\rm NH_2D$ in their relative abundance estimation, which may not be true. For example, OPR of $\rm NH_2D$ is reported by modeling of \citet{sipila15b} as $\sim2$, and 2.8 by \citet{harju17} from observations of the low-mass prestellar core. Nevertheless, this will only add an uncertainty of $<20\%$.

\end{enumerate}

In short, although the absolute value of the column density for a particular molecule in individual pixels may be underestimated or overestimated based on the above assumptions,  we can still  trust the gradient shown in the relative abundance ration maps of the deuterated isotopologues with respect to their hydrogenated isotopologues. 

\subsection{Variations in deuteration for different species}\label{variation}
The D-fraction varies  for different species. %, especially when the species are formed with different gas-grain paths. 
Theoretical studies predict that $\rm N_2H^+$ and $\rm HCO^+$ are  formed exclusively in the gas phase;  HNC and HCN are mainly formed in the gas phase except for very early times; 
$\rm NH_3$ and $\rm H_2CO$ are partially formed on the grain surface and partially in the gas phase; while  $\rm CH_3OH$ is exclusively form on grain surfaces  \citep[e.g., ][]{parise02,aikawa05,aikawa12,Garrod_ea07,graninger14}.
Table~\ref{tab:deuteration} lists the D-fraction of  $\rm N_2H^+$, HCN, HNC, $\rm HCO^+$, $\rm NH_3$, and  $\rm CH_3OH$ toward P1, S, and Soff. In considering the  beam at 1\,mm ($\rm \sim10\arcsec$) as well as the gridding kernel of the 3\,mm maps (one-third the size of the beam, see \citealp{mangum07}), we give the mean of five parameters, including the D-fraction ($D$) and its systematic uncertainty for each species, the intensity ratio $X$  and its uncertainty between the lines of deuterated-hydrogenated isotopologues, %between each deuterated-hydrogenated transition pair,  
 %{\color{red}(Siyi, I use $X$ here for the intensity ratio of the line and $\chi$ for the abundance ratio in between isotopologues, any other suggestion of the symbol here?)} 
the gas kinetic temperature $\rm T_{kin}$, $\rm NH_3$ column density, and the $\rm H_2$ column density, toward a 5\arcsec-radius region centered on P1, S, or Soff.  All species in this work, except for $\rm NH_3$ and $\rm CH_3OH$, show an enriched deuterium fractionation toward S compared with P1 of a factor of 2--3.

Compared with previous D-fraction studies of the above species toward high-/low-mass star-forming regions, we find that  $D\rm (N_2H^+)$ observed at an angular resolution of  30\arcsec~  is consistent with the results of observations at an angular resolution of 5\arcsec~  toward the same source \citep[][in which P1 is called MM4, and S is called MM9]{chen10}. Quantitatively, $D\rm (N_2H^+)$ ranging from $\rm 2\%$ to $\rm 6\%$ from P1 to S fits the correlation with CO depletion ($\rm \sim10$, see Section~\ref{codepletion}) at the source distance shown toward the low-mass prestellar cores in \citet{crapsi05} and  \citet{emprechtinger09}. However, the value is consistent with the mean $D\rm (N_2H^+)\sim4\%$ reported by \citet{fontani11} and \citet{gerner15} in  high-mass protostellar objects rather than the starless objects. This value is similar to the value found in the IRDC deuterium study by \citet{barnes16}, indicating not only that a protostellar object is embedded in both P1 and S, but also that S is chemically less far evolved than P1.
%{\color{red}Did you compare this with the results from Barnes, Caselli et al. ?}

$D\rm (HCO^+)$ is also consistent with those found in the high-mass starless clumps or very young high-mass protostellar objects given by \citet[][]{fontani14} and \citet{gerner15}. Moreover,  the relative abundance ratio of $\rm \chi(HCO^+/HC^{18}O^+)=\mathcal{R}_{\rm ^{12}C/^{13}C}\times \chi(H^{13}C^{16}O^+/H^{12}C^{18}O^+)=143_{-107}^{+214}$ is consistent with  $\rm \mathcal{R}_{\rm ^{16}O/^{18}O}\sim 300$ given by  \citet{giannetti14} at our source distance to the Galactic Center.

$D\rm (HCN)$ and $D\rm (HNC)$ in our study are consistent with the values toward other 70\,$\mu$m dark clouds  \citep[e.g., ][]{bergin99,turner01,lodders03,miettinen11,sakai12,fontani14,gerner15}. 
{ Although theoretical model predicted that gas-phase reactions dominate the formation of both species in a warm-temperature environment \citep{,graninger14}, 
we note that the integrated intensity ratio of DNC\,(1--0) with respect to DCN\,(1--0) is $\rm\sim$3, which is consistent with the results obtained from solid-state experiments on the reaction of D and CN at 10\,K by \citet{hiraoka06}, indicating that  our sources are chemically young. }
%{\color{red}paola, but you said before that these species are formed in the gas phase}
Moreover, assuming $\rm \mathcal{R}_{\rm ^{12}C/^{13}C} \sim43$, the relative abundance ratios of $\rm \chi(HNC/H^{15}NC)=\mathcal{R}_{\rm ^{12}C/^{13}C}\times \chi(H^{14}N^{13}C/H^{15}N^{12}C)$ and  $\rm \chi(HCN/HC^{15}N)=\mathcal{R}_{\rm ^{12}C/^{13}C}\times \chi(H^{13}C^{14}N/H^{12}C^{15}N)$ are $\rm 143_{107}^{+214}$, which is consistent with $\rm \mathcal{R}_{\rm ^{14}N/^{15}N}$ reported by \citet{adande12,zeng17} and \citet{colzi18}. 

The single $D\rm (NH_3)$  in the entire region is in general consistent with detections in dense ($\rm >10^5\,cm^{-3}$)  and cold ($\rm <20\,K$) cores \citep[e.g.,][$\rm 2.3\%$ toward MM9]{pillai11,fontani15}. The slightly lower value ($\rm 0.3\%-1.9\%$) in our study is likely due to the fact that the $\rm NH_3$ observed at an angular resolution of 5\arcsec~ traces more deeply embedded  gas than that traced by $\rm NH_2D$ at an angular resolution of 30\arcsec. This implies that deeper into clumps P1 and S, the gas temperature should be higher for such a low value of $D\rm (NH_3)$.
%{\color{red}Paola, and then what? You have to explain here that this implies that deeper into the selected object we are studying, the temperature should be higher, to explain this lower value of D-frac.}

The single $D\rm (CH_3OH)$ detection is $\rm 10\%-39\%$ in a region where $\rm >3\sigma$ emission of the only detected $\rm CH_2DOH$ line distributes compactly. This value is higher than the upper limit reported by  \citet[][]{fontani15} toward MM9. In particular, the D-fraction peak toward Soff is higher than that in low-mass prestellar cores \citep[$\sim10\%$, e.g.,][]{bizzocchi14}. Instead, it is closer to the value reported in the Class 0 protostars  \citep[e.g.,][]{parise06} and  gas-grain model results applied to protostellar objects \citep{awad14}. This may indicate a deeply embedded dense protostellar object(s) toward Soff.
An underestimate of the $\rm CH_3OH$ column density can also account for an overestimate of the $D\rm (CH_3OH)$, as discussed in Section~\ref{uncertainty}, but it is not likely the case for such high $D\rm (CH_3OH)$ here.
In fact,  checking the location where \citet{fontani15} carried out the pointing observation (south of S), we find the singly  $D\rm (CH_3OH)$ is $<0.3\%$  in our map, which is consistent with their upper limit of $0.4\%$  given by their nondetection of this line.  
This result also shows the advantage of mapping over pointing observations to study chemistry.

\begin{table*}
\caption{Deuterium Fraction of Six species at Different Locations}\label{tab:deuteration}
\scalebox{0.95}{
\begin{tabular}{l|cccccc}
\hline\hline

Parameters                            & $o$-$\rm NH_2D$/$p$-$\rm NH_3$     &$\rm  N_2D^+/N_2H^+$      &DCN/HCN$^a$    &DNC/HNC$^a$   &$\rm DCO^+/HCO^+$$^a$    &$\rm CH_2DOH/ CH_3OH$ \\

\hline
Hyd. lines used          &{\scalebox{0.8}{ $\rm NH_3$ ($J$, $K$)=(1,1), (2,2)}}    
&{\scalebox{0.8}{ $\rm N_2H^+\,(1-0)$}} 
&{\scalebox{0.8}{ $\rm H^{13}CN\,(1-0), (2-1)$}} 
&{\scalebox{0.8}{ $\rm HN^{13}C\,(1-0)$}} 
&{\scalebox{0.8}{ $\rm H^{13}CO^+\,(1-0), (2-1)$}} 
&{\scalebox{0.8}{ lines of $\rm CH_3OH$ list in Table~\ref{tab:dehyd} }} \\
\hline
  
$D_{\rm P1}$$^{b,d,h}$          &$\rm 0.8\%\pm 0.5\%$         &$\rm 2.1\%\pm 0.9\%$           &$\rm 0.7\%\pm 0.2\%$          &$\rm 0.8\%\pm 0.3\% $         &$\rm 0.4\%\pm 0.1\% $         &$\rm 10.3\%\pm 8.0\%$  \\
$D_{\rm S}$$^{d,h}$              &$\rm 0.4\%\pm 0.3\%$         &$\rm 6.1\%\pm 2.4\%$           &$\rm<0.3\%$$^i$          &$\rm 1.3\%\pm 0.5\% $         &$\rm 0.7\%\pm 0.2\% $         &$\rm <23.0\% $$^i$ \\
$D_{\rm Soff}$$^{c,d,h}$       &$\rm 0.3\%\pm 0.2\%$         &$\rm 5.9\%\pm 2.1\%$           &$\rm 1.6\%\pm 0.8\%$          &$\rm 1.4\%\pm 0.5\% $         &$\rm 0.6\%\pm 0.2\% $         &$\rm 33.3\%\pm 26.7\% $\\
$\rm P_{\it max}$$^{e}$ (w.r.t. S)  &$\rm [31\arcsec, 24\arcsec]$    &$\rm [-7\arcsec, 0\arcsec]$     &$\rm [11\arcsec, 8\arcsec]$   &$\rm [2\arcsec, -10\arcsec]$    &$\rm [-3\arcsec, 5\arcsec]$   &$\rm [17\arcsec, 9\arcsec]$\\
$D_{\rm P_{\it max}}$$^{d}$  &$\rm 2.2\%\pm 1.8\%$         &$\rm 6.8\%\pm 3.8\%$           &$\rm 1.5\%\pm 0.8\%$          &$\rm 1.5\%\pm 0.9\% $         &$\rm 0.7\%\pm 0.3\% $         &$\rm 41.3\%\pm 33.1\% $ \\

 $\rm N_{dust, P_{\it max}}$$^{d,f}$  ($\rm cm^{-2}$)   &$\rm 4.9\pm 1.2 (22)$          &$\rm 5.3\pm 0.8 (22)$          &$\rm 7.0\pm 2.6  (22)$         &$\rm 4.8\pm 0.8 (22)$          &$\rm 6.1\pm 1.3 (22)$        &$\rm 5.8\pm 1.7  (22)$\\
$\rm T_{kin, P_{\it max}}$$^{d}$   (K)                            &$\rm 19.0\pm 9.5 $            &$\rm 13.0\pm 5.5 $                  &$\rm 12.3\pm 3.4 $             &$\rm 9.2\pm 4.6 $        &$\rm 13.1\pm 3.6 $           &$\rm 11.6\pm 3.5 $\\
$\rm N_{NH_3,P_{\it max}}$$^{d,f,g}$  ($\rm cm^{-2}$)&$\rm 1.8\pm 1.1 (15)$          &$\rm 3.0\pm 2.4 (15) $              &$\rm 7.4\pm 5.9 (15) $         &$\rm 4.8\pm 3.8 (15)$              &$\rm 4.0\pm 3.2 (15) $       &$\rm 6.9\pm 5.5 (15)$\\ 

\hline

Line Ratio$^j$ &{\scalebox{0.8}{ $\rm \frac{ NH_2D\,(1_{1,1}\,0s\text{--}1_{0,1})}{NH_3\,(1_{1}\,0a\text{--}1_{1}\,0s)}$}}
&{\scalebox{0.8}{ $\rm \frac{ N_2D^+\,(1-0)}{N_2H^+\,(1-0)}$}}
&{\scalebox{0.8}{ $\rm \frac{ DCN\,(1-0)}{H^{13}CN\,(1-0)}$}}
&{\scalebox{0.8}{ $\rm \frac{ DNC\,(1-0)}{HN^{13}C\,(1-0)}$}}
&{\scalebox{0.8}{ $\rm \frac{ DCO^+\,(1-0)}{H^{13}CO^+\,(1-0)}$}}
&{\scalebox{0.8}{ $\rm \frac{ CH_2DOH\,(2_{0,2}\text{--}1_{0,1}\,e0)}{CH_3OH\,(2_{0,2}\text{--}1_{0,1}\,A)}$}}
\\
\hline
$\it X_{\rm P1}$   &$\rm  2.1\%\pm 0.3\% $   &$\rm  1.7\%\pm 0.5\% $   &$\rm  21.3\%\pm 4.8\%$    &$\rm  35.9\%\pm 8.9\%$   &$\rm  33.2\%\pm 15.7\% $  &$\rm  0.8\%\pm 0.2\% $ \\
$\it X_{\rm S}$    &$\rm  3.5\%\pm 0.5\% $   &$\rm  4.4\%\pm 1.1\% $   &$\rm  <10.0\% $$^i$   &$\rm  58.3\%\pm 15.3\% $  &$\rm  61.8\%\pm 23.4\% $  &$\rm  <1.8\% $$^i$ \\
$\it X_{\rm Soff}$ &$\rm  3.0\%\pm 0.5\% $   &$\rm  4.3\%\pm 0.9\% $   &$\rm  44.1\%\pm 14.6\% $   &$\rm  60.5\%\pm 12.7\% $  &$\rm  59.2\%\pm 24.1\% $  &$\rm  2.8\%\pm 0.8\% $ \\

%%%%%%%%%%%%%%%%%%%%%%%%%%%%%

\hline \hline
\multicolumn{7}{l}{\color{black} {\bf Note.} $^a$ Deuterium fraction is derived from the $\rm ^{13}C$ lines of the hydrogenated isotopologues}\\
\multicolumn{7}{l}{~~~~~~~~~~~~    by assuming optically thin lines, the same beam-filling factor, and a constant fraction of $\rm ^{12}C/{^{13}C}\sim 42.9$;}\\
\multicolumn{7}{l}{~~~~~~~~~~$^b$  The 870\,$\mu$m dust continuum peak P1  ($\rm 18^h42^m50^s.289$, $\rm -04^\circ03\arcmin16\arcsec.492$, J2000) is at the offset of [60\arcsec, 56\arcsec] }\\
\multicolumn{7}{l}{~~~~~~~~~~~~ with respect to S  ($\rm 18^h42^m46^s.358$, $\rm -04^\circ04\arcmin12\arcsec.559$, J2000);}\\
\multicolumn{7}{l}{~~~~~~~~~~$^c$  The 870\,$\mu$m dust continuum peak Soff ($\rm 18^h42^m47^s.240$, $\rm -04^\circ04\arcmin01\arcsec.37$, J2000) is at the offset of [12\arcsec, 11\arcsec] with respect to S;}\\

\multicolumn{7}{l}{~~~~~~~~~~$^d$  Values are derived within a region with a radius of $\rm 5\arcsec$ centered on P1, S, Soff, or $\rm P_{\it max}$;}\\

\multicolumn{7}{l}{~~~~~~~~~~$^e$   This location shows the maximum D-fraction for a certain species in the region where  870\,$\mu$m continuum, $\rm NH_3$\,(1,1), and lines from a}\\
\multicolumn{7}{l}{~~~~~~~~~~~~   particular hydrogenated-deuterated isotopologues all show $>3\sigma$ emission. We give its coordinate in the format of an offset to S.}\\
\multicolumn{7}{l}{~~~~~~~~~~$^f$  $\rm (x \pm y)\times 10^z$ is presented in the form of $\rm x \pm y(z)$;}\\

\multicolumn{7}{l}{~~~~~~~~~~$^g$ The gas column density $\rm N_{NH_3}$ is derived by assuming the filling factor $f\sim\rm 1$;}\\

\multicolumn{7}{l}{~~~~~~~~~~$^h$ $\rm T_{kin, P1}$, $\rm N_{H_2, P1}$, $\rm N_{NH_3, P1}$ within a radius of  $\rm 5\arcsec$ offset  around P1 is  $\rm 18.4\pm4.5\,K$, $\rm (9.8\pm1.1)\times10^{22}\,cm^{-2}$,  $\rm (4.2\pm2.6)\times10^{15}\,cm^{-2}$, respectively;}\\
\multicolumn{7}{l}{~~~~~~~~~~~~  $\rm T_{kin, Soff}$, $\rm N_{H_2, Soff}$, $\rm N_{NH_3, Soff}$ within a radius of $\rm 5\arcsec$ offset  around  Soff is  $\rm 13.7\pm4.1\,K$, $\rm (7.2\pm2.1)\times10^{22}\,cm^{-2}$,  $\rm (5.0\pm3.0)\times10^{15}\,cm^{-2}$, respectively;}\\
\multicolumn{7}{l}{~~~~~~~~~~~~ $\rm T_{kin, S}$, $\rm N_{H_2, S}$, $\rm N_{NH_3, S}$ within a radius of $\rm 5\arcsec$ offset  around  S  is $\rm 12.5\pm3.6\,K$, $\rm (6.2\pm2.1)\times10^{22}\,cm^{-2}$,  $\rm (6.7\pm5.4)\times10^{15}\,cm^{-2}$, respectively;}\\
\multicolumn{7}{l}{~~~~~~~~~~$^i$ The integrated intensity of the deuterated line toward S has an $\rm S/N<3$, we give the upper limit as $\rm 3\sigma$. }\\
\multicolumn{7}{l}{~~~~~~~~~~$^j$ Line ratio in main-beam temperature. }\\

\end{tabular}
}
\end{table*}

\section{The Spatial Correlation between Deuterium Fractionation and CO Depletion}\label{codepletion}
%{\color{red}(Referee: CO depletion map is, in my opinion, one of the most important results. This represents the second large-scale depletion map in high-mass star-forming regions after Hernandez+2012, but the authors do not stress too much this result. The importance of depletion is indeed not even discussed in the conclusions. I suggest the author to give more relevance to this result.

%\bf Siyi, I move this section to the discussion, ref Pon et al. 2016 and Feng et al. 2016 for CO depletion toward this source, write something about the correlation between the deuteration and CO depletion, but quantitative analysis is still missing unless we give some model prediction...Any suggestions or ref. to recommend?)
%}

%Most species show  D-fraction enrichment from P1 to S by a factor of 2--3. %Taking the systematic uncertainty into consideration, {\color{red}such enrichment does not show ``significance" in numbers. (rephrase)}
%Specifically, 
When comparing the D-fraction maps of each species (Figure~\ref{deutertionall}), we not only confirm the fact that the D-fraction of $\rm NH_3$ does not increase toward any place along the filamentary elongation from P1 to S, but we also  see the clear location where the D-fraction shows the most significant enrichment.
 Therefore, we define another species-dependent position $\rm P_{\it max}$, where we find the D-fraction maximum for individual species. Table~\ref{tab:deuteration} lists the mean of gas and dust parameters toward $\rm P_{\it max}$; the maxima of (single) deuterium fractionation for HCN and $\rm CH_3OH$ are toward Soff rather than S.

{It is predicted by models such as the model developed by \citet{vasyunin17} that $\rm CH_3OH$  is formed on grain surfaces as soon as CO freezes out, and hence more efficient deuterium fractionation of $\rm CH_3OH$ indicates a high abundance of CO on the grain surface.
 Moreover, the freeze-out temperature of CO is around 20\,K, %\citep[e.g., ][]{caselli99, fontani06, aikawa08}, 
 and the integrated intensity maps of low-$J$ as well as mid-$J$ $\rm ^{13}CO$, $\rm C^{18}O$, and $\rm C^{17}O$ show emission peaks 5\arcsec--10\arcsec offset from P1 and S \citep[][]{feng16a,pon16b}. Therefore,  it is likely that a large portion of CO resides on grain surfaces toward P1, S, and Soff.  }

According to \citet{frerking82}, \citet{wilson94}, and \citet{giannetti14}, the expected relative abundance of $\rm C^{18}O$ with respect to $\rm H_2$ (denoted as $\chi_{C^{18}O}^E$) in the  P1--S region should be $\chi_{C^{18}O}^E=\chi_{\rm ^{12}CO}^E/\mathcal{R}_{\rm ^{16}O/^{18}O}^E= \rm  9.50\times10^{-5} {\it e}^{(1.11-0.13{\it D}_{GC})}/ (58.80{\it D}_{GC}+37.10) \sim 4.97\times10^{-7}$.

\begin{figure*}
\begin{center}
\begin{minipage}[c]{1\textwidth}
\subfigure[]{\includegraphics[height=5cm] {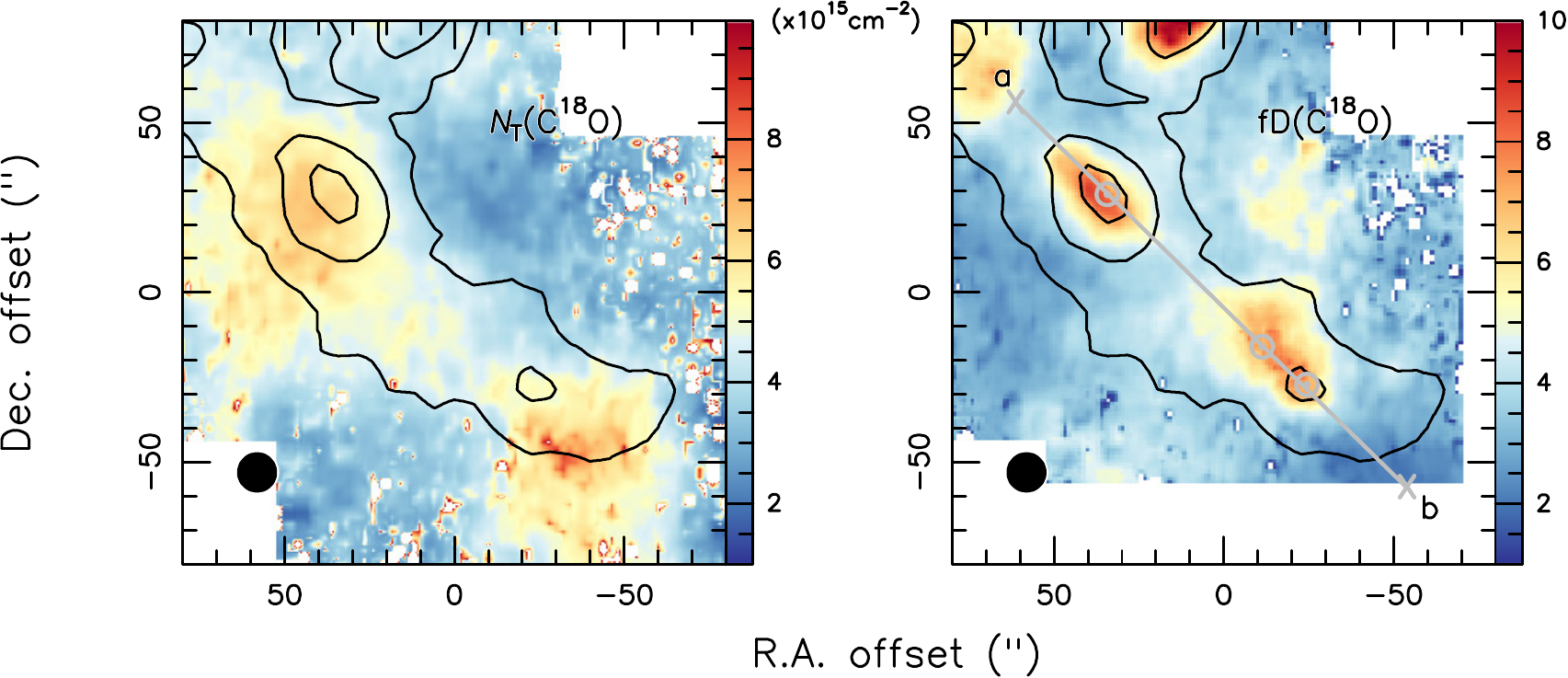}}
\subfigure[]{\includegraphics[height=5cm] {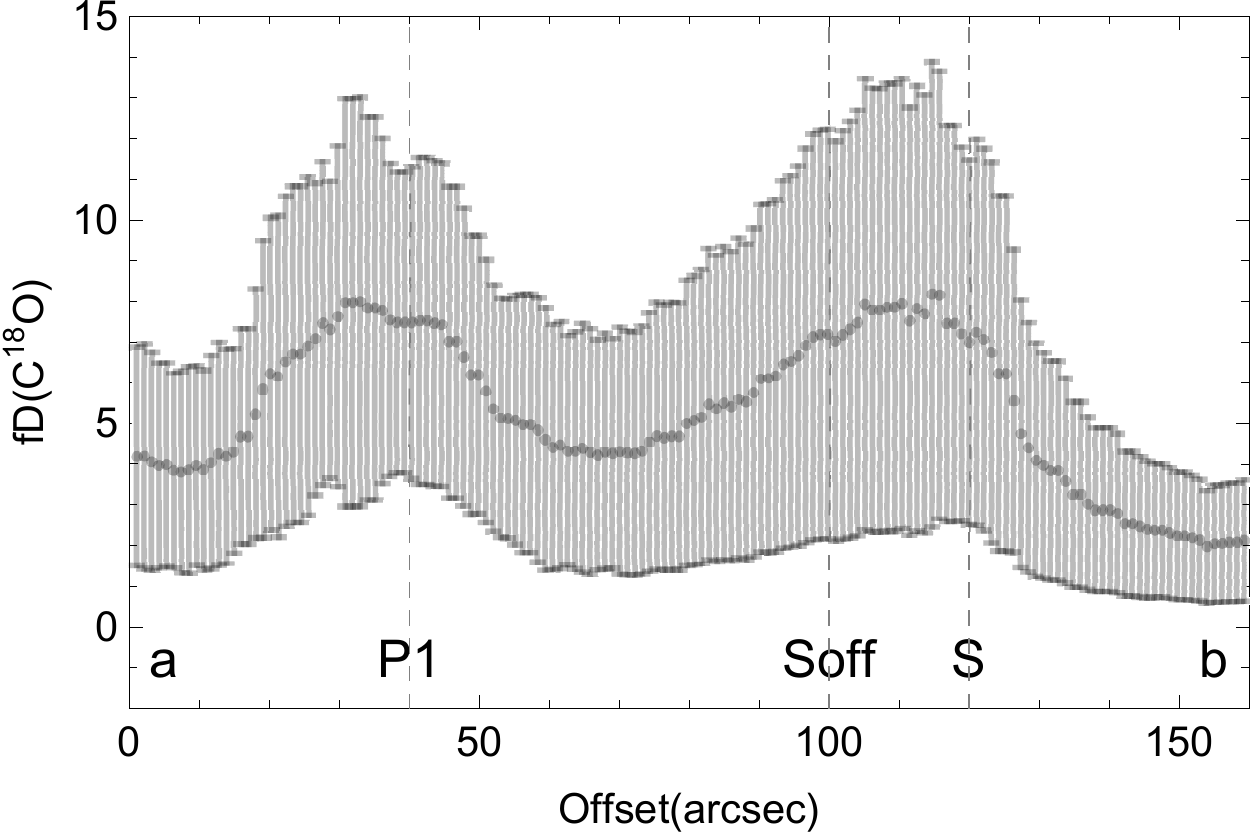}}
\end{minipage}

\end{center}
\caption{Maps for the $\rm C^{18}O$ column density and depletion.  
{\it Panel I}:  $\rm C^{18}O$ column density and depletion maps, derived by using the kinetic temperature of the gas and $\rm H_2$ column density maps (Figure~\ref{dust} and \ref{texnh3}).  The angular resolution of 11.8\arcsec~ is shown in black in the bottom left.  {$\rm C^{18}O$\,(2--1) shows $\rm >8\sigma$ emission ($\rm \sigma=0.11\,K$) at the system velocity ($\rm 79.5\,km\,s^{-1}$) toward all pixels on the map}.
{\it Panel II}: $\rm C^{18}O$ depletion profile along the directions of a--b (in black).
The marked positions are the same as those in Figure~\ref{dust}. 
%{\color{red} fD colour bar ranges from 1 (blue) to 10 (white). However in the text author write that the maximum value of fD is 14, but in the map, the max values do not even reach 10 (no white pixel). }
}\label{COmap}
\end{figure*}

{Gaussian fits to the beam-averaged line profiles of $\rm C^{18}O$\,(2--1) and $\rm C^{17}O$\,(2--1)  toward P1 and S show similar centroid velocity and FWHM as the lines we study here \citep{feng16a}. Therefore, we assume that the contamination from the foreground and background gas to CO is negligible. Low-$J$ CO isotopologue lines are not dense gas tracers, so $\rm C^{18}O$\,(2--1) shows emission with an $\rm S/N>8$ at the system velocity toward all pixels on the map. }
$\rm C^{17}O$\,(2--1) can be assumed to be optically thin  toward all the pixels within the  P1--S region, but its emission has a  low S/N ($\rm <3$ toward the map edge) in our observations. 
Given that the relative abundance of $\rm C^{17}O$ with respect to $\rm C^{18}O$ is homogeneous across the P1--S region (Figure~\ref{isoabundance}), we assume that $\rm C^{18}O$\,(2--1) is optically thin, or at least has the same level of optical depth  toward the pixels in the region we are interested in.  Using the gas kinetic temperature map $\rm T_{kin}$, we derived the observed column density map of $\rm C^{18}O$ (Figure~\ref{COmap}). Comparing the observed $\rm C^{18}O$ column density map with that obtained from the undepleted CO abundance ($\chi_{C^{18}O}^E$) and $\rm N(H_2)$ map, we present the $\rm C^{18}O$  depletion map in Figure~\ref{COmap} as well.

Although the $\rm C^{18}O$  depletion in general may be overestimated by our assumption of the gas-to-dust ratio (here we take 150 from \citealp{draine11})
% {\color{red}if you want to compare with previous work, you need to use the same assumptions. Please check.} 
and the  optical depth of the (2--1) line, its variation along the filament elongation should be real. That is to say, $\rm C^{18}O$ and then CO is frozen out onto the grains toward the center of P1 and S by a factor of $\rm \sim10$.  In particular, we find that the high depletion extends from S to Soff, and this value is consistent with that reported by pointing observations toward both locations \citep{pon16a}.

Although high CO depletion (up to $\sim$20) has commonly been detected by pointing observations toward the cold and dense cores of IRDCs \citep[with a gas-to-dust ratio assumed to be 100 in ][]{fontani12,giannetti14}, a  parcsec-scale CO depletion map and its spatial correlation with the widespread $D\rm(N_2H^+)$ map have only been reported in IRDC G035.39-00.33 \citep[e.g., ][]{hernandez11,barnes16}. Our current work,
followed by the study of a larger sample of 70\,$\mu$m dark clouds, unveils that parsec-scale CO depletion is widely detected, showing spatial anti-correlation with the deuterium peak of $\rm N_2H^+$ and $\rm HCO^+$  (S. Feng et al. in prep, Paper II).

%%%%%%%stop here
\section{Conclusions}\label{conclusion}
Using the IRAM-30\,m telescope, we  carried out an  line imaging survey at 1.3\,mm--4.3\,mm  toward two young high-mass star-forming clumps P1 and S, which are 1.8\,pc apart in the same 70\,$\mu$m dark filamentary IRDC G28.34+0.06. 
The broad bandwidth observations cover several ground-state  lines of deuterated and hydrogenated molecules simultaneously. 
The unique physical structures and evolutionary stages of our sources enable us to study the deuterium fractionation of six species ($\rm NH_3$, $\rm N_2H^+$, HCN, HNC, $\rm HCO^+$, and $\rm CH_3OH$) with improved robustness by canceling out calibration uncertainty. \\

Our conclusions are as follows: 

\begin{enumerate}
\item  Mapping is more advantageous over pointing observations for a detailed chemistry study. Maps of the D-fraction of different species with high dynamic range allow us
not only to verify a general increase/decrease trend along the filament elongation, but also to precisely determine the location where deuterium fractionation shows the most significant enrichment.

\item Along the filament elongation from P1 to S, the dust temperature and column density have small dynamic ranges. In contrast, the gas kinetic temperature shows a clear decrease and the $\rm NH_3$ column density shows a clear increase from P1 ($\rm \sim 20\,K$, $\rm \sim4\times10^{15}\,cm^{-2}$) to S  ($\rm \sim 14\,K$, $\rm \sim9\times10^{15}\,cm^{-2}$), indicating that the differentiations in physical properties dominate the chemical variations between P1 and S.

\item In the cold and dense P1 and S environments, the low-$J$ lines of $\rm H^{13}CN$, $\rm HN^{13}C$, $\rm H^{13}CO^+$, and $\rm N_2H^+$ are subthermally excited. Nevertheless, these  lines can reasonably be assumed to be optically thin. %Moreover, the total column density of the $\rm ^{13}C$- and deuterated isotopologues are insensitive to the temperature changes at $\rm <25\,K$.
For the $\rm CH_3OH$, the total column density derived from the RD is  consistent with that from LVG fitting. Therefore, the gradient or enrichment we see from P1 to S on the D-fraction maps of individual species is real.

\item  The D-fraction of $\rm N_2H^+$ is several percent in the filament from P1 to S, which is higher than the D-fraction of $\rm HCO^+$, HCN, and HNC by an order of magnitude. This is consistent with previous D-fraction studies toward the young protostellar objects  in  IRDCs. Specifically,  D-fraction of these species is higher toward S compared with P1, indicating that deuterium fractionations of these species favor the colder and denser environment, in agreement with previous findings. In contrast, single deuterium fractionation of $\rm NH_3$, which forms both in the gas phase and on dust grain surfaces, does not show significant enrichment anywhere on the map.
%{(\color{red}  SIyi, 1. Can we have a sound conclusion that the deuterium fractionation is species dependent, i.e., gas-forming species (N2H+, HCO+, HCN, HNC) have more efficient deuterium fractionation in low-temperature environment, while NH3 and CH2DOH do not follow such trend because they are partially formed on grain surface;
% 2.  Is there any other ref  to recommend for a quantification of the correlation between CO depletion and deuterium fractionation of different species, besides those I mentioned in Section 3.8 about N2D+ ?)}

{ 
\item A high level of CO depletion is mapped toward the P1--S region, with the depletion maximum toward P1, S, and Soff  $\rm \sim10$. This feature of widespread (parsec-scale) CO depletion is also detected in our follow-up large-sample study of the 70\,$\mu$m dark clouds.
}

\item We detect high level of a single D-fraction ($\rm >10\%$) of $\rm CH_3OH$ toward our source, which was not detected by pointing observations in previous studies. In particular, $D(\rm CH_3OH)$ shows significant enrichment (up to $\sim40\%$) toward the location 20\arcsec~ NE of the continuum peak S, which coincides with the location where CO has the maximum depletion. This is consistent with  modeling predictions, where $\rm CH_3OH$ is formed on grain surfaces as soon as CO starts to catastrophy called freezes out.

\item G28.34 P1 and S provide us a good space laboratory in which to study the initial chemical condition of HMSF. Because they are neighbors in the same natal cloud, the chemical differentiation is the result of their evolutionary stages instead of the  external environmental effects (e.g., interstellar UV heating).   This comparative study illustrates the potential of measuring the D-fraction of different species as a tool in diagnosing the evolutionary stage of an HMSF region.  Observations on a larger sample of IRDCs with similar 70\,$\mu$m dark/bright source pairs have been carried out to support this pilot study result. Moreover, high-spatial resolution observations are essential to eliminate  beam dilution problems, and are key to determine the the extremely young protostellar objects embedded in the IRDCs, as well as to understand the correlation between the source evolutionary stage and the deuterium fractionation of different species.

 \end{enumerate}

\acknowledgments
We would like to thank the IRAM-30\,m staff for
their helpful support during the performance of the 
IRAM-30\,m observations in service mode. 

We thank R. Estalella for help with the HfS program. 

We thank F. J. Du, Y. Wang, and Z. Y. Zhang  for helpful discussions.  

S. F. acknowledges the support of the EACOA fellowship from the East Asia Core Observatories Association, which consists of the National Astronomical Observatory of China, the National Astronomical Observatory of Japan, the Academia Sinica Institute of Astronomy and Astrophysics, and the Korea Astronomy and Space Science Institute. 

K. W. acknowledges support by
the National Key Research and Development Program of China (2017YFA0402702),
the National Science Foundation of China (11721303),
and the starting grant at the Kavli Institute for Astronomy and Astrophysics, Peking University (7101502016).

H. B. acknowledges support from the European Research Council under the Horizon 2020 Framework Program via the ERC Consolidator Grant CSF-648505.

\software{GILDAS/CLASS \citep{pety05}, HfS \citep{estalella17}, PySpecKit \citep{ginsburg11}, RADEX \citep{vandertak07}}

\bibliographystyle{aasjournal}
%\bibliography{G2834_deuteration-sub.bbl}
\bibliography{/Users/siyifeng/link2GD/HMSFR.bib}

\begin{thebibliography}{}
\expandafter\ifx\csname natexlab\endcsname\relax\def\natexlab#1{#1}\fi
\providecommand{\url}[1]{\href{#1}{#1}}
\providecommand{\dodoi}[1]{doi:~\href{http://doi.org/#1}{\nolinkurl{#1}}}
\providecommand{\doeprint}[1]{\href{http://ascl.net/#1}{\nolinkurl{http://ascl.net/#1}}}
\providecommand{\doarXiv}[1]{\href{https://arxiv.org/abs/#1}{\nolinkurl{https://arxiv.org/abs/#1}}}

\bibitem[{{Adande} \& {Ziurys}(2012)}]{adande12}
{Adande}, G.~R., \& {Ziurys}, L.~M. 2012, \apj, 744, 194,
  \dodoi{10.1088/0004-637X/744/2/194}

\bibitem[{{Aikawa} {et~al.}(2005){Aikawa}, {Herbst}, {Roberts}, \&
  {Caselli}}]{aikawa05}
{Aikawa}, Y., {Herbst}, E., {Roberts}, H., \& {Caselli}, P. 2005, \apj, 620,
  330, \dodoi{10.1086/427017}

\bibitem[{{Aikawa} {et~al.}(2008){Aikawa}, {Wakelam}, {Garrod}, \&
  {Herbst}}]{aikawa08}
{Aikawa}, Y., {Wakelam}, V., {Garrod}, R.~T., \& {Herbst}, E. 2008, \apj, 674,
  984, \dodoi{10.1086/524096}

\bibitem[{{Aikawa} {et~al.}(2012){Aikawa}, {Wakelam}, {Hersant}, {Garrod}, \&
  {Herbst}}]{aikawa12}
{Aikawa}, Y., {Wakelam}, V., {Hersant}, F., {Garrod}, R.~T., \& {Herbst}, E.
  2012, \apj, 760, 40, \dodoi{10.1088/0004-637X/760/1/40}

\bibitem[{{Albertsson} {et~al.}(2013){Albertsson}, {Semenov}, {Vasyunin},
  {Henning}, \& {Herbst}}]{Albertsson_ea13}
{Albertsson}, T., {Semenov}, D.~A., {Vasyunin}, A.~I., {Henning}, T., \&
  {Herbst}, E. 2013, \apjs, 207, 27, \dodoi{10.1088/0067-0049/207/2/27}

\bibitem[{{Ao} {et~al.}(2013){Ao}, {Henkel}, {Menten}, {Requena-Torres},
  {Stanke}, {Mauersberger}, {Aalto}, {M{\"u}hle}, \& {Mangum}}]{ao13}
{Ao}, Y., {Henkel}, C., {Menten}, K.~M., {et~al.} 2013, \aap, 550, A135,
  \dodoi{10.1051/0004-6361/201220096}

\bibitem[{{Awad} {et~al.}(2014){Awad}, {Viti}, {Bayet}, \& {Caselli}}]{awad14}
{Awad}, Z., {Viti}, S., {Bayet}, E., \& {Caselli}, P. 2014, \mnras, 443, 275,
  \dodoi{10.1093/mnras/stu1141}

\bibitem[{{Bacmann} {et~al.}(2003){Bacmann}, {Lefloch}, {Ceccarelli},
  {Steinacker}, {Castets}, \& {Loinard}}]{bacmann03}
{Bacmann}, A., {Lefloch}, B., {Ceccarelli}, C., {et~al.} 2003, \apjl, 585, L55,
  \dodoi{10.1086/374263}

\bibitem[{{Barnes} {et~al.}(2016){Barnes}, {Kong}, {Tan}, {Henshaw}, {Caselli},
  {Jim{\'e}nez-Serra}, \& {Fontani}}]{barnes16}
{Barnes}, A.~T., {Kong}, S., {Tan}, J.~C., {et~al.} 2016, \mnras, 458, 1990,
  \dodoi{10.1093/mnras/stw403}

\bibitem[{{Barone} {et~al.}(2015){Barone}, {Latouche}, {Skouteris}, {Vazart},
  {Balucani}, {Ceccarelli}, \& {Lefloch}}]{barone15}
{Barone}, V., {Latouche}, C., {Skouteris}, D., {et~al.} 2015, \mnras, 453, L31,
  \dodoi{10.1093/mnrasl/slv094}

\bibitem[{{Battersby} {et~al.}(2014){Battersby}, {Ginsburg}, {Bally},
  {Longmore}, {Dunham}, \& {Darling}}]{battersby14}
{Battersby}, C., {Ginsburg}, A., {Bally}, J., {et~al.} 2014, \apj, 787, 113,
  \dodoi{10.1088/0004-637X/787/2/113}

\bibitem[{{Bergin} {et~al.}(2002){Bergin}, {Alves}, {Huard}, \&
  {Lada}}]{bergin02}
{Bergin}, E.~A., {Alves}, J., {Huard}, T., \& {Lada}, C.~J. 2002, \apjl, 570,
  L101, \dodoi{10.1086/340950}

\bibitem[{{Bergin} {et~al.}(1999){Bergin}, {Plume}, {Williams}, \&
  {Myers}}]{bergin99}
{Bergin}, E.~A., {Plume}, R., {Williams}, J.~P., \& {Myers}, P.~C. 1999, \apj,
  512, 724, \dodoi{10.1086/306791}

\bibitem[{{Beuther} \& {Sridharan}(2007)}]{beuther07c}
{Beuther}, H., \& {Sridharan}, T.~K. 2007, \apj, 668, 348,
  \dodoi{10.1086/521142}

\bibitem[{{Bihr} {et~al.}(2015){Bihr}, {Beuther}, {Linz}, {Ragan}, {Hennemann},
  {Tackenberg}, {Smith}, {Krause}, \& {Henning}}]{bihr15}
{Bihr}, S., {Beuther}, H., {Linz}, H., {et~al.} 2015, \aap, 579, A51,
  \dodoi{10.1051/0004-6361/201321269}

\bibitem[{{Bizzocchi} {et~al.}(2014){Bizzocchi}, {Caselli}, {Spezzano}, \&
  {Leonardo}}]{bizzocchi14}
{Bizzocchi}, L., {Caselli}, P., {Spezzano}, S., \& {Leonardo}, E. 2014, \aap,
  569, A27, \dodoi{10.1051/0004-6361/201423858}

\bibitem[{{Butler} \& {Tan}(2009)}]{butler09}
{Butler}, M.~J., \& {Tan}, J.~C. 2009, \apj, 696, 484,
  \dodoi{10.1088/0004-637X/696/1/484}

\bibitem[{{Butler} \& {Tan}(2012)}]{butler12}
---. 2012, \apj, 754, 5, \dodoi{10.1088/0004-637X/754/1/5}

\bibitem[{{Carey} {et~al.}(2009){Carey}, {Noriega-Crespo}, {Mizuno}, {Shenoy},
  {Paladini}, {Kraemer}, {Price}, {Flagey}, {Ryan}, {Ingalls}, {Kuchar},
  {Pinheiro Gon{\c c}alves}, {Indebetouw}, {Billot}, {Marleau}, {Padgett},
  {Rebull}, {Bressert}, {Ali}, {Molinari}, {Martin}, {Berriman}, {Boulanger},
  {Latter}, {Miville-Deschenes}, {Shipman}, \& {Testi}}]{carey09}
{Carey}, S.~J., {Noriega-Crespo}, A., {Mizuno}, D.~R., {et~al.} 2009, \pasp,
  121, 76, \dodoi{10.1086/596581}

\bibitem[{{Carter} {et~al.}(2012){Carter}, {Lazareff}, {Maier}, {Chenu},
  {Fontana}, {Bortolotti}, {Boucher}, {Navarrini}, {Blanchet}, {Greve}, {John},
  {Kramer}, {Morel}, {Navarro}, {Pe{\~n}alver}, {Schuster}, \&
  {Thum}}]{carter12}
{Carter}, M., {Lazareff}, B., {Maier}, D., {et~al.} 2012, \aap, 538, A89,
  \dodoi{10.1051/0004-6361/201118452}

\bibitem[{{Caselli} {et~al.}(2002{\natexlab{a}}){Caselli}, {Benson}, {Myers},
  \& {Tafalla}}]{caselli02c}
{Caselli}, P., {Benson}, P.~J., {Myers}, P.~C., \& {Tafalla}, M.
  2002{\natexlab{a}}, \apj, 572, 238, \dodoi{10.1086/340195}

\bibitem[{{Caselli} {et~al.}(1993){Caselli}, {Hasegawa}, \&
  {Herbst}}]{caselli93}
{Caselli}, P., {Hasegawa}, T.~I., \& {Herbst}, E. 1993, \apj, 408, 548,
  \dodoi{10.1086/172612}

\bibitem[{{Caselli} {et~al.}(1999){Caselli}, {Walmsley}, {Tafalla}, {Dore}, \&
  {Myers}}]{caselli99}
{Caselli}, P., {Walmsley}, C.~M., {Tafalla}, M., {Dore}, L., \& {Myers}, P.~C.
  1999, \apjl, 523, L165, \dodoi{10.1086/312280}

\bibitem[{{Caselli} {et~al.}(2002{\natexlab{b}}){Caselli}, {Walmsley},
  {Zucconi}, {Tafalla}, {Dore}, \& {Myers}}]{caselli02b}
{Caselli}, P., {Walmsley}, C.~M., {Zucconi}, A., {et~al.} 2002{\natexlab{b}},
  \apj, 565, 344, \dodoi{10.1086/324302}

\bibitem[{{Caselli} {et~al.}(2002{\natexlab{c}}){Caselli}, {Walmsley},
  {Zucconi}, {Tafalla}, {Dore}, \& {Myers}}]{caselli12}
---. 2002{\natexlab{c}}, \apj, 565, 331, \dodoi{10.1086/324301}

\bibitem[{{Ceccarelli} {et~al.}(2014){Ceccarelli}, {Caselli},
  {Bockel{\'e}e-Morvan}, {Mousis}, {Pizzarello}, {Robert}, \&
  {Semenov}}]{ceccarelli14}
{Ceccarelli}, C., {Caselli}, P., {Bockel{\'e}e-Morvan}, D., {et~al.} 2014,
  Protostars and Planets VI, 859,
  \dodoi{10.2458/azu_uapress_9780816531240-ch037}

\bibitem[{{Ceccarelli} {et~al.}(2007){Ceccarelli}, {Caselli}, {Herbst},
  {Tielens}, \& {Caux}}]{ceccarelli07}
{Ceccarelli}, C., {Caselli}, P., {Herbst}, E., {Tielens}, A.~G.~G.~M., \&
  {Caux}, E. 2007, in Protostars and Planets V, ed. B.~{Reipurth}, D.~{Jewitt},
  \& K.~{Keil}, 47.
\newblock \doarXiv{astro-ph/0603018}

\bibitem[{{Chen} {et~al.}(2010{\natexlab{a}}){Chen}, {Liu}, {Su}, \&
  {Zhang}}]{chenhr10}
{Chen}, H.-R., {Liu}, S.-Y., {Su}, Y.-N., \& {Zhang}, Q. 2010{\natexlab{a}},
  \apjl, 713, L50, \dodoi{10.1088/2041-8205/713/1/L50}

\bibitem[{{Chen} {et~al.}(2010{\natexlab{b}}){Chen}, {Liu}, {Su}, \&
  {Zhang}}]{chen10}
---. 2010{\natexlab{b}}, \apjl, 713, L50, \dodoi{10.1088/2041-8205/713/1/L50}

\bibitem[{{Chira} {et~al.}(2013){Chira}, {Beuther}, {Linz}, {Schuller},
  {Walmsley}, {Menten}, \& {Bronfman}}]{chira13}
{Chira}, R.-A., {Beuther}, H., {Linz}, H., {et~al.} 2013, \aap, 552, A40,
  \dodoi{10.1051/0004-6361/201219567}

\bibitem[{{Colzi} {et~al.}(2018){Colzi}, {Fontani}, {Caselli}, {Ceccarelli},
  {Hily-Blant}, \& {Bizzocchi}}]{colzi18}
{Colzi}, L., {Fontani}, F., {Caselli}, P., {et~al.} 2018, \aap, 609, A129,
  \dodoi{10.1051/0004-6361/201730576}

\bibitem[{{Crapsi} {et~al.}(2005){Crapsi}, {Caselli}, {Walmsley}, {Myers},
  {Tafalla}, {Lee}, \& {Bourke}}]{crapsi05}
{Crapsi}, A., {Caselli}, P., {Walmsley}, C.~M., {et~al.} 2005, \apj, 619, 379,
  \dodoi{10.1086/426472}

\bibitem[{{Crapsi} {et~al.}(2007){Crapsi}, {Caselli}, {Walmsley}, \&
  {Tafalla}}]{crapsi07}
{Crapsi}, A., {Caselli}, P., {Walmsley}, M.~C., \& {Tafalla}, M. 2007, \aap,
  470, 221, \dodoi{10.1051/0004-6361:20077613}

\bibitem[{{Csengeri} {et~al.}(2017){Csengeri}, {Bontemps}, {Wyrowski}, {Motte},
  {Menten}, {Beuther}, {Bronfman}, {Commer{\c c}on}, {Chapillon},
  {Duarte-Cabral}, {Fuller}, {Henning}, {Leurini}, {Longmore}, {Palau},
  {Peretto}, {Schuller}, {Tan}, {Testi}, {Traficante}, \&
  {Urquhart}}]{csengeri17}
{Csengeri}, T., {Bontemps}, S., {Wyrowski}, F., {et~al.} 2017, \aap, 600, L10,
  \dodoi{10.1051/0004-6361/201629754}

\bibitem[{{Cyganowski} {et~al.}(2008){Cyganowski}, {Whitney}, {Holden},
  {Braden}, {Brogan}, {Churchwell}, {Indebetouw}, {Watson}, {Babler},
  {Benjamin}, {Gomez}, {Meade}, {Povich}, {Robitaille}, \&
  {Watson}}]{cyganowski08}
{Cyganowski}, C.~J., {Whitney}, B.~A., {Holden}, E., {et~al.} 2008, \aj, 136,
  2391, \dodoi{10.1088/0004-6256/136/6/2391}

\bibitem[{{Draine}(2003)}]{draine03}
{Draine}, B.~T. 2003, \araa, 41, 241,
  \dodoi{10.1146/annurev.astro.41.011802.094840}

\bibitem[{{Draine}(2011)}]{draine11}
---. 2011, {Physics of the Interstellar and Intergalactic Medium}

\bibitem[{{Emprechtinger} {et~al.}(2009){Emprechtinger}, {Caselli}, {Volgenau},
  {Stutzki}, \& {Wiedner}}]{emprechtinger09}
{Emprechtinger}, M., {Caselli}, P., {Volgenau}, N.~H., {Stutzki}, J., \&
  {Wiedner}, M.~C. 2009, \aap, 493, 89, \dodoi{10.1051/0004-6361:200810324}

\bibitem[{{Estalella}(2017)}]{estalella17}
{Estalella}, R. 2017, \pasp, 129, 025003,
  \dodoi{10.1088/1538-3873/129/972/025003}

\bibitem[{{Feng} {et~al.}(2015){Feng}, {Beuther}, {Henning}, {Semenov},
  {Palau}, \& {Mills}}]{feng15}
{Feng}, S., {Beuther}, H., {Henning}, T., {et~al.} 2015, \aap, 581, A71,
  \dodoi{10.1051/0004-6361/201322725}

\bibitem[{{Feng} {et~al.}(2016{\natexlab{a}}){Feng}, {Beuther}, {Semenov},
  {Henning}, {Linz}, {Mills}, \& {Teague}}]{feng16c}
{Feng}, S., {Beuther}, H., {Semenov}, D., {et~al.} 2016{\natexlab{a}}, \aap,
  593, A46, \dodoi{10.1051/0004-6361/201424912}

\bibitem[{{Feng} {et~al.}(2016{\natexlab{b}}){Feng}, {Beuther}, {Zhang},
  {Henning}, {Linz}, {Ragan}, \& {Smith}}]{feng16a}
{Feng}, S., {Beuther}, H., {Zhang}, Q., {et~al.} 2016{\natexlab{b}}, \aap, 592,
  A21, \dodoi{10.1051/0004-6361/201526864}

\bibitem[{{Feng} {et~al.}(2016{\natexlab{c}}){Feng}, {Beuther}, {Zhang}, {Liu},
  {Zhang}, {Wang}, \& {Qiu}}]{feng16b}
---. 2016{\natexlab{c}}, \apj, 828, 100, \dodoi{10.3847/0004-637X/828/2/100}

\bibitem[{{Feroz} \& {Hobson}(2008)}]{feroz07}
{Feroz}, F., \& {Hobson}, M.~P. 2008, \mnras, 384, 449,
  \dodoi{10.1111/j.1365-2966.2007.12353.x}

\bibitem[{{Feroz} {et~al.}(2009){Feroz}, {Hobson}, \& {Bridges}}]{feroz08}
{Feroz}, F., {Hobson}, M.~P., \& {Bridges}, M. 2009, \mnras, 398, 1601,
  \dodoi{10.1111/j.1365-2966.2009.14548.x}

\bibitem[{{Feroz} {et~al.}(2013){Feroz}, {Hobson}, {Cameron}, \&
  {Pettitt}}]{feroz13}
{Feroz}, F., {Hobson}, M.~P., {Cameron}, E., \& {Pettitt}, A.~N. 2013, arXiv
  e-prints.
\newblock \doarXiv{1306.2144}

\bibitem[{{Fontani} {et~al.}(2015){Fontani}, {Busquet}, {Palau}, {Caselli},
  {S{\'a}nchez-Monge}, {Tan}, \& {Audard}}]{fontani15}
{Fontani}, F., {Busquet}, G., {Palau}, A., {et~al.} 2015, \aap, 575, A87,
  \dodoi{10.1051/0004-6361/201424753}

\bibitem[{{Fontani} {et~al.}(2006){Fontani}, {Caselli}, {Crapsi}, {Cesaroni},
  {Molinari}, {Testi}, \& {Brand}}]{fontani06}
{Fontani}, F., {Caselli}, P., {Crapsi}, A., {et~al.} 2006, \aap, 460, 709,
  \dodoi{10.1051/0004-6361:20066105}

\bibitem[{{Fontani} {et~al.}(2012){Fontani}, {Giannetti}, {Beltr{\'a}n},
  {Dodson}, {Rioja}, {Brand}, {Caselli}, \& {Cesaroni}}]{fontani12}
{Fontani}, F., {Giannetti}, A., {Beltr{\'a}n}, M.~T., {et~al.} 2012, \mnras,
  423, 2342, \dodoi{10.1111/j.1365-2966.2012.21043.x}

\bibitem[{{Fontani} {et~al.}(2014){Fontani}, {Sakai}, {Furuya}, {Sakai},
  {Aikawa}, \& {Yamamoto}}]{fontani14}
{Fontani}, F., {Sakai}, T., {Furuya}, K., {et~al.} 2014, \mnras, 440, 448,
  \dodoi{10.1093/mnras/stu298}

\bibitem[{{Fontani} {et~al.}(2011){Fontani}, {Palau}, {Caselli},
  {S{\'a}nchez-Monge}, {Butler}, {Tan}, {Jim{\'e}nez-Serra}, {Busquet},
  {Leurini}, \& {Audard}}]{fontani11}
{Fontani}, F., {Palau}, A., {Caselli}, P., {et~al.} 2011, \aap, 529, L7,
  \dodoi{10.1051/0004-6361/201116631}

\bibitem[{{Frerking} {et~al.}(1982){Frerking}, {Langer}, \&
  {Wilson}}]{frerking82}
{Frerking}, M.~A., {Langer}, W.~D., \& {Wilson}, R.~W. 1982, \apj, 262, 590,
  \dodoi{10.1086/160451}

\bibitem[{{Furuya} {et~al.}(2011){Furuya}, {Aikawa}, {Sakai}, \&
  {Yamamoto}}]{furuya11}
{Furuya}, K., {Aikawa}, Y., {Sakai}, N., \& {Yamamoto}, S. 2011, \apj, 731, 38,
  \dodoi{10.1088/0004-637X/731/1/38}

\bibitem[{{Garrod} {et~al.}(2007){Garrod}, {Wakelam}, \&
  {Herbst}}]{Garrod_ea07}
{Garrod}, R.~T., {Wakelam}, V., \& {Herbst}, E. 2007, \aap, 467, 1103,
  \dodoi{10.1051/0004-6361:20066704}

\bibitem[{{Gerner} {et~al.}(2015){Gerner}, {Shirley}, {Beuther}, {Semenov},
  {Linz}, {Albertsson}, \& {Henning}}]{gerner15}
{Gerner}, T., {Shirley}, Y.~L., {Beuther}, H., {et~al.} 2015, \aap, 579, A80,
  \dodoi{10.1051/0004-6361/201423989}

\bibitem[{{Giannetti} {et~al.}(2017){Giannetti}, {Leurini}, {Wyrowski},
  {Urquhart}, {Csengeri}, {Menten}, {K{\"o}nig}, \& {G{\"u}sten}}]{giannetti17}
{Giannetti}, A., {Leurini}, S., {Wyrowski}, F., {et~al.} 2017, \aap, 603, A33,
  \dodoi{10.1051/0004-6361/201630048}

\bibitem[{{Giannetti} {et~al.}(2014){Giannetti}, {Wyrowski}, {Brand},
  {Csengeri}, {Fontani}, {Walmsley}, {Nguyen Luong}, {Beuther}, {Schuller},
  {G{\"u}sten}, \& {Menten}}]{giannetti14}
{Giannetti}, A., {Wyrowski}, F., {Brand}, J., {et~al.} 2014, \aap, 570, A65,
  \dodoi{10.1051/0004-6361/201423692}

\bibitem[{{Ginsburg} \& {Mirocha}(2011)}]{ginsburg11}
{Ginsburg}, A., \& {Mirocha}, J. 2011, {PySpecKit: Python Spectroscopic
  Toolkit}, Astrophysics Source Code Library.
\newblock \doeprint{1109.001}

\bibitem[{{Ginsburg} {et~al.}(2016){Ginsburg}, {Henkel}, {Ao}, {Riquelme},
  {Kauffmann}, {Pillai}, {Mills}, {Requena-Torres}, {Immer}, {Testi}, {Ott},
  {Bally}, {Battersby}, {Darling}, {Aalto}, {Stanke}, {Kendrew}, {Kruijssen},
  {Longmore}, {Dale}, {Guesten}, \& {Menten}}]{ginsburg16}
{Ginsburg}, A., {Henkel}, C., {Ao}, Y., {et~al.} 2016, \aap, 586, A50,
  \dodoi{10.1051/0004-6361/201526100}

\bibitem[{{Graninger} {et~al.}(2014){Graninger}, {Herbst}, {{\"O}berg}, \&
  {Vasyunin}}]{graninger14}
{Graninger}, D.~M., {Herbst}, E., {{\"O}berg}, K.~I., \& {Vasyunin}, A.~I.
  2014, \apj, 787, 74, \dodoi{10.1088/0004-637X/787/1/74}

\bibitem[{{Harju} {et~al.}(2017){Harju}, {Daniel}, {Sipil{\"a}}, {Caselli},
  {Pineda}, {Friesen}, {Punanova}, {G{\"u}sten}, {Wiesenfeld}, {Myers},
  {Faure}, {Hily-Blant}, {Rist}, {Rosolowsky}, {Schlemmer}, \&
  {Shirley}}]{harju17}
{Harju}, J., {Daniel}, F., {Sipil{\"a}}, O., {et~al.} 2017, \aap, 600, A61,
  \dodoi{10.1051/0004-6361/201628463}

\bibitem[{{Henshaw} {et~al.}(2016){Henshaw}, {Caselli}, {Fontani},
  {Jim{\'e}nez-Serra}, {Tan}, {Longmore}, {Pineda}, {Parker}, \&
  {Barnes}}]{henshaw16}
{Henshaw}, J.~D., {Caselli}, P., {Fontani}, F., {et~al.} 2016, \mnras, 463,
  146, \dodoi{10.1093/mnras/stw1794}

\bibitem[{{Hernandez} {et~al.}(2011){Hernandez}, {Tan}, {Caselli}, {Butler},
  {Jim{\'e}nez-Serra}, {Fontani}, \& {Barnes}}]{hernandez11}
{Hernandez}, A.~K., {Tan}, J.~C., {Caselli}, P., {et~al.} 2011, \apj, 738, 11,
  \dodoi{10.1088/0004-637X/738/1/11}

\bibitem[{{Hidaka} {et~al.}(2004){Hidaka}, {Watanabe}, {Shiraki}, {Nagaoka}, \&
  {Kouchi}}]{hidaka04}
{Hidaka}, H., {Watanabe}, N., {Shiraki}, T., {Nagaoka}, A., \& {Kouchi}, A.
  2004, \apj, 614, 1124, \dodoi{10.1086/423889}

\bibitem[{{Hiraoka} {et~al.}(2006){Hiraoka}, {Ushiama}, {Enoura}, {Unagiike},
  {Mochizuki}, \& {Wada}}]{hiraoka06}
{Hiraoka}, K., {Ushiama}, S., {Enoura}, T., {et~al.} 2006, \apj, 643, 917,
  \dodoi{10.1086/501517}

\bibitem[{{Ho} \& {Townes}(1983)}]{ho83}
{Ho}, P.~T.~P., \& {Townes}, C.~H. 1983, \araa, 21, 239,
  \dodoi{10.1146/annurev.aa.21.090183.001323}

\bibitem[{{Jim{\'e}nez-Serra} {et~al.}(2010){Jim{\'e}nez-Serra}, {Caselli},
  {Tan}, {Hernandez}, {Fontani}, {Butler}, \& {van Loo}}]{jimenez10}
{Jim{\'e}nez-Serra}, I., {Caselli}, P., {Tan}, J.~C., {et~al.} 2010, \mnras,
  406, 187, \dodoi{10.1111/j.1365-2966.2010.16698.x}

\bibitem[{{Johnstone} {et~al.}(2003){Johnstone}, {Boonman}, \& {van
  Dishoeck}}]{johnston03}
{Johnstone}, D., {Boonman}, A.~M.~S., \& {van Dishoeck}, E.~F. 2003, \aap, 412,
  157, \dodoi{10.1051/0004-6361:20031370}

\bibitem[{{J{\o}rgensen} {et~al.}(2004){J{\o}rgensen}, {Sch{\"o}ier}, \& {van
  Dishoeck}}]{jorgensen04}
{J{\o}rgensen}, J.~K., {Sch{\"o}ier}, F.~L., \& {van Dishoeck}, E.~F. 2004,
  \aap, 416, 603, \dodoi{10.1051/0004-6361:20034440}

\bibitem[{{Juvela} \& {Ysard}(2011)}]{juvela11}
{Juvela}, M., \& {Ysard}, N. 2011, \apj, 739, 63,
  \dodoi{10.1088/0004-637X/739/2/63}

\bibitem[{{Kalenskii} \& {Kurtz}(2016)}]{kalenskii16}
{Kalenskii}, S.~V., \& {Kurtz}, S. 2016, Astronomy Reports, 60, 702,
  \dodoi{10.1134/S1063772916080047}

\bibitem[{{Kong} {et~al.}(2015){Kong}, {Caselli}, {Tan}, {Wakelam}, \&
  {Sipil{\"a}}}]{kong15}
{Kong}, S., {Caselli}, P., {Tan}, J.~C., {Wakelam}, V., \& {Sipil{\"a}}, O.
  2015, \apj, 804, 98, \dodoi{10.1088/0004-637X/804/2/98}

\bibitem[{{Kong} {et~al.}(2018){Kong}, {Tan}, {Caselli}, {Fontani}, {Wang}, \&
  {Butler}}]{kong18}
{Kong}, S., {Tan}, J.~C., {Caselli}, P., {et~al.} 2018, \apj, 867, 94,
  \dodoi{10.3847/1538-4357/aae1b2}

\bibitem[{{Leurini} {et~al.}(2016){Leurini}, {Menten}, \&
  {Walmsley}}]{leurini16}
{Leurini}, S., {Menten}, K.~M., \& {Walmsley}, C.~M. 2016, \aap, 592, A31,
  \dodoi{10.1051/0004-6361/201527974}

\bibitem[{{Leurini} {et~al.}(2004){Leurini}, {Schilke}, {Menten}, {Flower},
  {Pottage}, \& {Xu}}]{leurini04}
{Leurini}, S., {Schilke}, P., {Menten}, K.~M., {et~al.} 2004, \aap, 422, 573,
  \dodoi{10.1051/0004-6361:20047046}

\bibitem[{{Leurini} {et~al.}(2007){Leurini}, {Schilke}, {Wyrowski}, \&
  {Menten}}]{leurini07}
{Leurini}, S., {Schilke}, P., {Wyrowski}, F., \& {Menten}, K.~M. 2007, \aap,
  466, 215, \dodoi{10.1051/0004-6361:20054245}

\bibitem[{{Lin} {et~al.}(2017){Lin}, {Liu}, {Dale}, {Li}, {Busquet}, {Zhang},
  {Ginsburg}, {Galv{\'a}n-Madrid}, {Kov{\'a}cs}, {Koch}, {Qian}, {Wang},
  {Longmore}, {Chen}, \& {Walker}}]{lin17}
{Lin}, Y., {Liu}, H.~B., {Dale}, J.~E., {et~al.} 2017, \apj, 840, 22,
  \dodoi{10.3847/1538-4357/aa6c67}

\bibitem[{{Linsky} {et~al.}(2006){Linsky}, {Draine}, {Moos}, {Jenkins}, {Wood},
  {Oliveira}, {Blair}, {Friedman}, {Gry}, {Knauth}, {Kruk}, {Lacour}, {Lehner},
  {Redfield}, {Shull}, {Sonneborn}, \& {Williger}}]{linsky06}
{Linsky}, J.~L., {Draine}, B.~T., {Moos}, H.~W., {et~al.} 2006, \apj, 647,
  1106, \dodoi{10.1086/505556}

\bibitem[{{Lodders}(2003)}]{lodders03}
{Lodders}, K. 2003, \apj, 591, 1220, \dodoi{10.1086/375492}

\bibitem[{{Lucy}(1974)}]{lucy74}
{Lucy}, L.~B. 1974, \aj, 79, 745, \dodoi{10.1086/111605}

\bibitem[{{Mangum} {et~al.}(2007){Mangum}, {Emerson}, \& {Greisen}}]{mangum07}
{Mangum}, J.~G., {Emerson}, D.~T., \& {Greisen}, E.~W. 2007, \aap, 474, 679,
  \dodoi{10.1051/0004-6361:20077811}

\bibitem[{{Mangum} \& {Shirley}(2015)}]{mangum15}
{Mangum}, J.~G., \& {Shirley}, Y.~L. 2015, \pasp, 127, 266,
  \dodoi{10.1086/680323}

\bibitem[{{Mangum} \& {Wootten}(1993)}]{mangum93}
{Mangum}, J.~G., \& {Wootten}, A. 1993, \apjs, 89, 123, \dodoi{10.1086/191841}

\bibitem[{{Maret} {et~al.}(2009){Maret}, {Faure}, {Scifoni}, \&
  {Wiesenfeld}}]{maret09}
{Maret}, S., {Faure}, A., {Scifoni}, E., \& {Wiesenfeld}, L. 2009, \mnras, 399,
  425, \dodoi{10.1111/j.1365-2966.2009.15294.x}

\bibitem[{{Maret} {et~al.}(2011){Maret}, {Hily-Blant}, {Pety}, {Bardeau}, \&
  {Reynier}}]{maret11}
{Maret}, S., {Hily-Blant}, P., {Pety}, J., {Bardeau}, S., \& {Reynier}, E.
  2011, \aap, 526, A47, \dodoi{10.1051/0004-6361/201015487}

\bibitem[{{McKee} \& {Tan}(2003)}]{mckee03}
{McKee}, C.~F., \& {Tan}, J.~C. 2003, \apj, 585, 850, \dodoi{10.1086/346149}

\bibitem[{{Miettinen} {et~al.}(2011){Miettinen}, {Hennemann}, \&
  {Linz}}]{miettinen11}
{Miettinen}, O., {Hennemann}, M., \& {Linz}, H. 2011, \aap, 534, A134,
  \dodoi{10.1051/0004-6361/201117187}

\bibitem[{{Millar} {et~al.}(1989){Millar}, {Bennett}, \& {Herbst}}]{millar89}
{Millar}, T.~J., {Bennett}, A., \& {Herbst}, E. 1989, \apj, 340, 906,
  \dodoi{10.1086/167444}

\bibitem[{{Molinari} {et~al.}(2010){Molinari}, {Swinyard}, {Bally}, {Barlow},
  {Bernard}, {Martin}, {Moore}, {Noriega-Crespo}, {Plume}, {Testi}, {Zavagno},
  {Abergel}, {Ali}, {Anderson}, {Andr{\'e}}, {Baluteau}, {Battersby},
  {Beltr{\'a}n}, {Benedettini}, {Billot}, {Blommaert}, {Bontemps}, {Boulanger},
  {Brand}, {Brunt}, {Burton}, {Calzoletti}, {Carey}, {Caselli}, {Cesaroni},
  {Cernicharo}, {Chakrabarti}, {Chrysostomou}, {Cohen}, {Compiegne}, {de
  Bernardis}, {de Gasperis}, {di Giorgio}, {Elia}, {Faustini}, {Flagey},
  {Fukui}, {Fuller}, {Ganga}, {Garcia-Lario}, {Glenn}, {Goldsmith}, {Griffin},
  {Hoare}, {Huang}, {Ikhenaode}, {Joblin}, {Joncas}, {Juvela}, {Kirk},
  {Lagache}, {Li}, {Lim}, {Lord}, {Marengo}, {Marshall}, {Masi}, {Massi},
  {Matsuura}, {Minier}, {Miville-Desch{\^e}nes}, {Montier}, {Morgan}, {Motte},
  {Mottram}, {M{\"u}ller}, {Natoli}, {Neves}, {Olmi}, {Paladini}, {Paradis},
  {Parsons}, {Peretto}, {Pestalozzi}, {Pezzuto}, {Piacentini}, {Piazzo},
  {Polychroni}, {Pomar{\`e}s}, {Popescu}, {Reach}, {Ristorcelli}, {Robitaille},
  {Robitaille}, {Rod{\'o}n}, {Roy}, {Royer}, {Russeil}, {Saraceno}, {Sauvage},
  {Schilke}, {Schisano}, {Schneider}, {Schuller}, {Schulz}, {Sibthorpe},
  {Smith}, {Smith}, {Spinoglio}, {Stamatellos}, {Strafella}, {Stringfellow},
  {Sturm}, {Taylor}, {Thompson}, {Traficante}, {Tuffs}, {Umana}, {Valenziano},
  {Vavrek}, {Veneziani}, {Viti}, {Waelkens}, {Ward-Thompson}, {White},
  {Wilcock}, {Wyrowski}, {Yorke}, \& {Zhang}}]{molinari10}
{Molinari}, S., {Swinyard}, B., {Bally}, J., {et~al.} 2010, \aap, 518, L100,
  \dodoi{10.1051/0004-6361/201014659}

\bibitem[{{M{\"u}ller} {et~al.}(2005){M{\"u}ller}, {Schl{\"o}der}, {Stutzki},
  \& {Winnewisser}}]{muller05}
{M{\"u}ller}, H.~S.~P., {Schl{\"o}der}, F., {Stutzki}, J., \& {Winnewisser}, G.
  2005, Journal of Molecular Structure, 742, 215,
  \dodoi{10.1016/j.molstruc.2005.01.027}

\bibitem[{{M{\"u}ller} {et~al.}(2002){M{\"u}ller}, {Shirley}, {Evans}, \&
  {Jacobson}}]{mueller02}
{M{\"u}ller}, K.~E., {Shirley}, Y.~L., {Evans}, II, N.~J., \& {Jacobson}, H.~R.
  2002, \apjs, 143, 469, \dodoi{10.1086/342881}

\bibitem[{{Oliveira} {et~al.}(2003){Oliveira}, {H{\'e}brard}, {Howk}, {Kruk},
  {Chayer}, \& {Moos}}]{oliveira03}
{Oliveira}, C.~M., {H{\'e}brard}, G., {Howk}, J.~C., {et~al.} 2003, \apj, 587,
  235, \dodoi{10.1086/368019}

\bibitem[{{Ossenkopf} \& {Henning}(1994)}]{ossenkopf94}
{Ossenkopf}, V., \& {Henning}, T. 1994, \aap, 291, 943

\bibitem[{{Padovani} {et~al.}(2011){Padovani}, {Walmsley}, {Tafalla},
  {Hily-Blant}, \& {Pineau Des For{\^e}ts}}]{padovani11}
{Padovani}, M., {Walmsley}, C.~M., {Tafalla}, M., {Hily-Blant}, P., \& {Pineau
  Des For{\^e}ts}, G. 2011, \aap, 534, A77, \dodoi{10.1051/0004-6361/201117134}

\bibitem[{{Parise} {et~al.}(2006){Parise}, {Ceccarelli}, {Tielens}, {Castets},
  {Caux}, {Lefloch}, \& {Maret}}]{parise06}
{Parise}, B., {Ceccarelli}, C., {Tielens}, A.~G.~G.~M., {et~al.} 2006, \aap,
  453, 949, \dodoi{10.1051/0004-6361:20054476}

\bibitem[{{Parise} {et~al.}(2002){Parise}, {Ceccarelli}, {Tielens}, {Herbst},
  {Lefloch}, {Caux}, {Castets}, {Mukhopadhyay}, {Pagani}, \&
  {Loinard}}]{parise02}
---. 2002, \aap, 393, L49, \dodoi{10.1051/0004-6361:20021131}

\bibitem[{{Peretto} {et~al.}(2013){Peretto}, {Fuller}, {Duarte-Cabral},
  {Avison}, {Hennebelle}, {Pineda}, {Andr{\'e}}, {Bontemps}, {Motte},
  {Schneider}, \& {Molinari}}]{peretto13}
{Peretto}, N., {Fuller}, G.~A., {Duarte-Cabral}, A., {et~al.} 2013, \aap, 555,
  A112, \dodoi{10.1051/0004-6361/201321318}

\bibitem[{{Pety}(2005)}]{pety05}
{Pety}, J. 2005, in SF2A-2005: Semaine de l'Astrophysique Francaise, ed.
  F.~{Casoli}, T.~{Contini}, J.~M. {Hameury}, \& L.~{Pagani}, 721

\bibitem[{{Pickett} {et~al.}(1998){Pickett}, {Poynter}, {Cohen}, {Delitsky},
  {Pearson}, \& {M{\"u}ller}}]{pickett98}
{Pickett}, H.~M., {Poynter}, R.~L., {Cohen}, E.~A., {et~al.} 1998, \jqsrt, 60,
  883, \dodoi{10.1016/S0022-4073(98)00091-0}

\bibitem[{{Pillai} {et~al.}(2011){Pillai}, {Kauffmann}, {Wyrowski}, {Hatchell},
  {Gibb}, \& {Thompson}}]{pillai11}
{Pillai}, T., {Kauffmann}, J., {Wyrowski}, F., {et~al.} 2011, \aap, 530, A118,
  \dodoi{10.1051/0004-6361/201015899}

\bibitem[{{Pillai} {et~al.}(2006){Pillai}, {Wyrowski}, {Carey}, \&
  {Menten}}]{pillai06}
{Pillai}, T., {Wyrowski}, F., {Carey}, S.~J., \& {Menten}, K.~M. 2006, \aap,
  450, 569, \dodoi{10.1051/0004-6361:20054128}

\bibitem[{{Pon} {et~al.}(2016{\natexlab{a}}){Pon}, {Kaufman}, {Johnstone},
  {Caselli}, {Fontani}, {Butler}, {Jim{\'e}nez-Serra}, {Palau}, \&
  {Tan}}]{pon16b}
{Pon}, A., {Kaufman}, M.~J., {Johnstone}, D., {et~al.} 2016{\natexlab{a}},
  \apj, 827, 107, \dodoi{10.3847/0004-637X/827/2/107}

\bibitem[{{Pon} {et~al.}(2016{\natexlab{b}}){Pon}, {Johnstone}, {Caselli},
  {Fontani}, {Palau}, {Butler}, {Kaufman}, {Jim{\'e}nez-Serra}, \&
  {Tan}}]{pon16a}
{Pon}, A., {Johnstone}, D., {Caselli}, P., {et~al.} 2016{\natexlab{b}}, \aap,
  587, A96, \dodoi{10.1051/0004-6361/201527154}

\bibitem[{{Prodanovi{\'c}} {et~al.}(2010){Prodanovi{\'c}}, {Steigman}, \&
  {Fields}}]{prodanovic10}
{Prodanovi{\'c}}, T., {Steigman}, G., \& {Fields}, B.~D. 2010, \mnras, 406,
  1108, \dodoi{10.1111/j.1365-2966.2010.16734.x}

\bibitem[{{Rabli} \& {Flower}(2010)}]{rabli10}
{Rabli}, D., \& {Flower}, D.~R. 2010, \mnras, 406, 95,
  \dodoi{10.1111/j.1365-2966.2010.16671.x}

\bibitem[{{Ragan} {et~al.}(2012){Ragan}, {Henning}, {Krause}, {Pitann},
  {Beuther}, {Linz}, {Tackenberg}, {Balog}, {Hennemann}, {Launhardt}, {Lippok},
  {Nielbock}, {Schmiedeke}, {Schuller}, {Steinacker}, {Stutz}, \&
  {Vasyunina}}]{ragan12}
{Ragan}, S., {Henning}, T., {Krause}, O., {et~al.} 2012, \aap, 547, A49,
  \dodoi{10.1051/0004-6361/201219232}

\bibitem[{{Ragan} {et~al.}(2009){Ragan}, {Bergin}, \& {Gutermuth}}]{ragan09}
{Ragan}, S.~E., {Bergin}, E.~A., \& {Gutermuth}, R.~A. 2009, \apj, 698, 324,
  \dodoi{10.1088/0004-637X/698/1/324}

\bibitem[{{Ragan} {et~al.}(2011){Ragan}, {Bergin}, \& {Wilner}}]{ragan11}
{Ragan}, S.~E., {Bergin}, E.~A., \& {Wilner}, D. 2011, \apj, 736, 163,
  \dodoi{10.1088/0004-637X/736/2/163}

\bibitem[{{Ragan} {et~al.}(2013){Ragan}, {Henning}, \& {Beuther}}]{ragan13}
{Ragan}, S.~E., {Henning}, T., \& {Beuther}, H. 2013, \aap, 559, A79,
  \dodoi{10.1051/0004-6361/201321869}

\bibitem[{{Rathborne} {et~al.}(2006){Rathborne}, {Jackson}, \&
  {Simon}}]{rathborne06}
{Rathborne}, J.~M., {Jackson}, J.~M., \& {Simon}, R. 2006, \apj, 641, 389,
  \dodoi{10.1086/500423}

\bibitem[{Richardson(1972)}]{richardson72}
Richardson, W.~H. 1972, J. Opt. Soc. Am., 62, 55,
  \dodoi{10.1364/JOSA.62.000055}

\bibitem[{{Robitaille} {et~al.}(2008){Robitaille}, {Meade}, {Babler},
  {Whitney}, {Johnston}, {Indebetouw}, {Cohen}, {Povich}, {Sewilo}, {Benjamin},
  \& {Churchwell}}]{robitaille08}
{Robitaille}, T.~P., {Meade}, M.~R., {Babler}, B.~L., {et~al.} 2008, \aj, 136,
  2413, \dodoi{10.1088/0004-6256/136/6/2413}

\bibitem[{{Rosolowsky} {et~al.}(2008){Rosolowsky}, {Pineda}, {Foster},
  {Borkin}, {Kauffmann}, {Caselli}, {Myers}, \& {Goodman}}]{rosolowsky08}
{Rosolowsky}, E.~W., {Pineda}, J.~E., {Foster}, J.~B., {et~al.} 2008, \apjs,
  175, 509, \dodoi{10.1086/524299}

\bibitem[{{Sakai} {et~al.}(2012){Sakai}, {Sakai}, {Furuya}, {Aikawa}, {Hirota},
  \& {Yamamoto}}]{sakai12}
{Sakai}, T., {Sakai}, N., {Furuya}, K., {et~al.} 2012, \apj, 747, 140,
  \dodoi{10.1088/0004-637X/747/2/140}

\bibitem[{{Sakai} {et~al.}(2008){Sakai}, {Sakai}, {Kamegai}, {Hirota},
  {Yamaguchi}, {Shiba}, \& {Yamamoto}}]{sakai08}
{Sakai}, T., {Sakai}, N., {Kamegai}, K., {et~al.} 2008, \apj, 678, 1049,
  \dodoi{10.1086/587050}

\bibitem[{{Sanhueza} {et~al.}(2017){Sanhueza}, {Jackson}, {Zhang},
  {Guzm{\'a}n}, {Lu}, {Stephens}, {Wang}, \& {Tatematsu}}]{sanhueza17}
{Sanhueza}, P., {Jackson}, J.~M., {Zhang}, Q., {et~al.} 2017, \apj, 841, 97,
  \dodoi{10.3847/1538-4357/aa6ff8}

\bibitem[{{Sch{\"o}ier} {et~al.}(2005){Sch{\"o}ier}, {van der Tak}, {van
  Dishoeck}, \& {Black}}]{schoier05}
{Sch{\"o}ier}, F.~L., {van der Tak}, F.~F.~S., {van Dishoeck}, E.~F., \&
  {Black}, J.~H. 2005, \aap, 432, 369, \dodoi{10.1051/0004-6361:20041729}

\bibitem[{{Schuller} {et~al.}(2009){Schuller}, {Menten}, {Contreras},
  {Wyrowski}, {Schilke}, {Bronfman}, {Henning}, {Walmsley}, {Beuther},
  {Bontemps}, {Cesaroni}, {Deharveng}, {Garay}, {Herpin}, {Lefloch}, {Linz},
  {Mardones}, {Minier}, {Molinari}, {Motte}, {Nyman}, {Reveret}, {Risacher},
  {Russeil}, {Schneider}, {Testi}, {Troost}, {Vasyunina}, {Wienen}, {Zavagno},
  {Kovacs}, {Kreysa}, {Siringo}, \& {Wei{\ss}}}]{schuller09}
{Schuller}, F., {Menten}, K.~M., {Contreras}, Y., {et~al.} 2009, \aap, 504,
  415, \dodoi{10.1051/0004-6361/200811568}

\bibitem[{{Shirley}(2015)}]{shirley15}
{Shirley}, Y.~L. 2015, \pasp, 127, 299, \dodoi{10.1086/680342}

\bibitem[{Shirley {et~al.}(2013)Shirley, Ellsworth-Bowers, Svoboda, Schlingman,
  Ginsburg, Rosolowsky, Gerner, Mairs, Battersby, Stringfellow, Dunham, Glenn,
  \& Bally}]{shirley13}
Shirley, Y.~L., Ellsworth-Bowers, T.~P., Svoboda, B., {et~al.} 2013, The
  Astrophysical Journal Supplement Series, 209, 2,
  \dodoi{10.1088/0067-0049/209/1/2}

\bibitem[{{Sipil{\"a}} {et~al.}(2015){Sipil{\"a}}, {Harju}, {Caselli}, \&
  {Schlemmer}}]{sipila15b}
{Sipil{\"a}}, O., {Harju}, J., {Caselli}, P., \& {Schlemmer}, S. 2015, \aap,
  581, A122, \dodoi{10.1051/0004-6361/201526468}

\bibitem[{{Sokolov} {et~al.}(2018){Sokolov}, {Wang}, {Pineda}, {Caselli},
  {Henshaw}, {Barnes}, {Tan}, {Fontani}, {Jim{\'e}nez-Serra}, \&
  {Zhang}}]{sokolov18}
{Sokolov}, V., {Wang}, K., {Pineda}, J.~E., {et~al.} 2018, \aap, 611, L3,
  \dodoi{10.1051/0004-6361/201832746}

\bibitem[{{Sridharan} {et~al.}(2005){Sridharan}, {Beuther}, {Saito},
  {Wyrowski}, \& {Schilke}}]{sridharan05}
{Sridharan}, T.~K., {Beuther}, H., {Saito}, M., {Wyrowski}, F., \& {Schilke},
  P. 2005, \apjl, 634, L57, \dodoi{10.1086/498644}

\bibitem[{{Svoboda} {et~al.}(2016){Svoboda}, {Shirley}, {Battersby},
  {Rosolowsky}, {Ginsburg}, {Ellsworth-Bowers}, {Pestalozzi}, {Dunham},
  {Evans}, {Bally}, \& {Glenn}}]{svoboda16}
{Svoboda}, B.~E., {Shirley}, Y.~L., {Battersby}, C., {et~al.} 2016, \apj, 822,
  59, \dodoi{10.3847/0004-637X/822/2/59}

\bibitem[{{Tan} {et~al.}(2014){Tan}, {Beltr{\'a}n}, {Caselli}, {Fontani},
  {Fuente}, {Krumholz}, {McKee}, \& {Stolte}}]{tan14}
{Tan}, J.~C., {Beltr{\'a}n}, M.~T., {Caselli}, P., {et~al.} 2014, Protostars
  and Planets VI, 149, \dodoi{10.2458/azu_uapress_9780816531240-ch007}

\bibitem[{{Tan} {et~al.}(2016){Tan}, {Kong}, {Zhang}, {Fontani}, {Caselli}, \&
  {Butler}}]{tan16}
{Tan}, J.~C., {Kong}, S., {Zhang}, Y., {et~al.} 2016, \apjl, 821, L3,
  \dodoi{10.3847/2041-8205/821/1/L3}

\bibitem[{{Tang} {et~al.}(2018){Tang}, {Henkel}, {Menten}, {Wyrowski},
  {Brinkmann}, {Zheng}, {Gong}, {Lin}, {Esimbek}, {Zhou}, {Yuan}, {Li}, \&
  {He}}]{tang18}
{Tang}, X.~D., {Henkel}, C., {Menten}, K.~M., {et~al.} 2018, \aap, 609, A16,
  \dodoi{10.1051/0004-6361/201731849}

\bibitem[{{Teyssier} {et~al.}(2002){Teyssier}, {Hennebelle}, \&
  {P{\'e}rault}}]{teyssier02}
{Teyssier}, D., {Hennebelle}, P., \& {P{\'e}rault}, M. 2002, \aap, 382, 624,
  \dodoi{10.1051/0004-6361:20011646}

\bibitem[{{Turner}(2001)}]{turner01}
{Turner}, B.~E. 2001, \apjs, 136, 579, \dodoi{10.1086/322536}

\bibitem[{{van der Tak} {et~al.}(2007){van der Tak}, {Black}, {Sch{\"o}ier},
  {Jansen}, \& {van Dishoeck}}]{vandertak07}
{van der Tak}, F.~F.~S., {Black}, J.~H., {Sch{\"o}ier}, F.~L., {Jansen}, D.~J.,
  \& {van Dishoeck}, E.~F. 2007, \aap, 468, 627,
  \dodoi{10.1051/0004-6361:20066820}

\bibitem[{{Vastel} {et~al.}(2006){Vastel}, {Phillips}, {Caselli}, {Ceccarelli},
  \& {Pagani}}]{vastel06}
{Vastel}, C., {Phillips}, T.~G., {Caselli}, P., {Ceccarelli}, C., \& {Pagani},
  L. 2006, Philosophical Transactions of the Royal Society of London Series A,
  364, 3081, \dodoi{10.1098/rsta.2006.1880}

\bibitem[{{Vasyunin} {et~al.}(2017){Vasyunin}, {Caselli}, {Dulieu}, \&
  {Jim{\'e}nez-Serra}}]{vasyunin17}
{Vasyunin}, A.~I., {Caselli}, P., {Dulieu}, F., \& {Jim{\'e}nez-Serra}, I.
  2017, \apj, 842, 33, \dodoi{10.3847/1538-4357/aa72ec}

\bibitem[{{Vasyunina} {et~al.}(2011){Vasyunina}, {Linz}, {Henning},
  {Zinchenko}, {Beuther}, \& {Voronkov}}]{vasyunin11}
{Vasyunina}, T., {Linz}, H., {Henning}, T., {et~al.} 2011, \aap, 527, A88,
  \dodoi{10.1051/0004-6361/201014974}

\bibitem[{{Walmsley} \& {Ungerechts}(1983)}]{walmsley83}
{Walmsley}, C.~M., \& {Ungerechts}, H. 1983, \aap, 122, 164

\bibitem[{{Wang}(2015)}]{wangk15b}
{Wang}, K. 2015, {The Earliest Stages of Massive Clustered Star Formation:
  Fragmentation of Infrared Dark Clouds}, \dodoi{10.1007/978-3-662-44969-1}

\bibitem[{{Wang}(2018)}]{wangk18}
---. 2018, Research Notes of the American Astronomical Society, 2, 52,
  \dodoi{10.3847/2515-5172/aacb29}

\bibitem[{{Wang} {et~al.}(2012){Wang}, {Zhang}, {Wu}, {Li}, \&
  {Zhang}}]{wangk12}
{Wang}, K., {Zhang}, Q., {Wu}, Y., {Li}, H.-b., \& {Zhang}, H. 2012, \apjl,
  745, L30, \dodoi{10.1088/2041-8205/745/2/L30}

\bibitem[{{Wang} {et~al.}(2011){Wang}, {Zhang}, {Wu}, \& {Zhang}}]{wangk11}
{Wang}, K., {Zhang}, Q., {Wu}, Y., \& {Zhang}, H. 2011, \apj, 735, 64,
  \dodoi{10.1088/0004-637X/735/1/64}

\bibitem[{{Wang} {et~al.}(2014){Wang}, {Zhang}, {Testi}, {Tak}, {Wu}, {Zhang},
  {Pillai}, {Wyrowski}, {Carey}, {Ragan}, \& {Henning}}]{wangk14}
{Wang}, K., {Zhang}, Q., {Testi}, L., {et~al.} 2014, \mnras, 439, 3275,
  \dodoi{10.1093/mnras/stu127}

\bibitem[{{Wang} {et~al.}(2008){Wang}, {Zhang}, {Pillai}, {Wyrowski}, \&
  {Wu}}]{wangy08}
{Wang}, Y., {Zhang}, Q., {Pillai}, T., {Wyrowski}, F., \& {Wu}, Y. 2008, \apjl,
  672, L33, \dodoi{10.1086/524949}

\bibitem[{{Watanabe} \& {Kouchi}(2002)}]{watanabe02}
{Watanabe}, N., \& {Kouchi}, A. 2002, \apj, 571, L173, \dodoi{10.1086/341412}

\bibitem[{{Wienen} {et~al.}(2012){Wienen}, {Wyrowski}, {Schuller}, {Menten},
  {Walmsley}, {Bronfman}, \& {Motte}}]{wienen12}
{Wienen}, M., {Wyrowski}, F., {Schuller}, F., {et~al.} 2012, \aap, 544, A146,
  \dodoi{10.1051/0004-6361/201118107}

\bibitem[{{Wilson} \& {Rood}(1994)}]{wilson94}
{Wilson}, T.~L., \& {Rood}, R. 1994, \araa, 32, 191,
  \dodoi{10.1146/annurev.aa.32.090194.001203}

\bibitem[{{Woon}(2002)}]{woon02}
{Woon}, D.~E. 2002, \apj, 569, 541, \dodoi{10.1086/339279}

\bibitem[{{Zeng} {et~al.}(2017){Zeng}, {Jim{\'e}nez-Serra}, {Cosentino},
  {Viti}, {Barnes}, {Henshaw}, {Caselli}, {Fontani}, \& {Hily-Blant}}]{zeng17}
{Zeng}, S., {Jim{\'e}nez-Serra}, I., {Cosentino}, G., {et~al.} 2017, \aap, 603,
  A22, \dodoi{10.1051/0004-6361/201630210}

\bibitem[{{Zhang} {et~al.}(2015){Zhang}, {Wang}, {Lu}, \&
  {Jim{\'e}nez-Serra}}]{zhang15}
{Zhang}, Q., {Wang}, K., {Lu}, X., \& {Jim{\'e}nez-Serra}, I. 2015, \apj, 804,
  141, \dodoi{10.1088/0004-637X/804/2/141}

\end{thebibliography}

%%%%%%%%%%%%%%%%%%%%%%%%%%%%%%%%%%%%%%%
%%%%%%%%%%%%%%%%%%%%%%%%%%%%%%%%%%%%%%%%
%%%%%%%%%%%%%%%%%%%%%%%%%%%%%%%%%%%%%%%% 
\newpage
\setcounter{section}{0}
\renewcommand{\thetable}{A\arabic{section}}
\setcounter{table}{0}
\renewcommand{\thetable}{A\arabic{table}}
\setcounter{figure}{0}
\renewcommand{\thefigure}{A\arabic{figure}}
\appendix

\section{Modified Blackbody Fitting for $\rm N(H_2)$ and $\rm T(dust)$ Maps}\label{blackbody}
We assume a dust opacity law of $\kappa_{\nu} = \kappa_{230}(\frac{\nu}{\rm 230\,GHz})^{\beta}$, where $ \rm \kappa_{230} = 0.899\,cm^{2}\,g^{-1}$ \citep{ossenkopf94}. The flux density $\rm S_{\nu}$ at the frequency of $\nu$ can be given as $\rm S_{\nu} = \Omega_{m}B_{\nu}(T_{d})(1-e^{-\tau_{\nu}})$ from the single-component modified blackbody model, where $\rm \Omega_{m}$ is the solid angle of the beam. The $\rm H_2$ column density can therefore be derived as $\rm N(H_{2}) =\it R\frac{\tau_{\nu}}{\kappa_{\nu}\mu_{ISM} \rm m_{H}}$, where $\mu_{ISM}$ is the is the mean molecular weight in the ISM, which is assumed to be 2.33;  $\rm m_H$ is the mass of an hydrogen atom ($\rm 1.67\times10^{-24}\,g$); and the isothermal gas-to-dust mass ratio $R$ is taken to be 150 \citep{draine03}.

To derive the $\rm H_2$ column density and dust temperature maps, we adopt the following iterative procedure pixel by pixel: \\
(1) Assuming $\rm \beta\sim1.8$, we derive a continuum image at  850\,$\mu$m  from the  extrapolation of {\it Herschel} PACS 70\,$\mu$m, 160\,$\mu$m, and SPIRE 250\,$\mu$m, 350\,$\mu$m, and 500\,$\mu$m images.\\
(2) Using this extrapolated 850\,$\mu$m continuum image as a model, we deconvolved the Planck 353\,GHz image using Lucy-Richardson method \citep{richardson72,lucy74}; the deconvolved 
image has an angular resolution close to the SPIRE 500\,$\mu$m image (37\arcsec) . The deconvolved image is 
combined with the {\it JCMT -SCUBA2}  850\,$\mu$m map, achieving an angular resolution of 14\arcsec.\\
(3) We fit the SED by using the  {\it Herschel} PACS 70\,$\mu$m, 160\,$\mu$m, SPIRE 250\,$\mu$m maps, the 850\,$\mu$m map derived from step (2), and the 350\,$\mu$m map by combining  {\it Herschel}  SPIRE and  {\it CSO/SHARC-II}.  At an angular resolution of 22\arcsec, we obtain a $\beta$ map.\\
(4) Assuming the $\beta$ map has no local variation from 22\arcsec~ to 10\arcsec~ resolution, we apply the  Monte Carlo method and fit $\rm N(H_{2})$ and $\rm T(dust)$ simultaneously by using the {\it CSO/SHARC-II}  350\,$\mu$m and {\it Herschel} PACS 70\,$\mu$m emission maps at an angular resolution of 10\arcsec.
Assuming a global flux error as 5$\sigma$ for each continuum map, we take $\rm N(H_{2})$ and $\rm T(dust)$ as well as their $\rm 1\sigma$ as uncertainty from the PDF.

The detailed combination and SED fitting procedures are given in \citet{lin17}.

\section{Figures}
\begin{figure}
\begin{center}
\includegraphics[height=10cm] {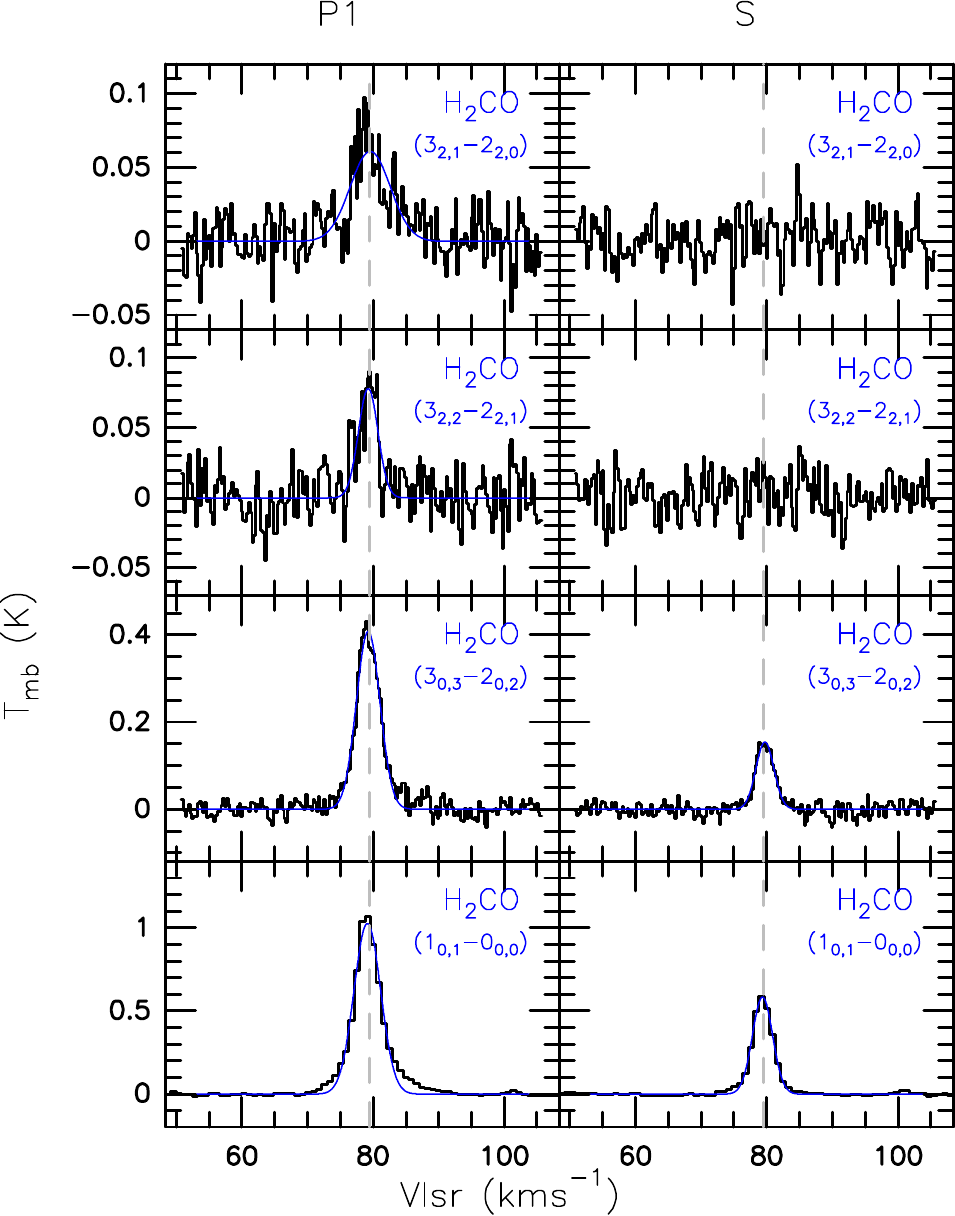}
\end{center}
\caption{Line profiles of identified $p$-$\rm H_2CO$ lines toward G28.34\,P1--S, averaged from a beam-sized region with the center on P1 and S  in the plane of the sky. All $p$-$\rm H_2CO$ lines are extracted from images smoothed to the same pixel size and angular resolution (35.6\arcsec). 
The best-fit parameters from GAUSS method are given in Table~\ref{tab:lineh2coch3oh}. 
The system velocity ($\rm 79.5\,km\,s^{-1}$) is shown as gray dashed line in each panel.
}\label{h2cospec}
\end{figure}

\begin{figure*}
\begin{center}
\includegraphics[height=10cm] {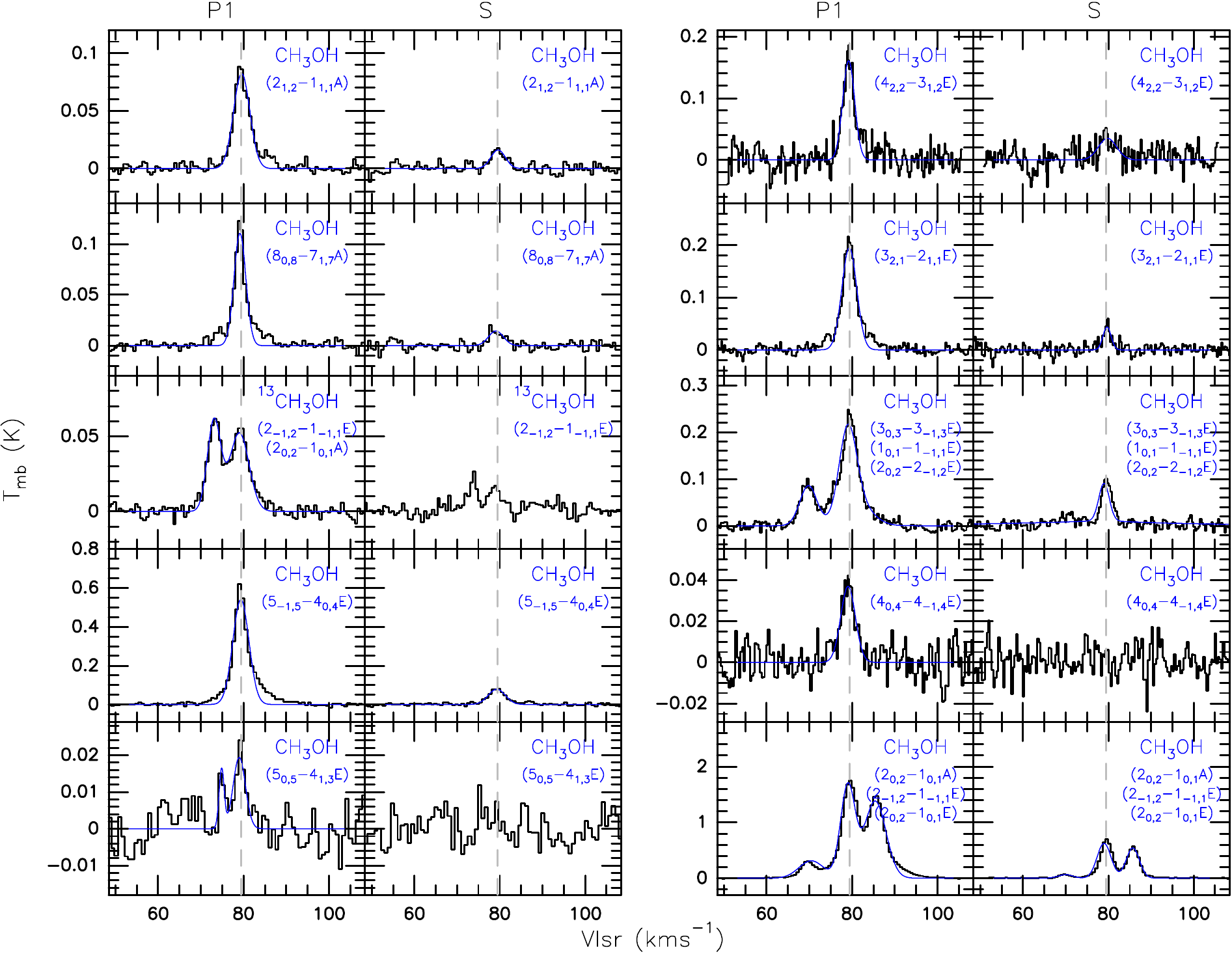}
\end{center}
\caption{Line profiles of the identified $\rm CH_3OH$ lines toward G28.34\,P1--S. The line intensity given in the main-beam temperature is averaged from a beam-sized region with the center on P1 or S  in the plane of the sky. All $\rm CH_3OH$ line images are smoothed to the same pixel size and angular resolution (33.9\arcsec).
 The best-fit parameters are given by using the GAUSS method, and the parameters are given in Table~\ref{tab:lineh2coch3oh}.
 The system velocity ($\rm 79.5\,km\,s^{-1}$) is shown as a gray dashed line in each panel.
}\label{ch3ohspec}
\end{figure*}

\begin{figure*}
\begin{center}

\subfigure[]{  \includegraphics[height=15cm] {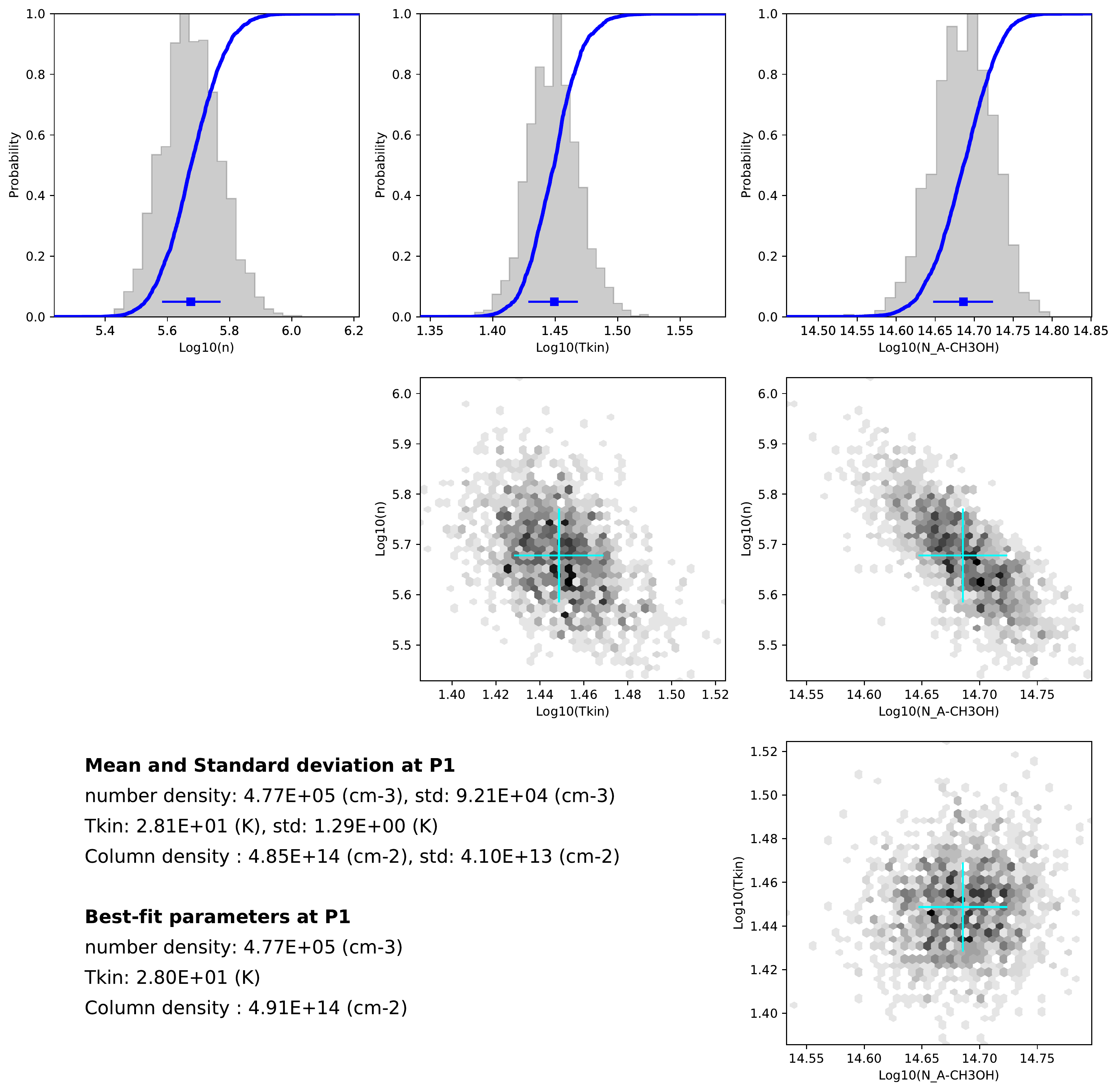}}
%\subfigure[]{  \includegraphics[height=8.7cm] {TkdV-S-ach3oh-marg.pdf}}
%\subfigure[]{ \includegraphics[height=8.7cm] {TkdV-Soff-ach3oh-marg.pdf}}
\end{center}
\caption{The probability density function (PDF) of A-$\rm CH_3OH$ parameters toward P1, derived by using RADEX and MultiNest Algorithm. The fitting parameters, including the $\rm H_2$ number density $n$, $\rm CH_3OH$ column density N(A-$\rm CH_3OH$),  the gas kinetic temperature $\rm T_{kin}$  are listed. 
}\label{a-ch3ohlvg}
\end{figure*}

\begin{figure*}
\begin{center}
%\begin{minipage}[c]{.28\textwidth}
%\subfigure[]{\includegraphics[height=4cm] {Trot-H2CO_rdg_sm-eps-converted-to.pdf}}
%\end{minipage}
%\begin{minipage}[c]{.28\textwidth}
%\subfigure[]{\includegraphics[height=4cm] {errTrot-H2CO_rdg_sm-eps-converted-to.pdf}}
%\end{minipage}
\begin{minipage}[c]{.8\textwidth}
\subfigure[]{\includegraphics[height=4cm] {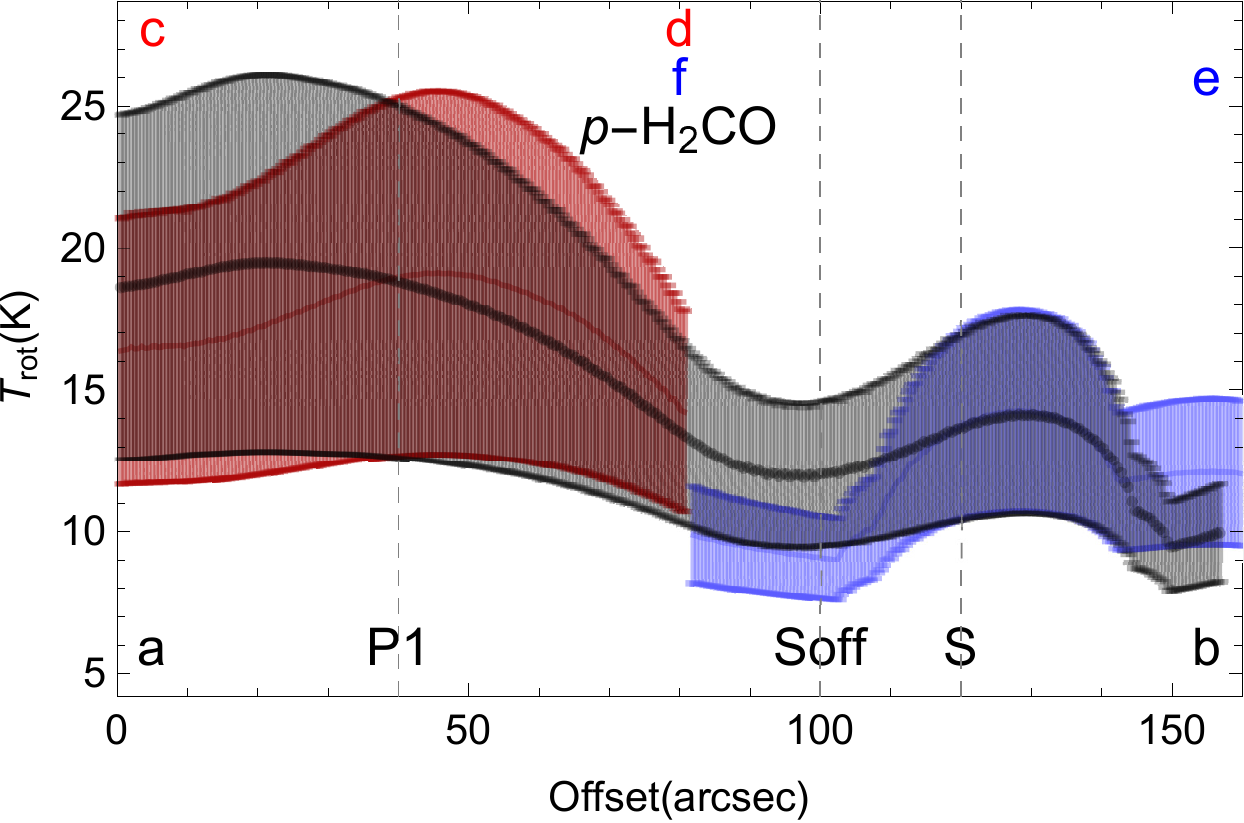}}
\subfigure[]{\includegraphics[height=4cm] {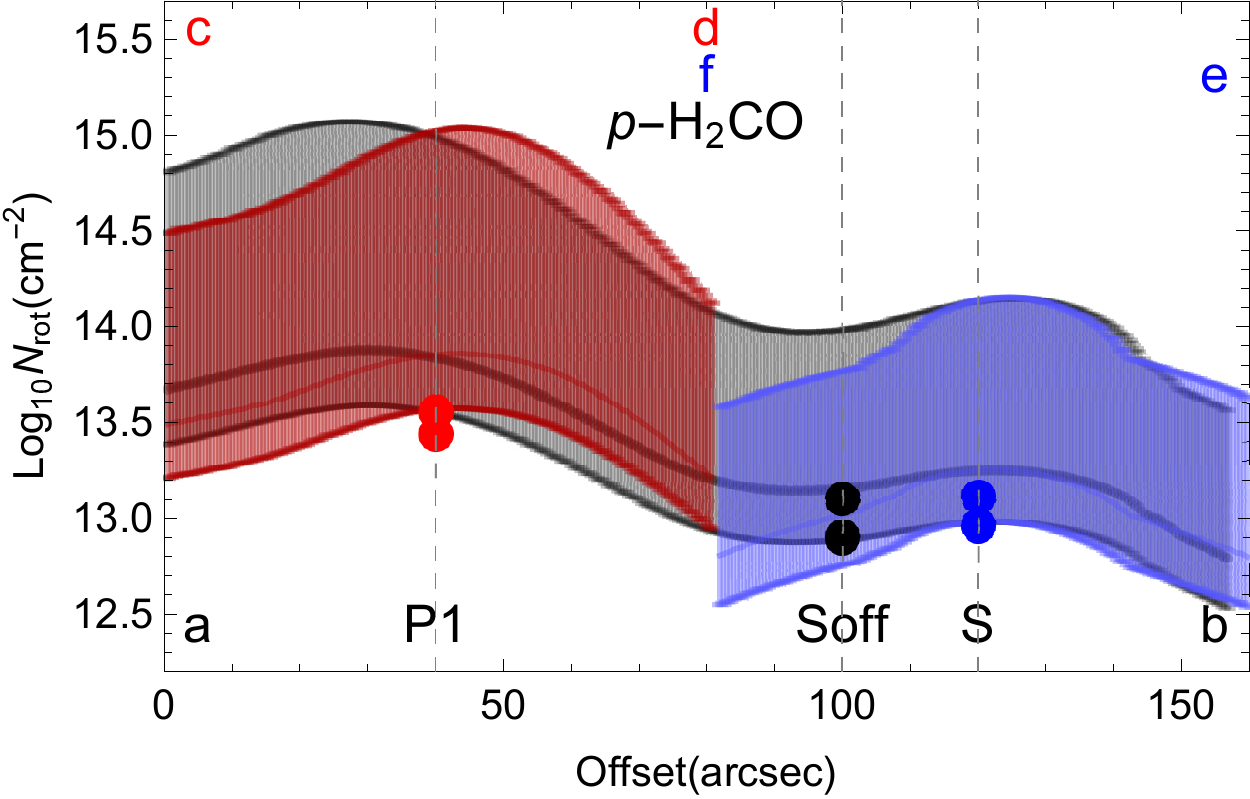}}
\end{minipage}
\\
%\begin{minipage}[c]{.28\textwidth}
%\subfigure[]{\includegraphics[height=4cm] {Nrot-H2CO_rdg_sm-eps-converted-to.pdf}}
%\end{minipage}
%\begin{minipage}[c]{.28\textwidth}
%\subfigure[]{\includegraphics[height=4cm] {errNTl-H2CO_rdg_sm-eps-converted-to.pdf}}
%\end{minipage}
\begin{minipage}[c]{.8\textwidth}
\subfigure[]{\includegraphics[height=4cm] {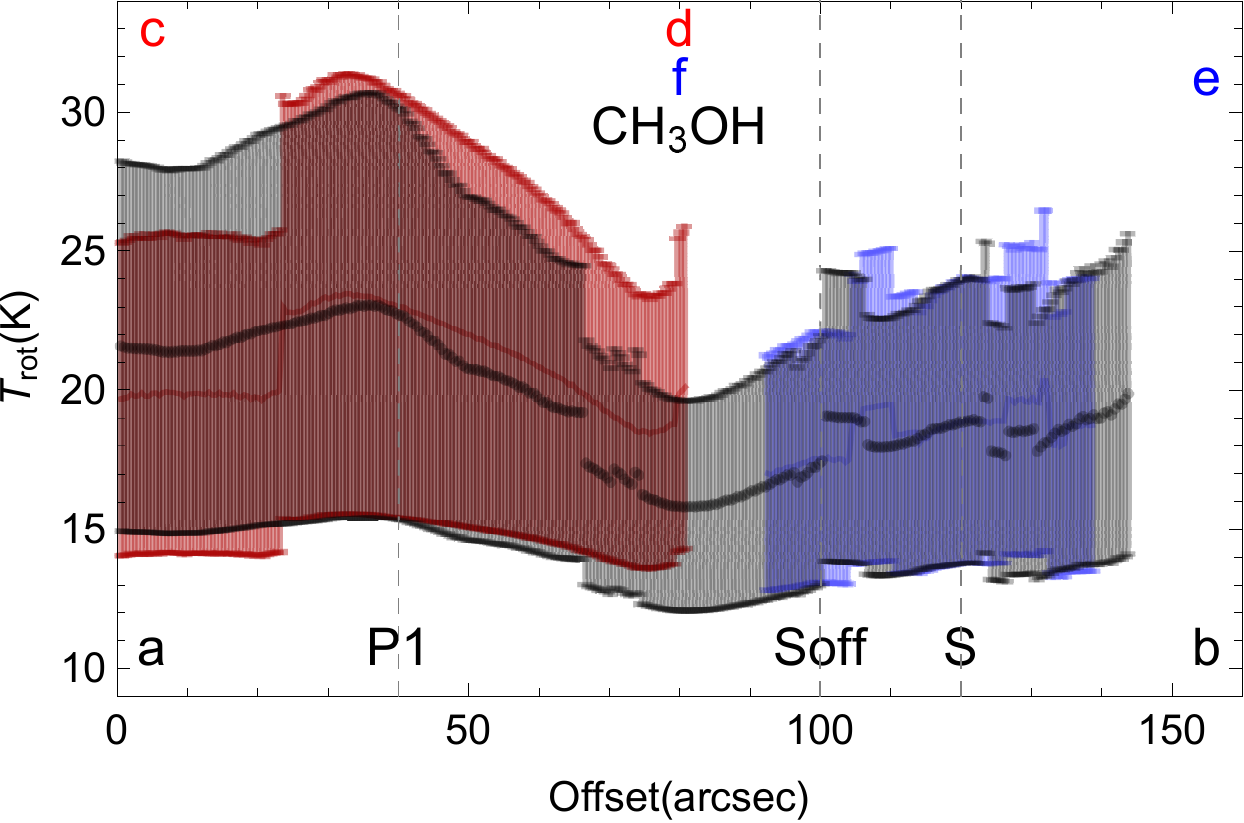}}
\subfigure[]{\includegraphics[height=4cm] {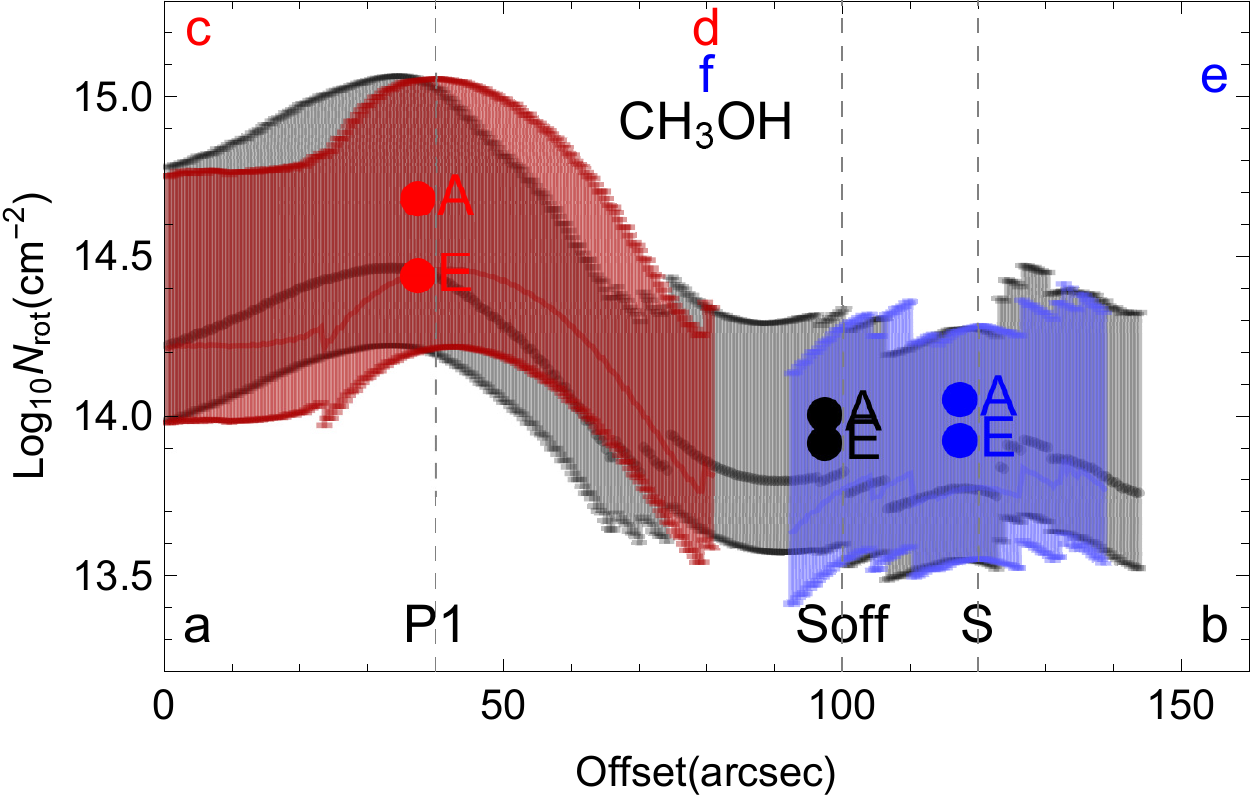}}
\end{minipage}
\\

\end{center}
\caption{{\it Panel I--II}: Profiles of the rotation temperature  $\rm T_{rot}$ and column density $\rm N_{rot}$ of $p$-$\rm H_2CO$, extracted along the directions of a--b (in black), c--d (in red), and e--f (in blue) in their maps. 
{\it Panel III--IV}: Profiles of the rotation temperature  $\rm T_{rot}$ and column density $\rm N_{rot}$ of $\rm CH_3OH$, extracted along the directions of a--b (in black), c--d (in red), and e--f (in blue) in their maps. 
The labeled positions are the same as those in Figure~\ref{dust}.  
The red, black, and blue dots indicate the upper and lower limits of column density toward P1, S, and Soff given by LVG fitting.
The pixels where any of the $p$-$\rm H_2CO$ or $\rm CH_3OH$ lines show $\rm <3\sigma$ integrated intensity  are blanked.
}\label{h2colte}
\end{figure*}

\begin{figure*}
\begin{center}
\begin{minipage}[c]{.45\textwidth}
\subfigure[]{\includegraphics[height=4cm] {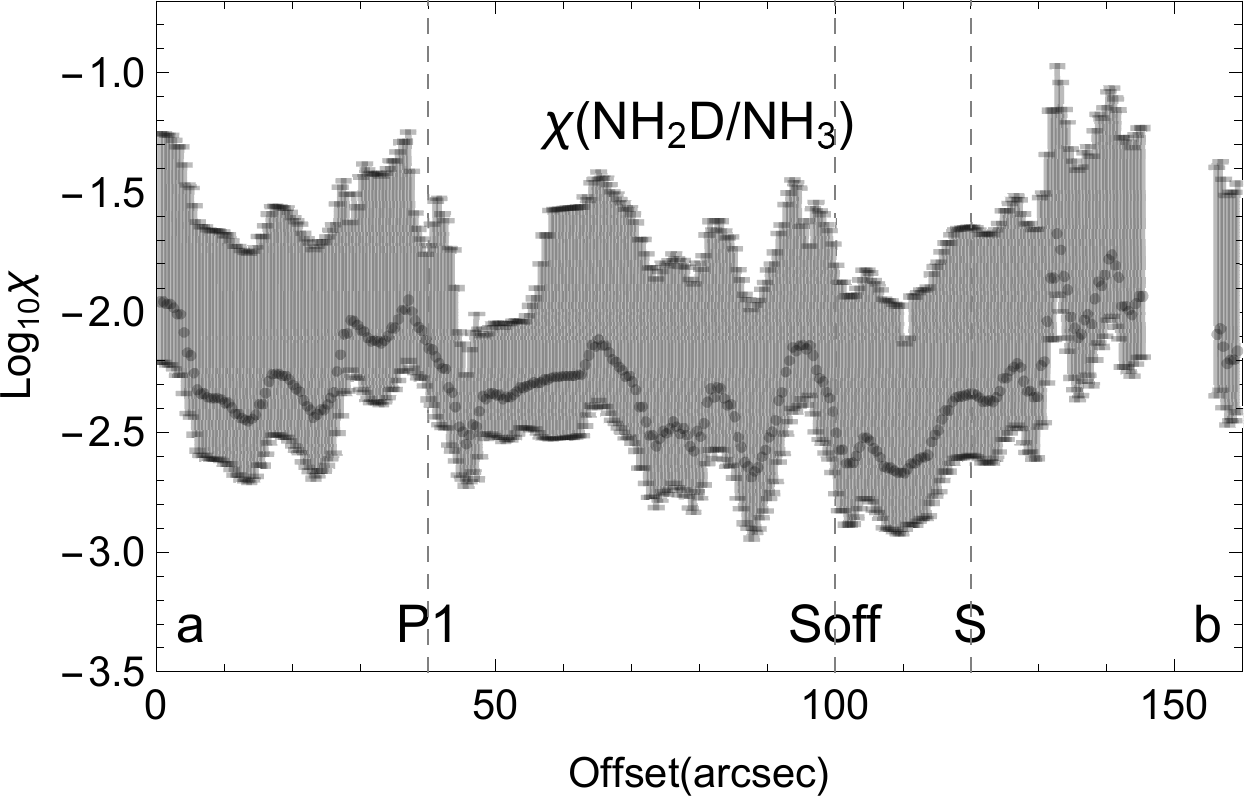}}
\end{minipage}
\begin{minipage}[c]{.45\textwidth}
\subfigure[]{\includegraphics[height=4cm] {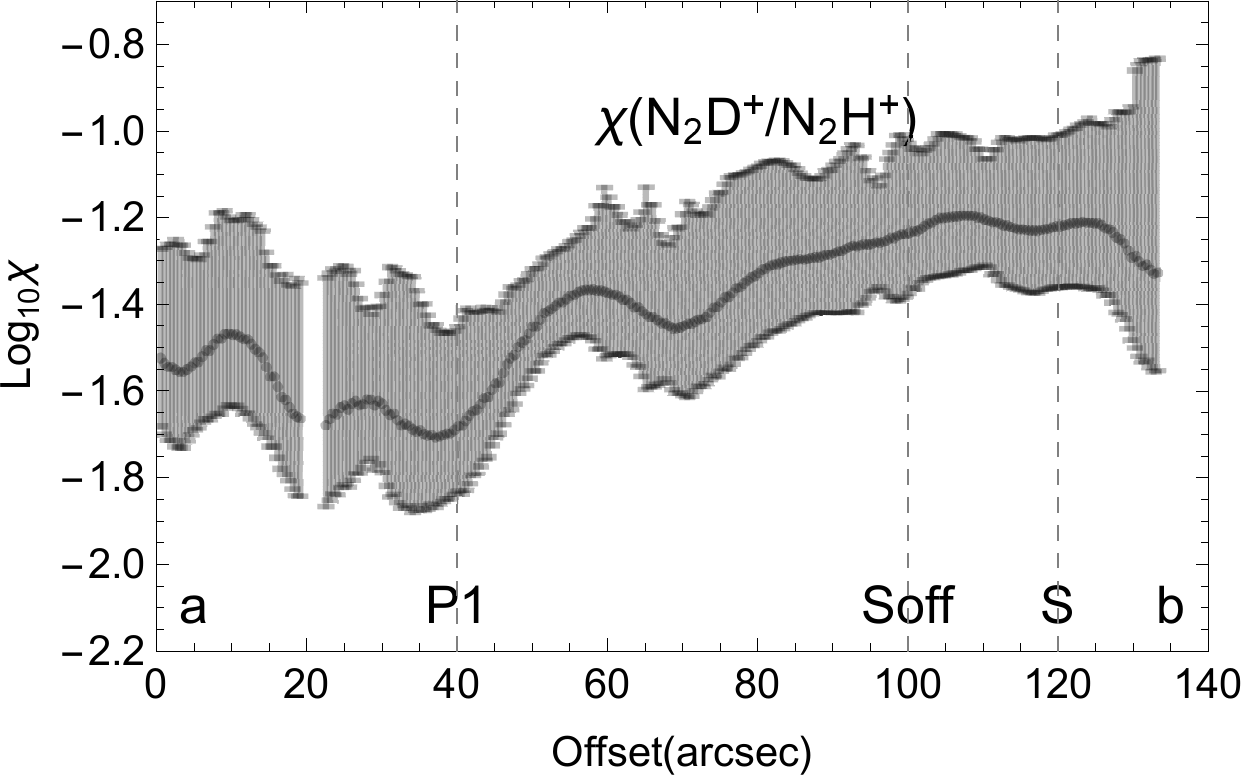}}
\end{minipage}
\\
\begin{minipage}[c]{.45\textwidth}
\subfigure[]{\includegraphics[height=4cm] {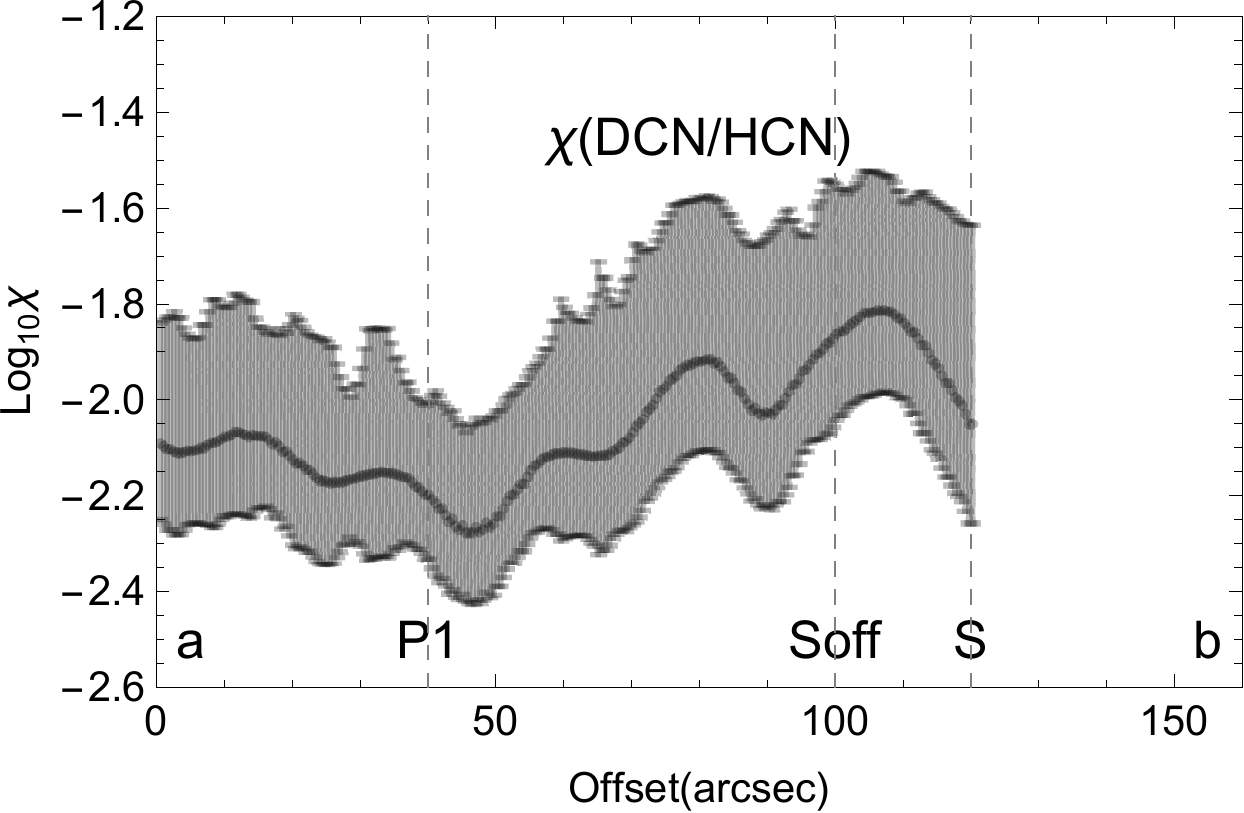}}
\end{minipage}
\begin{minipage}[c]{.45\textwidth}
\subfigure[]{\includegraphics[height=4cm] {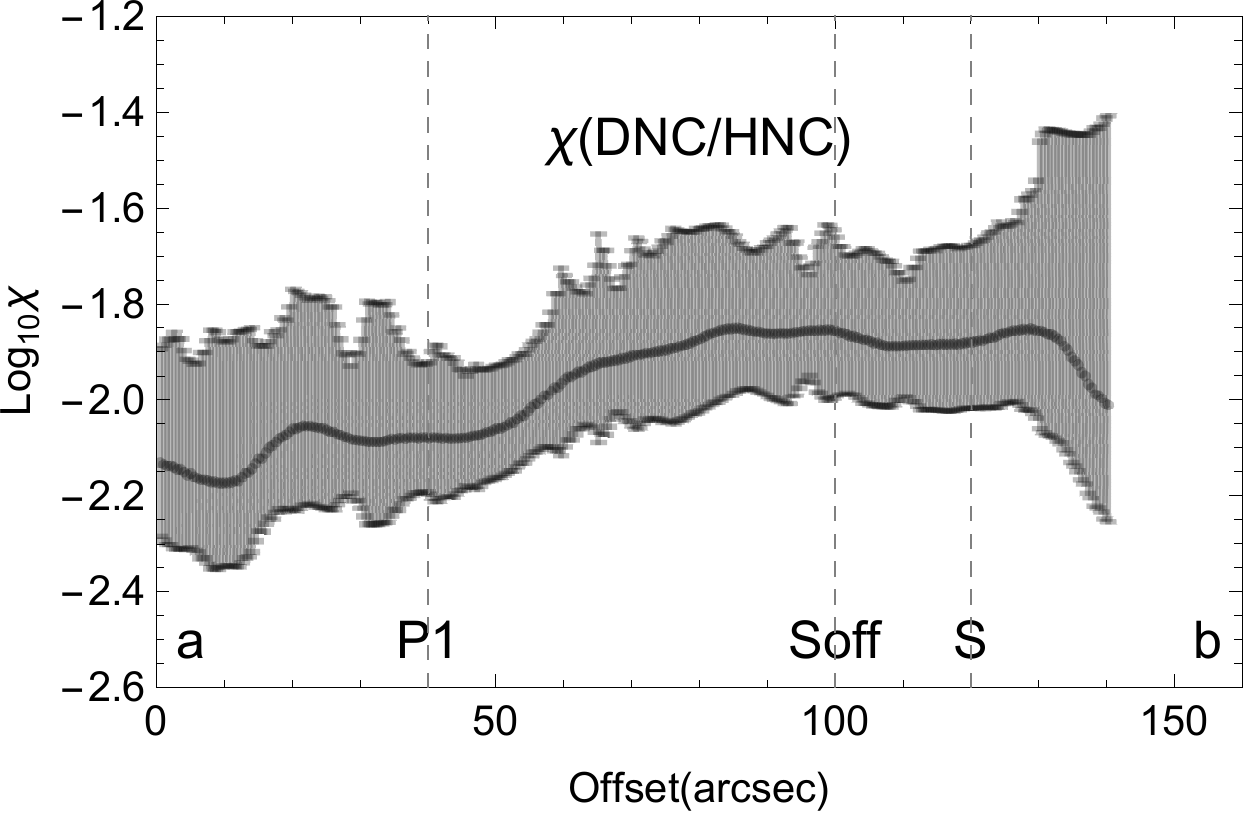}}
\end{minipage}
\\
\begin{minipage}[c]{.45\textwidth}
\subfigure[]{\includegraphics[height=4cm] {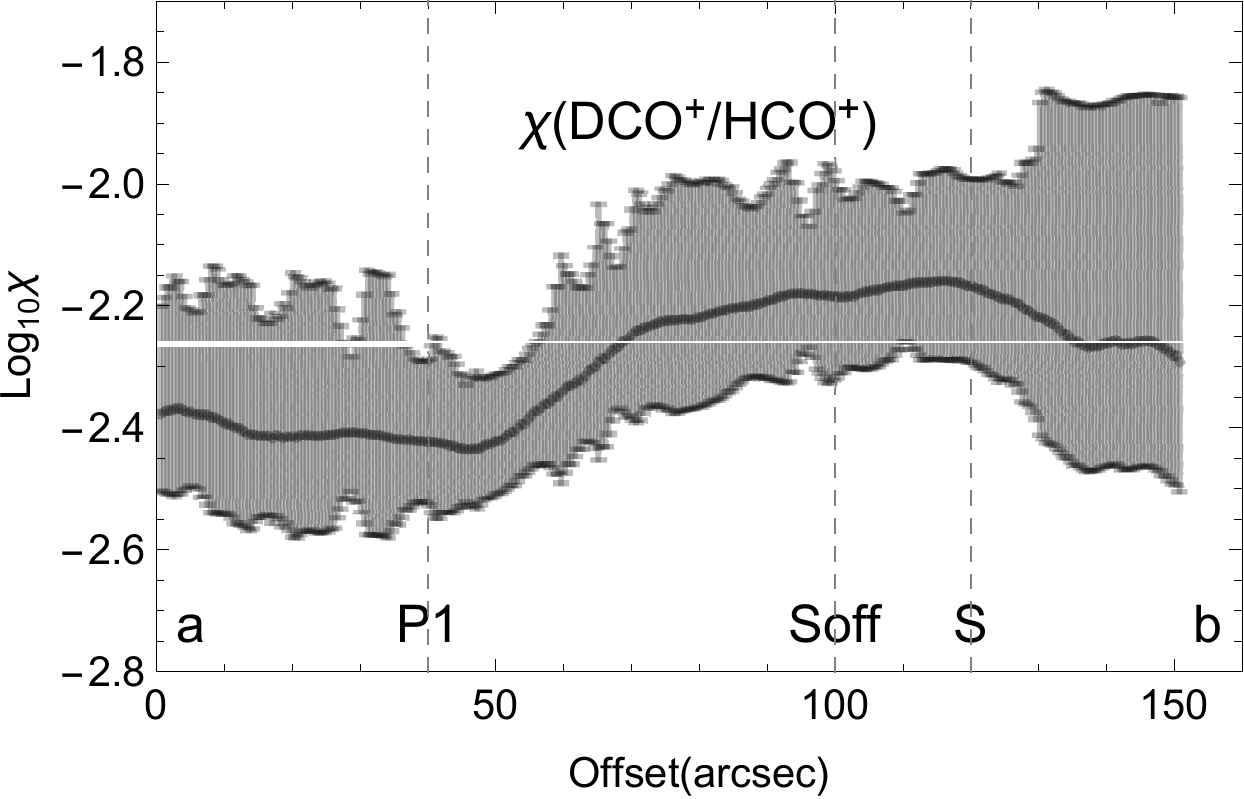}}
\end{minipage}
\begin{minipage}[c]{.45\textwidth}
\subfigure[]{\includegraphics[height=4cm] {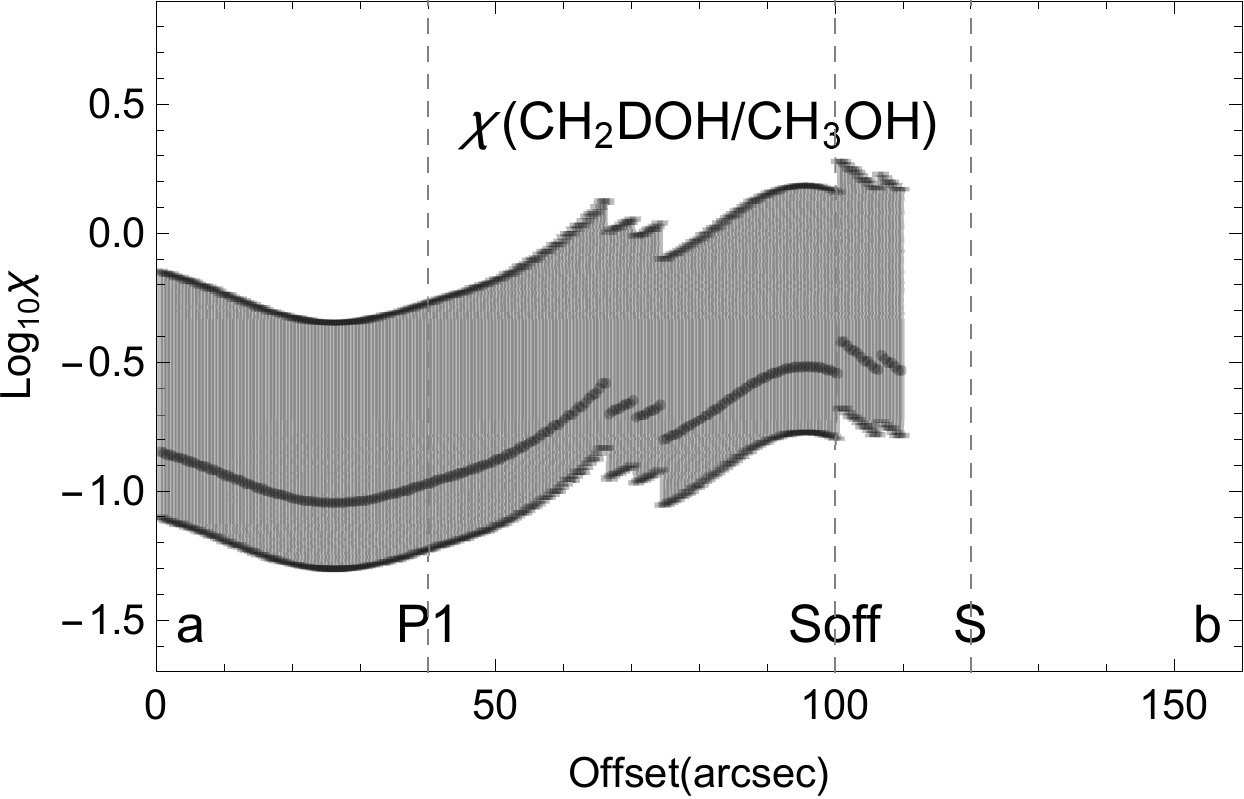}}
\end{minipage}

\end{center}
\caption{Profile of the relative abundance ratio of the deuterated species with respect to their hydrogenated isotopologues, extracted along the filamentary elongation from a to b in their maps (shown as the gray line in Figure~\ref{deutertionall}). 
The labeled positions are the same as those in Figure~\ref{dust}.  
The pixels where the lines used for calculation show $\rm <3\sigma$ integrated intensity are blanked.
}\label{deutertionallprofile}
\end{figure*}

\begin{figure*}
\begin{center}
\includegraphics[width=15cm] {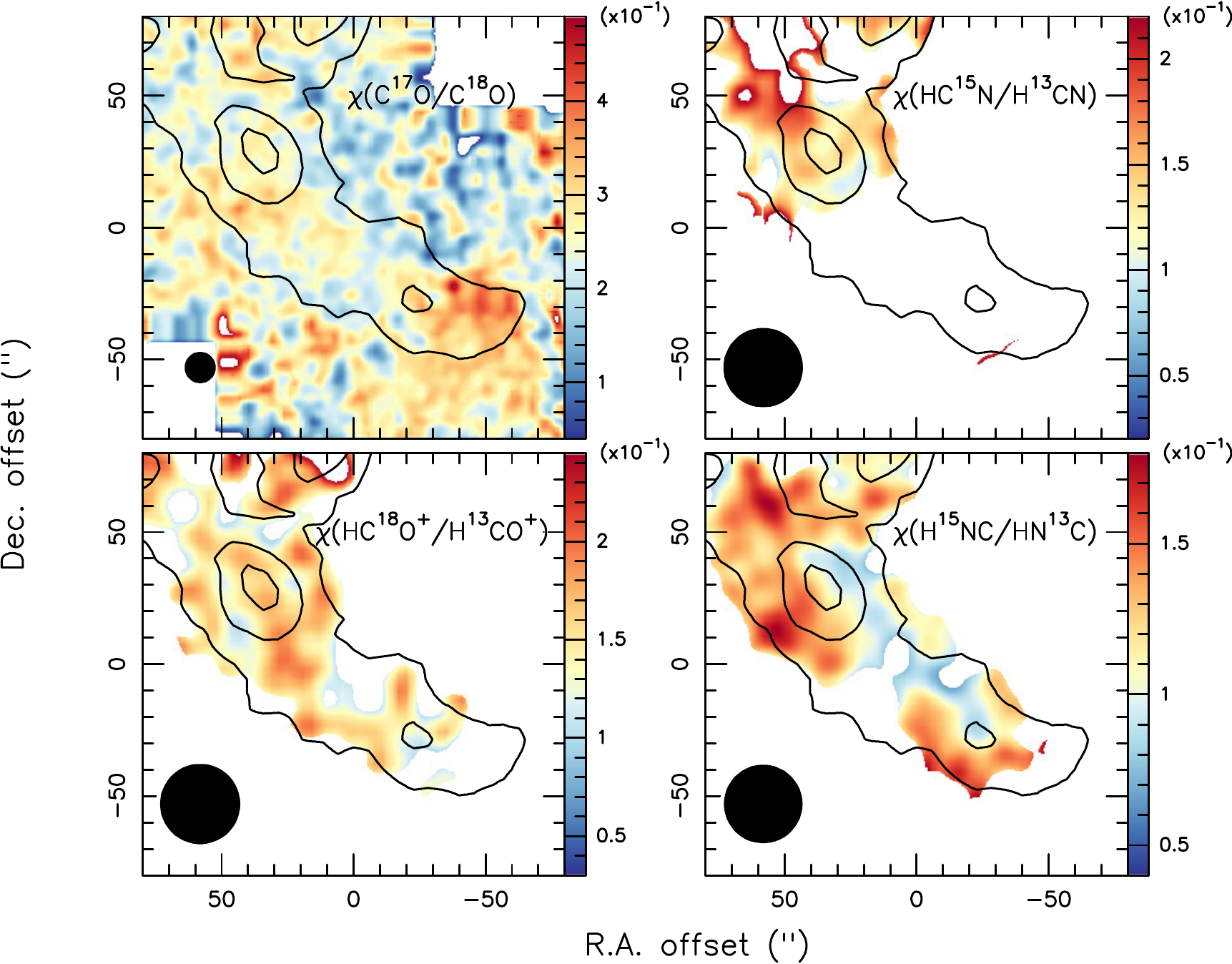}
\end{center}
\caption{Relative abundance ratio maps of the $\rm ^{15}N$ or $\rm ^{18}O$ isotopologues with respect to their  $\rm ^{13}C$ isotopologues.  The values derived by using the temperature map of $\rm T_{kin}$ (given here) or $\rm T_{\Delta J}$ are the same. 
The pixels where the $\rm HC^{15}N$\,(1--0), $\rm H^{15}NC$\,(1--0), or $\rm HC^{18}O^+$\,(1--0) line show $\rm<3\sigma$ integrated intensity are blanked. 
}\label{isoabundance}
\end{figure*}

Figure~\ref{h2cospec} and \ref{ch3ohspec} show the $\rm H_2CO$ lines and the $\rm CH_3OH$ lines we use to measure the gas temperature in G28.34\,P1--S. 
Figure~\ref{a-ch3ohlvg} takes the A-$\rm CH_3OH$ lines toward P1 as an example, and illustrates the probability density function (PDF) of parameters we derive using the LVG method. 
Figure~\ref{h2colte} shows the temperature profiles and column density profiles of $p$-$\rm H_2CO$ and $\rm CH_3OH$, extracted from the elongation and perpendicular directions of the G28.34\,P1--S filament.
Figure~\ref{deutertionallprofile}  shows the profiles of D-fraction for six species, extracted from the G28.34\,P1--S filament elongation.
Figure~\ref{isoabundance} gives the relative abundance ratios between the $\rm ^{18}O$ and $\rm ^{15}N$  isotopologues with respect to their $\rm ^{16}O$ and $\rm ^{14}N$ isotopologues.

\section{Tables}
Table~\ref{tab:linedeu} and \ref{tab:lineh2coch3oh} list the line profile fitting results  using GAUSS or HFS method toward P1 and S. 
Table~\ref{tab:column} lists the measured molecular column densities toward P1, S, and Soff.

\newpage

\iffalse

 \begin{figure*}
\begin{center}

\subfigure[]{\includegraphics[height=8.7cm,angle=0] {TkdV-P1-ph2co-marg.pdf}}
\subfigure[]{\includegraphics[height=8.7cm,angle=0] {TkdV-S-ph2co-marg.pdf}}
\subfigure[]{\includegraphics[height=8.7cm,angle=0] {TkdV-Soff-ph2co-marg.pdf}}
\end{center}
\caption{The probability density function (PDF) of $p$-$\rm H_2 CO$ parameters toward P1, S, and Soff, derived using RADEX and the MultiNest algorithm. The fitting parameters, including the $\rm H_2$ number density $n$, $p$-$\rm H_2CO$ column density N($p$-$\rm H_2 CO$),  the gas kinetic temperature $\rm T_{kin}$  are listed. 
}\label{h2colvg}
\end{figure*} 

\fi

\iffalse
  \begin{figure*}
\begin{center}

\subfigure[]{ \includegraphics[height=8.7cm] {TkdV-P1-ech3oh-marg.pdf}}
\subfigure[]{ \includegraphics[height=8.7cm] {TkdV-S-ech3oh-marg.pdf}}
\subfigure[]{ \includegraphics[height=8.7cm] {TkdV-Soff-ech3oh-marg.pdf}}
\end{center}
\caption{The probability density function (PDF) of E-$\rm CH_3OH$ parameters toward P1, S, and Soff, derived by using RADEX and MultiNest Algorithm. The fitting parameters, including the $\rm H_2$ number density $n$, $\rm CH_3OH$ column density N(E-$\rm CH_3OH$),  the gas kinetic temperature $\rm T_{kin}$  are listed. 
}\label{e-ch3ohlvg}
\end{figure*} 

\fi

%%%%%%%%%%%%%%%%%

%%%%%%%%%%%%%%%%%%%Table

\newpage

%\begin{landscape}

\begin{table*}
\caption{The best-fit parameters for lines in Figure \ref{velpro}, given by GILDAS.
}\label{tab:linedeu}
\scalebox{0.95}{
\begin{tabular}{c|c|p{2cm}p{2cm}|p{2cm}p{2cm}|p{2cm}}
\hline\hline
Line  &Freq       &\multicolumn{2}{c|}{P1}   &\multicolumn{2}{c|}{S}   &P1--S\\
\hline
\multicolumn{6}{c}{I. line width $\rm \Delta \upsilon\,(km~s^{-1})$ and integrated intensity $\rm \int T_B(\upsilon)d\upsilon\,(K\, km~s^{-1})$$^{a,d,f}$ } \\
\hline

&(GHz)     &$\rm (km~s^{-1})$    &$\rm (K\,km~s^{-1})$    &$\rm (km~s^{-1})$    &$\rm (K\,km~s^{-1})$   &$\rm [km~s^{-1},km~s^{-1}]$$^{g}$\\
 \hline
$\rm HC^{15}N\,(1-0)$ &86.055 &$\rm     3.3 \pm     0.2 $    &$\rm    0.18 \pm    0.01 $   %  &$\rm    79.4 \pm     0.1 $\n
&$\rm     3.0 \pm     0.4 $    &$\rm    0.08 \pm    0.01 $      &[76, 90]\\%  &$\rm    79.5 \pm     0.2 $\n
%$\rm  HNC \, (1-0) $ &  90.664   &$\rm     3.8 \pm     0.0 $    &$\rm    8.24 \pm    0.08 $   %  &$\rm    78.3 \pm     0.0 $\n
%&$\rm     2.9 \pm     0.0 $    &$\rm    4.77 \pm    0.06 $   &[74, 83]\\%  &$\rm    78.7 \pm     0.0 $\n
$\rm  HN^{13}C \, (1-0) $ &  87.091     &$\rm     3.6 \pm     0.0 $    &$\rm    1.51 \pm    0.01 $   %  &$\rm    79.5 \pm     0.0 $\n
&$\rm     3.0 \pm     0.0 $    &$\rm    0.99 \pm    0.01 $     &[74, 83] \\%  &$\rm    80.0 \pm     0.0 $\n
$\rm  H^{15}NC \, (1-0) $ &  88.866   &$\rm     3.3 \pm     0.2 $    &$\rm    0.18 \pm    0.01 $   %  &$\rm    79.4 \pm     0.1 $\n
&$\rm     2.9 \pm     0.2 $    &$\rm    0.11 \pm    0.01 $   &[74, 83] \\%  &$\rm    79.8 \pm     0.1 $\n
$\rm  DNC \, (1-0) $ &  76.306   &$\rm     3.6 \pm     0.1 $    &$\rm    0.55 \pm    0.01 $   %  &$\rm    79.7 \pm     0.0 $\n
&$\rm     3.1 \pm     0.1 $    &$\rm    0.49 \pm    0.01 $   &[74, 83] \\%  &$\rm    80.1 \pm     0.0 $\n
%$\rm  HCO^+ \, (1-0) $ &  89.189   &$\rm     3.6 \pm     0.1 $    &$\rm    8.56 \pm    0.20 $   %  &$\rm    77.3 \pm     0.0 $\n
%&$\rm     2.8 \pm     0.1 $    &$\rm    4.62 \pm    0.13 $  &[74, 83] \\%  &$\rm    77.8 \pm     0.0 $\n
$\rm  H^{13}CO^+ \, (1-0) $$^c$ &  86.754    &$\rm     3.7 \pm     0.0 $    &$\rm    1.76 \pm    0.01 $   %  &$\rm    79.2 \pm     0.0 $\n
&$\rm     3.2 \pm     0.0 $    &$\rm    1.26 \pm    0.01 $   &[74, 83] \\%  &$\rm    79.5 \pm     0.0 $\n
$\rm  HC^{18}O^+ \, (1-0) $ &  85.162     &$\rm     3.4 \pm     0.2 $    &$\rm    0.25 \pm    0.01 $  %  &$\rm    79.1 \pm     0.1 $\n
&$\rm     2.2 \pm     0.2 $    &$\rm    0.14 \pm    0.01 $     &[74, 83]\\%  &$\rm    79.8 \pm     0.1 $\n
$\rm  DCO^+ \, (1-0) $$^c$ &72.039  &$\rm     3.7 \pm     0.1 $    &$\rm    0.61 \pm    0.01 $   %  &$\rm    79.6 \pm     0.0 $\n
&$\rm     3.0 \pm     0.1 $    &$\rm    0.62 \pm    0.01 $      &[74, 83]\\%  &$\rm    79.7 \pm     0.0 $\n
$\rm  H^{13}CO^+ \, (2-1) $$^c$ &  173.507  &$\rm     2.9 \pm     0.1 $    &$\rm    1.26 \pm    0.03 $   %  &$\rm    79.1 \pm     0.0 $\n
&$\rm     3.0 \pm     0.2 $    &$\rm    0.69 \pm    0.04 $   &[74, 83]\\%  &$\rm    79.5 \pm     0.1 $\n
$\rm  CH_3OH \, (2_{0,2}-1_{0,1}A) $ &  96.741   &$\rm     3.9 \pm     0.1 $    &$\rm    8.07 \pm    0.17 $   %  &$\rm    79.0 \pm     0.0 $\n
&$\rm     3.8 \pm     0.0 $    &$\rm    2.95 \pm    0.02 $  &[77, 83] \\%  &$\rm    79.0 \pm     0.0 $\n
$\rm  CH_2DOH \, (2_{0,2}-1_{0,1}e0) $ &  89.408    &$\rm     4.0 \pm     0.6 $    &$\rm    0.10 \pm    0.01 $   %  &$\rm    79.0 \pm     0.2 $\n
&$\rm     3.6 \pm     0.7 $    &$\rm    0.06 \pm    0.01 $     &[77, 83] \\%  &$\rm    79.9 \pm     0.2 $\n

\hline
% \hline
%\end{tabular}
%}

%\vspace{0.5cm}
\multicolumn{6}{c}{II. line width $\rm \Delta \upsilon\,(km~s^{-1})$ and optical-depth-corrected integrated intensity $\rm A\tau_0\Delta\upsilon\,(K\,km~s^{-1})$ of the main hyperfine line$^{b,d,f}$} \\
\hline
   &$\rm (GHz)$$^{e}$   &$\rm (km~s^{-1})$   &$\rm (K\,km~s^{-1})$   &$\rm (km~s^{-1})$   &$\rm (K\,km~s^{-1})$  &$\rm [km~s^{-1},km~s^{-1}]$$^{g}$ \\
 \hline

$\rm  NH_3 \, (J,K=1,1) $ &  23.694 &$\rm     2.3 \pm     0.0 $    &$\rm   16.51 \pm    0.51 $  %  &$\rm    79.1 \pm     0.0 $     &$\rm     2.0    0.1 $\n
&$\rm     2.0 \pm     0.0 $    &$\rm   18.65 \pm    0.43 $   &[66, 88] \\%  &$\rm    79.6 \pm     0.0 $     &$\rm     3.2    0.1 $\n   
$\rm  NH_2D \, (J,K=1,0) $ &  85.926 &$\rm     2.5 \pm     0.3 $    &$\rm    0.08 \pm    0.02 $   %  &$\rm    79.5 \pm     0.1 $     &$\rm     0.4    0.0 $\n
 &$\rm     3.0 \pm     0.2 $    &$\rm    0.10 \pm    0.01 $   &[66, 88]\\%  &$\rm    80.0 \pm     0.0 $     &$\rm     0.9    0.2 $\n
%$\rm  HCN \, (1-0) $ &  88.632 &$\rm     2.9 \pm     0.1$$^f$    &$\rm    1.03 \pm    0.05$$^f$   %  &$\rm    77.7 \pm     0.0 $     &$\rm     0.1 0.0 $\n
%&$\rm     2.6 \pm     0.1$$^f$    &$\rm    0.88 \pm    0.02$$^f$   &[66, 88]\\%  &$\rm    78.3 \pm     0.0 $     &$\rm     0.1 0.0 $\n
$\rm  H^{13}CN \, (1-0) $ &  86.340 &$\rm     4.1 \pm     0.1 $    &$\rm    0.24 \pm    0.00 $   %  &$\rm    79.1 \pm     0.0 $     &$\rm     0.1 0.0 $\n
&$\rm     3.2 \pm     0.1 $    &$\rm    0.14 \pm    0.00 $    &[66, 88] \\%  &$\rm    79.8 \pm     0.0 $     &$\rm     0.1 0.0 $\n
$\rm  H^{13}CN \, (2-1) $ &  172.678 &$\rm     5.6 \pm     0.4 $    &$\rm    0.10 \pm    0.01 $   %  &$\rm    78.9 \pm     0.1 $     &$\rm     0.1 0.0 $\n
&$\rm     3.4 \pm     0.6 $    &$\rm    0.05 \pm    0.01 $   &[66, 88] \\%  &$\rm    79.3 \pm     0.0 $     &$\rm     0.1 0.0 $\n
$\rm  DCN \, (1-0) $ &  72.415 &$\rm     3.4 \pm     0.2 $    &$\rm    0.06 \pm    0.00 $   %  &$\rm    79.5 \pm     0.1 $     &$\rm     0.2    0.1 $\n
&$\rm     2.4 \pm     0.3 $    &$\rm    0.06 \pm    0.02 $    &[66, 88] \\%  &$\rm    79.8 \pm     0.0 $     &$\rm     0.8    0.1 $\n
$\rm  N_2H^+ \, (1-0) $ &  93.174 &$\rm     3.6 \pm     0.0 $    &$\rm    1.24 \pm    0.00 $   %  &$\rm    79.0 \pm     0.0 $     &$\rm     0.1 0.0 $\n
&$\rm     3.0 \pm     0.0 $    &$\rm    0.86 \pm    0.00 $     &[68, 91]\\%  &$\rm    79.4 \pm     0.0 $     &$\rm     0.3    0.0 $\n
$\rm  N_2D^+ \, (1-0) $ &  77.110 &$\rm     2.4 \pm     0.2 $    &$\rm    0.05 \pm    0.01 $   %  &$\rm    79.6 \pm     0.1 $     &$\rm     0.5    0.2 $\n
&$\rm     2.7 \pm     0.2 $    &$\rm    0.03 \pm    0.00 $    &[68, 91]\\%  &$\rm    80.0 \pm     0.0 $     &$\rm     0.1 0.0 $\n
$\rm  N_2D^+ \, (2-1) $ &  154.217 &$\rm     2.7 \pm     0.6 $    &$\rm    0.03 \pm    0.01 $   %  &$\rm    80.0 \pm     0.2 $     &$\rm     0.5    0.4 $\n
&$\rm     2.3 \pm     0.3 $    &$\rm    0.04 \pm    0.01 $   &[68, 91] \\%  &$\rm    79.7 \pm     0.0 $     &$\rm     0.5    0.3 $\n
$\rm  H^{13}CO^+ \, (1-0) $$^c$ & 86.754    &$\rm     3.5 \pm     0.1 $    &$\rm    0.21 \pm    0.00 $   %  &$\rm    79.2 \pm     0.0 $     &$\rm     0.1 0.0 $\n
&$\rm     2.8 \pm     0.1 $    &$\rm    0.25 \pm    0.02 $    &[74, 83]\\%  &$\rm    79.5 \pm     0.0 $     &$\rm     0.5    0.1 $\n
$\rm  DCO^+ \, (1-0) $$^c$ &72.039  &$\rm     2.3 \pm     0.1 $    &$\rm    0.51 \pm    0.12 $   %  &$\rm    79.6 \pm     0.0 $     &$\rm     3.6    0.9 $\n
&$\rm     2.5 \pm     0.2 $    &$\rm    0.20 \pm    0.05 $    &[74, 83]\\%  &$\rm    79.7 \pm     0.0 $     &$\rm     0.9    0.4 $\n
$\rm  H^{13}CO^+ \, (2-1) $$^c$ &  173.506 &$\rm     2.8 \pm     0.1 $    &$\rm    0.16 \pm    0.00 $   %  &$\rm    79.1 \pm     0.0 $     &$\rm     0.1 0.0 $\n
&$\rm     1.9 \pm     0.1 $    &$\rm    0.43 \pm    0.12 $   &[74, 83]\\%  &$\rm    79.7 \pm     0.0 $     &$\rm     2.2    0.6 $\n

\hline
 \hline

  \multicolumn{7}{l}{{\bf Note.} $a$. Subtable I lists the parameters of a single unblended line or the strongest line when several lines at different transitions}\\
        \multicolumn{7}{l}{~~~~~~~~~~~~~~ are not completely resolved by our observations. Lines are fitted using the GAUSS method.}\\
      \multicolumn{7}{l}{~~~~~~~~~~$b$.  Subtable II lists  the parameters of the main line taking into account the hyperfine structure.  Lines are fitted using the HFS method.}\\
      \multicolumn{7}{l}{~~~~~~~~~~~~~~ 
The excitation temperature of the hyperfine splitting $\rm T_{\Delta F}$ can be derived from $\rm A\tau_0$ and $\rm \tau_0$, as $\rm T_{\Delta F}=\frac{h\nu/k_B}{ln[\frac{h\nu/k_B}{{\it A\tau_0/(\tau_0f)+J_\nu}(T_{bg})}+1]}$.}.\\
      \multicolumn{7}{l}{~~~~~~~~~~$c$.  Hyperfine splittings are recorded in JPL or CDMS but are not resolved in observations. Line is fitted by using both GAUSS and HFS.}\\
    \multicolumn{7}{l}{~~~~~~~~~~$d$.  Lines are extracted from images by averaging a beam-sized region centered at 870\,$\mu$m continuum peak P1 or S  in the plane of the sky. }\\
  \multicolumn{7}{l}{~~~~~~~~~~~~~~ All  line images have the same pixel size, but their angular resolutions were the same as in the observations without smoothing (Table \ref{tab:dehyd}).}\\
      \multicolumn{7}{l}{~~~~~~~~~~$e$.  The rest frequency is given from  the main line of the hyperfine splittings.}\\
   \multicolumn{7}{l}{~~~~~~~~~~$f$. Uncertainties on the measured intensities are typically $\le10\%$.}\\
% \multicolumn{7}{l}{~~~~~~~~~~$f$. Hyperfine splitting of HCN\,(1--0) is not well fitted by one velocity component,}\\
%  \multicolumn{7}{l}{~~~~~~~~~~~~~~ see the discussion in \citet[][]{feng16a} and \citet{beuther07c};}\\
 \multicolumn{7}{l}{~~~~~~~~~~$g$. This column lists the velocity range we integrate for individual lines to obtain their intensity maps in Figure~\ref{molint}.}
\end{tabular}
}
\end{table*}

%\end{landscape}

\newpage

%\begin{landscape}

\begin{table*}
\caption{The best-fit parameters for $\rm H_2CO$ lines in Figure \ref{h2cospec} and for $\rm CH_3OH$ lines in Figure \ref{ch3ohspec}, given by GILDAS}\label{tab:lineh2coch3oh}
\scalebox{0.95}{
\begin{tabular}{c|c|p{2cm}p{2.8cm}|p{2cm}p{2.8cm}|p{2cm}}
\hline\hline
Line$^{a,b}$   &Freq       &\multicolumn{2}{c|}{P1}   &\multicolumn{2}{c|}{S}   &P1--S\\
\hline

&(GHz)     &$\rm \Delta \upsilon~ (km~s^{-1})$$^c$     &$\rm \int T_B(\upsilon)d\upsilon~ (K\, km~s^{-1})$$^c$     &$\rm \Delta \upsilon~ (km~s^{-1})$$^{c,d}$     &$\rm \int T_B(\upsilon)d\upsilon~ (K\, km~s^{-1})$$^{c,d}$   &$\rm [km~s^{-1},km~s^{-1}]$$^{e,f,i}$ \\
 \hline
$\rm  H_2CO \, (1_{0,1}-0_{0,0}) $ &  72.838 &$\rm     4.8 \pm     0.1 $    &$\rm    5.25 \pm    0.05 $   %  &$\rm    79.1 \pm     0.0 $\n
&$\rm     3.6 \pm     0.0 $    &$\rm    2.25 \pm    0.02 $    &[74, 83]\\%  &$\rm    79.4 \pm     0.0 $\n
$\rm  H_2CO \, (3_{0,3}-2_{0,2}) $ &  218.222 &$\rm     4.2 \pm     0.1 $    &$\rm    1.84 \pm    0.03 $   %  &$\rm    79.3 \pm     0.0 $\n
&$\rm     3.2 \pm     0.2 $    &$\rm    0.53 \pm    0.02 $  &[74, 83] \\%  &$\rm    79.7 \pm     0.1 $\n
$\rm  H_2CO \, (3_{2,2}-2_{2,1})  $ &  218.476 &$\rm     3.5 \pm     0.4 $    &$\rm    0.29 \pm    0.03 $  %  &$\rm    79.3 \pm     0.1 $\n
&$\rm --  $     &$\rm \sigma= 0.09 $ &[74, 83] \\%   NAN NAN NAN NAN \n
$\rm  H_2CO \, (3_{2,1}-2_{2,0}) $ &  218.760 &$\rm     7.2 \pm     1.0 $    &$\rm    0.47 \pm    0.04 $  %  &$\rm    79.5 \pm     0.3 $\n
&$\rm --  $     &$\rm \sigma= 0.06 $  &[74, 83]\\%   NAN NAN NAN NAN \n
\hline

$\rm  CH_3OH \, (5_{0,5}-4_{1,3}E) $ &  76.510 &$\rm     3.4 \pm     1.6 $    &$\rm    0.07 \pm    0.02 $   %  &$\rm    79.0 \pm     0.0 $\n
&$\rm --  $     &$\rm \sigma= 0.01 $   &$\rm --  $$^f$\\%   NAN NAN NAN NAN \n
$\rm  CH_3OH \, (5_{-1,5}-4_{0,4}E) $ &  84.521 &$\rm     4.5 \pm     0.1 $    &$\rm    2.56 \pm    0.05 $   %  &$\rm    79.4 \pm     0.0 $\n
&$\rm     4.7 \pm     0.3 $    &$\rm    0.39 \pm    0.02 $   &[74, 84]\\%  &$\rm    79.3 \pm     0.1 $\n
$\rm  ^{13}CH_3OH \, (2_{-1,2}-1_{-1,1}E) $ &  94.405 &$\rm     5.3 \pm     0.2 $    &$\rm    0.30 \pm    0.01 $   %  &$\rm    79.0 \pm     0.0 $\n
&$\rm --  $     &$\rm \sigma=0 .11 $  &$\rm --  $$^f$\\%   NAN NAN NAN NAN \n
$\rm  ^{13}CH_3OH \, (2_{0,2}-1_{0,1}A) $ &  94.407 &$\rm     3.7 \pm     0.1 $    &$\rm    0.24 \pm    0.00 $   %  &$\rm    73.1 \pm     0.1 $\n
&$\rm --  $     &$\rm \sigma=0 .11 $   &$\rm --  $$^f$\\%   NAN NAN NAN NAN \n
$\rm  CH_3OH \, (8_{0,8}-7_{1,7}A) $ &  95.169 &$\rm     3.3 \pm     0.2 $    &$\rm    0.38 \pm    0.01 $   %  &$\rm    79.2 \pm     0.0 $\n
&$\rm     4.1 \pm     0.7 $    &$\rm    0.06 \pm    0.01 $    &[74, 84]\\%  &$\rm    79.1 \pm     0.3 $\n
$\rm  CH_3OH \, (2_{1,2}-1_{1,1}A) $ &  95.914 &$\rm     4.5 \pm     0.2 $    &$\rm    0.39 \pm    0.01 $  %  &$\rm    79.6 \pm     0.1 $\n
&$\rm     4.0 \pm     0.6 $    &$\rm    0.07 \pm    0.01 $    &[74, 84]\\%  &$\rm    79.6 \pm     0.3 $\n

$\rm  CH_3OH \, (2_{-1,2}-1_{-1,1}E) $ &  96.739 &$\rm     6.8 \pm     0.6 $    &$\rm    2.28 \pm    0.35 $    %  &$\rm    70.5 \pm     0.2 $\n
&$\rm     3.9 \pm     0.3 $    &$\rm    0.27 \pm    0.01 $  &[82, 95]$^{g}$ \\%  &$\rm    72.0 \pm     0.6 $\n
$\rm  CH_3OH \, (2_{0,2}-1_{0,1}A) $ &  96.741 &$\rm     4.0 \pm     0.6 $    &$\rm    7.08 \pm    0.35 $    %  &$\rm    79.0 \pm     0.0 $\n
&$\rm     3.8 \pm     0.0 $    &$\rm    2.56 \pm    0.01 $    &[73, 82]$^{g}$\\%  &$\rm    79.0 \pm     0.0 $\n
$\rm  CH_3OH \, (2_{0,2}-1_{0,1}E) $ &  96.745 &$\rm     5.9 \pm     0.6 $   &$\rm   8.74 \pm    0.35 $    %  &$\rm    85.3 \pm     0.1 $\n
&$\rm     3.6 \pm     0.0 $   &$\rm   2.09 \pm    0.02 $    &[65, 73]$^{g}$\\%  &$\rm    85.7 \pm     0.6 $\n
$\rm  CH_3OH \, (4_{0,4}-4_{-1,4}E) $ &  157.246 &$\rm     4.1 \pm     0.5 $    &$\rm    0.16 \pm    0.01 $   %  &$\rm    79.0 \pm     0.2 $\n
&$\rm --  $     &$\rm \sigma= 0.04 $   &$\rm --  $\\%   NAN NAN NAN NAN \n
$\rm  CH_3OH \, (3_{0,3}-3_{-1,3}E) $ &  157.270 &$\rm     4.8 \pm     0.1 $    &$\rm    0.89 \pm    0.02 $   %  &$\rm    79.0 \pm     0.0 $\n
&$\rm     2.9 \pm     0.2 $    &$\rm    0.25 \pm    0.02 $    &[77, 88]$^{h}$\\%  &$\rm    79.0 \pm     0.0 $\n
$\rm  CH_3OH \, (1_{0,1}-1_{-1,1}E) $ &  157.271 &$\rm     4.7 \pm     0.3 $    &$\rm    0.43 \pm    0.02 $   %  &$\rm    69.7 \pm     0.1 $\n
&$\rm    5.71 \pm    3.98 $    &$\rm    0.43 \pm    0.00 $   &[72, 77]$^{h}$\\%  &$\rm    75.8 \pm     0.4 $\n
$\rm  CH_3OH \, (2_{0,2}-2_{-1,2}E) $ &  157.276 &$\rm     8.3 \pm     0.8 $   &$\rm   0.43 \pm    0.02 $   %  &$\rm    80.6 \pm     0.3 $\n
&$\rm    5.6 \pm     2.1 $   &$\rm   0.05 \pm    0.00 $   &[64, 72]$^{h}$ \\%  &$\rm    94.4 \pm     0.4 $\n
$\rm  CH_3OH \, (3_{2,1}-2_{1,1}E) $ &  170.061 &$\rm     4.1 \pm     0.1 $    &$\rm    0.85 \pm    0.02 $    %  &$\rm    79.3 \pm     0.0 $\n
&$\rm     2.3 \pm     0.5 $    &$\rm    0.10 \pm    0.02 $    &[74, 84]\\%  &$\rm    79.7 \pm     0.1 $\n
$\rm  CH_3OH \, (4_{2,2}-3_{1,2}E) $ &  218.440 &$\rm     3.3 \pm     0.2 $    &$\rm    0.56 \pm    0.03 $   %  &$\rm    79.2 \pm     0.1 $\n
&$\rm     5.0 \pm     1.0 $    &$\rm    0.19 \pm    0.03 $    &[74, 84]\\%  &$\rm    79.7 \pm     0.4 $\n

\hline
 \hline

  \multicolumn{7}{l}{{\bf Note.} $a$.  We list the parameters given from the GAUSS fit, by assuming one velocity component in the line of sight.}\\
    \multicolumn{7}{l}{~~~~~~~~~~$b$.  Lines are extracted from images by averaging a beam-sized region centered at P1 or S  in the plane of the sky (Figure~\ref{h2cospec}--\ref{ch3ohspec}). }\\
  \multicolumn{7}{l}{~~~~~~~~~~~~~~  All  line images from the same species have the same pixel size and angular resolution  (35.6\arcsec for $\rm H_2CO$ and 33.9\arcsec for $\rm CH_3OH$).}\\
   \multicolumn{7}{l}{~~~~~~~~~~$c$. Uncertainties on the measured intensities are typically $\le10\%$.}\\
  \multicolumn{7}{l}{~~~~~~~~~~$d$. For line  with $\rm <3\sigma$ emission, ``$--$'' is given in the line width, and a $\rm \sigma$ rms is given.}\\
  \multicolumn{7}{l}{~~~~~~~~~~$e$. This column lists the velocity range we integrate for individual lines in order to estimate the molecular rotation temperature}\\
   \multicolumn{7}{l}{~~~~~~~~~~~~~and column density using both RD and LVG fits.}\\
       \multicolumn{7}{l}{~~~~~~~~~~$f$. The line shows emission with $\rm S/N<4$ toward S, so we do not use it for the RD or LVG fits.}\\
    \multicolumn{7}{l}{~~~~~~~~~~$g$.  At the rest frequency of 96.741\,GHz, three transitions are blended in the velocity range of $\rm 65\sim90\,km\,s^{-1}$. In the velocity range}\\
    \multicolumn{7}{l}{~~~~~~~~~~~~~  where line wings of two transitions overlap, we assume that the transition with its line center close to  the $\rm V_{lsr}$ dominates the }\\
    \multicolumn{7}{l}{~~~~~~~~~~~~~integrated intensity. The velocity range over which we integrate each line is given with respect to the rest frequency of 96.741\,GHz.}\\
      \multicolumn{7}{l}{~~~~~~~~~~$h$.  At the rest frequency of 157.270\,GHz, three transitions are blended in the velocity range of $\rm 70\sim80\,km\,s^{-1}$. In the velocity range}\\
    \multicolumn{7}{l}{~~~~~~~~~~~~~where line wings of two transitions overlap, we assume that the transition with its line center close to the $\rm V_{lsr}$ dominates the}\\
    \multicolumn{7}{l}{~~~~~~~~~~~~~  integrated intensity. The velocity range over which we integrate each line is given with respect to the rest frequency of  157.270\,GHz.}\\
 \multicolumn{7}{l}{~~~~~~~~~~$i$. The velocity range we select  adds  20\% uncertainty to the integrated intensity, but this is within the uncertainty from RD and LVG fits.}     
\end{tabular}
}
\end{table*}

%%%%%%%%%%%%%%%%%%%
\newpage

\begin{table*}
\caption{Molecular column density toward P1, S, and Soff}\label{tab:column}
\centering
\scalebox{1}{
\begin{tabular}{c|p{2cm}p{2cm}p{2cm}}
\hline\hline
Mol.         &P1   &S   &Soff\\
                 &$\rm (cm^{-2})$$^{a}$   &$\rm (cm^{-2})$$^a$  &$\rm (cm^{-2})$$^a$   \\
 \hline
 $\rm  H^{13}CN $$^b$ &$\rm  6.2\pm 1.4(12)$   &$\rm  2.0\pm 0.7(12) $   &$\rm  1.9\pm 0.7(12) $  \\
 $\rm  HC^{15}N $$^b$ &$\rm 8.6\pm 3.7(11) $  &$--$   &$--$\\
 DCN$^b$ &$\rm  1.8\pm 0.6(12) $  &$\rm  0.9\pm 0.4(12) $  &$\rm  1.1\pm 0.4(12) $\\
 \hline
  $\rm  HN^{13}C $$^b$ &$\rm  5.8\pm 1.2(12) $  &$\rm  3.1\pm 0.8(12)$  &$\rm  3.6\pm 0.9(12) $\\
 $\rm  H^{15}NC $$^b$ &$\rm  7.0\pm 2.2(11) $  &$\rm  3.3\pm 1.5(11) $  &$\rm  4.4\pm 1.6(11)$  \\
 DNC$^b$   &$\rm  2.1\pm 0.6(12)$   &$\rm  1.8\pm 0.5(12)$  &$\rm  2.1\pm 0.6(12) $ \\
 \hline
$\rm  H^{13}CO^+ $$^b$  &$\rm  3.2\pm 0.7(12)$  &$\rm  1.9\pm 0.4(12) $   &$\rm  1.7\pm 0.4(12) $     \\ 
$\rm  HC^{18}O^+ $$^b$   &$\rm  5.8\pm 1.6(11) $ &$\rm  2.5\pm 1.0(11) $ &$\rm  2.4\pm 1.0(11) $  \\
$\rm  DCO^+ $$^b$  &$\rm  5.2\pm 1.3(11) $   &$\rm  5.5\pm 1.4(11) $  &$\rm  4.8\pm 1.2(11)$\\
\hline
$\rm  N_2H^+ $$^b$ &$\rm  3.4\pm 0.7(13) $ &$\rm  1.5\pm 0.4(13)$ &$\rm  1.7\pm 0.4(13)$ \\
$\rm  N_2D^+ $$^b$ &$\rm  0.7\pm 0.3(12) $  &$\rm  0.9\pm 0.3(12)$  &$\rm  1.0\pm 0.3(12) $  \\
\hline
$\rm  NH_3 $$^c$ &$\rm  3.9\pm 2.6(15) $  &$\rm  5.2\pm 4.2(15) $ &$\rm  7.0\pm 5.6(15)$ \\
$\rm  NH_2D $$^d$  &$\rm  2.7\pm 1.1(13)$ &$\rm  2.2\pm 1.4(13)$ &$\rm  2.1\pm 1.4(13) $ \\
\hline
$\rm  CH_3OH $$^e$  &$\rm 2.8\pm 2.1(14)$  &$\rm 0.6\pm 0.4(14) $  &$\rm 0.6\pm 0.4(14) $\\
$\rm  CH_2DOH $$^e$   &$\rm 2.9\pm 1.7(13) $   &$--$  &$\rm 2.0\pm 1.0(13) $  \\
\hline
 \hline

  \multicolumn{4}{l}{{\bf Note.} $a$. Molecular column density is derived within a region with a radius of 5\arcsec}\\
   \multicolumn{4}{l}{~~~~~~~~~~~~  centered on P1, S, Soff, in the form of $\rm x\pm y (z)=(x\pm y)\times 10^z\,cm^{-2}$;}\\
      \multicolumn{4}{l}{~~~~~~~~~~~~  ``$--$" stands for the intensity of $\rm S/N<3$.}\\
  \multicolumn{4}{l}{~~~~~~~~~ $b$. Using $\rm T_{kin}$ derived from $p$-$\rm NH_3$\,(1,1) and (2,2) lines;}\\
    \multicolumn{4}{l}{~~~~~~~~~ $c$. Assuming OPR of $\rm  NH_3 $ is 1;}\\
       \multicolumn{4}{l}{~~~~~~~~~ $d$. Assuming OPR of $\rm  NH_2D $ is 3;}\\
  \multicolumn{4}{l}{~~~~~~~~~ $e$. Using $\rm T_{rot}$ derived from the 
  rotational diagram of $\rm CH_3OH$ lines.}

\end{tabular}
}
\end{table*}

\end{document}